\documentclass[aps,prd,floatfix,10pt,nofootinbib,showpacs,notitlepage]{revtex4-1}
\usepackage{amsmath}
\usepackage{amsfonts}
\usepackage[colorlinks=true,linkcolor=blue,citecolor=blue,urlcolor=blue]{hyperref}
\usepackage{amssymb}
\usepackage{graphicx}
\usepackage{enumerate}
\usepackage{csquotes}
\usepackage[caption=false]{subfig}
\usepackage{dsfont}
\usepackage{appendix}
\usepackage{color}
\usepackage{bbold}
\usepackage{mathtools}
\usepackage{tensor}

\newcommand{\be}{\begin{eqnarray}} 
\newcommand{\ee}{\end{eqnarray}}
\newcommand{\om}{\ensuremath{\omega}}

\newcommand{\eq}[1]{Eq.~(\ref{#1})}

\newcommand{\beq}{\begin{equation}}
\newcommand{\eeq}{\end{equation}}
\newcommand{\bsub}{\begin{subequations}}
\newcommand{\esub}{\end{subequations}}
\newcommand{\bi} {\begin{itemize}}
\newcommand{\ei} {\end{itemize}}
\newcommand{\ben} {\begin{enumerate}}
\newcommand{\een} {\end{enumerate}}
\newcommand{\bmat} {\begin{pmatrix}}
\newcommand{\emat} {\end{pmatrix}} 
\newcommand{\bal} {\begin{aligned}}
\newcommand{\eal} {\end{aligned}}
\newcommand{\btab}{\begin{tabular}}
\newcommand{\etab}{\end{tabular}}

\begin{document}

\title{Hawking radiation in the presence of high-momentum dissipation} 

\author{Scott Robertson}
\email{scott.robertson@th.u-psud.fr}
\affiliation{Laboratoire de Physique Th\'eorique, CNRS UMR 8627, B\^atiment 210, Universit\'e Paris-Sud 11, 91405 Orsay Cedex, France}
\author{Renaud Parentani}
\email{renaud.parentani@th.u-psud.fr}
\affiliation{Laboratoire de Physique Th\'eorique, CNRS UMR 8627, B\^atiment 210, Universit\'e Paris-Sud 11, 91405 Orsay Cedex, France}

\date{\today}

\begin{abstract}
We study the Hawking radiation in field theories which break Lorentz invariance via dissipative effects above a certain energy scale. 
We assume that the additional degrees of freedom which cause dissipation are Gaussian and freely falling.
The asymptotic spectrum and the correlations are extracted from the anticommutator of the radiation field.
The singular behavior of the Green function found for relativistic fields as one point crosses the horizon is completely suppressed by dissipation.
Yet, when the dissipative frequency scale is much larger than the surface gravity of the black hole, we show that the asymptotic observables acquire their standard (relativistic) vacuum expectation values.
We explicitly compute the effects of dissipation on the spectrum and on the nonseparable character of the correlations when varying the dissipative scale, the extension of the near-horizon geometry, and the temperature of the environment. 
\end{abstract}

\pacs{04.62.+v, 04.70.Dy, 05.70.Ln, 42.50.Lc}  

\maketitle


\section{Introduction}

Analogue black hole radiation~\cite{Unruh-1981,Unruh-1995} is on the verge of experimental verification, with an abundance of media now being studied~\cite{LivingReview}.
These analogue systems mimic Lorentz invariance over large distances, while providing a natural mechanism for the breaking of Lorentz invariance on microscopic scales.
They can thus also be used to model and test possible modifications of phenomena sensitive to ultrahigh-energy physics and quantum gravity effects, see e.g.~\cite{Jacobsonreview,Liberati-dissip,MichelRP-Horava}. 
High-frequency dispersion -- that is, a frequency-dependent modification of the wave group velocity -- has been well studied theoretically~\cite{Macher-Parentani-2009,Robertson-2012,Finazzi-Parentani-2012,Coutant-RP-SF}. 
On the one hand, the robustness of Hawking radiation has been confirmed in this scenario when the dispersive frequency scale $\Lambda$ is much larger than the surface gravity $\kappa$ determining the black hole temperature.
On the other hand, the properties of the spectral deviations due to high-frequency dispersion have now been computed. 
We can thus expect that forthcoming experiments will observe the modified spectrum rather than the relativistic one computed by Hawking. 
In fact, the recent observations made in a water tank~\cite{Rousseaux-et-al-2008,Weinfurtner-et-al-2011} and in an atomic BEC~\cite{Jeff-nature-physics2014} can only be explained by taking dispersive effects into account~\cite{MichelRP2013,Finazzi-Parentani-2010,MichelRP2015}.
 
As yet, however, the related phenomenon of dissipation -- the loss of energy to an environment -- is much less understood, even though many systems, such as polaritons~\cite{Gerace-Carusotto-2012,prl-Marcoussis}, water waves~\cite{Rousseaux-et-al-2008,Weinfurtner-et-al-2011} and sound waves~\cite{Y.Auregan-slowsound,Y.Auregan2}, all display dissipation.~\footnote{It has also been proposed that quantum gravity could give rise to dissipative effects in the near vicinity of a black hole horizon~\cite{Parentani-2001}. We conjecture that the chaotic behavior recently discussed in~\cite{Polchinski-2015} should effectively engender dissipation which will damp 2-point functions in the near-horizon region in a manner similar to that studied in the present work.}
It is thus imperative to study theoretically the impact of dissipation on power spectra and correlations. 
It is not clear that we can infer robustness under dissipation from robustness under dispersion, since dissipation alters the past propagation of the outgoing modes even more drastically, and we might therefore expect that the deviations due to dissipation could differ considerably from those due to dispersion.
Indeed, in a recent theoretical study of a modulated Dynamical Casimir Effect~\cite{Busch-Parentani-Robertson-2014}, it was found that a relatively small rate of dissipation can have a significant effect on the properties of the emitted radiation, in particular on the entanglement of the pairs produced, i.e., on the nonseparable character of their state, which is the property that allows us to distinguish the stimulated (classical) effect from the spontaneous effect due to vacuum fluctuations~\cite{Zapata,Finazzi-Carusotto-2014,Busch-Parentani-2014}. 

In this paper, our first aim is to provide a general formalism which could be adapted so as to properly describe the required specific aspects of any particular system. 
To preserve unitarity of the system as a whole, dissipation is introduced through coupling to an environment.  Its effects are twofold: energy is lost to the environment, but also gained from the environment through stimulation.  
Using a simple model for the environment,which allows us to choose how the dissipative rate varies with momentum, we shall analyze the propagation of a dissipative field in an analogue black hole flow.

Our second aim is to study the spectrum of the emitted radiation and the strength of the long-distance correlations in order to identify the relevant parameters which govern the deviations due to dissipation. 
As expected, the UV dissipative scale $\Lambda_{\rm diss}$ and the temperature of the environment are both important. More surprisingly, the parameter which fixes the spatial extension of the near-horizon geometry also plays a crucial role.
When establishing this, we shall flesh out a prediction of \cite{Adamek-Busch-Parentani-2013}, that the effects caused by dissipation in the UV sector should be similar to those in de Sitter space since the near-horizon region can be mapped onto a portion of de Sitter space, with the surface gravity $\kappa$ identified with the Hubble constant $H$.

The layout of this paper is as follows.
In Section \ref{sec:Settings} we describe the general settings of the system we are considering, through the action and the resulting equations of motion.
In Section \ref{sec:Modes}, we construct the retarded Green function for the radiation field in a stationary inhomogeneous situation.
In Section \ref{sec:State}, we calculate the anticommutator which encodes the state of the field, i.e., the spectrum and correlations between emitted quasiparticles.
We then conclude with Section \ref{sec:Conclusions}.
In appendices, we detail the calculations presented in the main body of the text.
In particular, in Appendix~\ref{app:Gret}, we give the general structure of the stationary retarded Green function when the effective dispersion relation is a polynomial in the wave vector. 


\section{Settings
\label{sec:Settings}}

In this section, we describe the structure of the models we are considering, as well as the assumptions and simplifications we make for ease of calculation.
We shall follow the procedure and philosophy outlined in \cite{Parentani-2007}.
In particular, the specifics of the environment and its coupling to the field do not matter so much as their desired {\it effects} once the environment degrees of freedom have been traced over.
We therefore choose the simplest possible system compatible with our desire for a quantum field which exhibits dissipative effects at large momenta, and which thus reproduces approximate Lorentz invariance in the infrared regime.
We shall mainly work with a dissipative rate which grows like the square of the spatial momentum, $\Gamma \sim P^2/\Lambda_{\rm diss}$. 
The dissipative UV scale $\Lambda_{\rm diss}$ will be treated as an adjustable parameter to see how increasing dissipation, i.e., reducing $\Lambda_{\rm diss}$, affects the observables. 


\subsection{Metric and dispersion relation}

We work with a $1+1$-dimensional system, and assume that the field propagates in a stationary effective metric which takes the form
\begin{equation}
ds^{2} = -c^{2} dt^{2} + \left( dx - v \, dt \right)^2 \,.
\label{eq:PG_metric}
\end{equation}
The analogy with condensed matter systems is through the interpretation of $v(x)$ as the local flow velocity of a moving medium and $c$ as the low-frequency group velocity of the waves with respect to the medium~\cite{Unruh-1981,LivingReview}. 
Note that $K=\partial_{t}$ is a Killing vector field.
Its squared norm is equal to $K^{2} = g_{tt} = -(c^{2}-v^{2})$, so there is a Killing horizon where $\left|v\right|=c$.
We shall assume $v<0$, describing a flow to the left, and that the magnitude of $v$ increases in the direction of flow; therefore, the point where $v=-c$ corresponds to a black hole horizon.
We shall also assume $c$ is constant, and we set it to $1$.
Note that Eq. (\ref{eq:PG_metric}) is a generalization of the (radial part of the) Schwarzschild metric in Painlev\'{e}-Gullstrand coordinates, where $v \propto -1/\sqrt{x}$ in accordance with the Einstein field equations; here, $v$ is left arbitrary, to be treated as a fixed background.

Dissipative and/or dispersive properties pick out a preferred frame~\cite{Jacobson-1991} -- typically, that in which the medium is at rest -- which does not coincide with the stationary frame $(t,x)$ since the flow velocity $v \neq 0$.
Geometrically, the preferred frame is associated with a unit timelike vector field $u^{\mu}$, and in $1+1$ dimensions this also fixes (up to a sign) the unit spacelike vector field $s^{\mu}$ orthogonal to $u^{\mu}$.
If these are indeed to be associated with the rest frame of the medium whose flow velocity is $v$, then we have explicitly
\begin{alignat}{2}
u = \partial_{t} + v\,\partial_{x} \,, & \qquad s = \partial_{x} \,.
\label{eq:preferred_unit_vectors}
\end{alignat}
These vectors are shown in Figure \ref{fig:freefall}.
For a given momentum $p_{\mu}=(-\omega,k)$, they define the ``proper'' frequency and momentum
\begin{alignat}{2}
\Omega = - u^{\mu}p_{\mu} = \omega-v\,k \,, & \qquad P = s^{\mu}p_{\mu} = k \,.
\label{eq:Om_P}
\end{alignat}
The proper frequency $\Omega$ is thus related to the Killing frequency $\omega$ simply by a Doppler shift.

Following \cite{Adamek-Busch-Parentani-2013}, the high-frequency behavior of the field of interest can be described phenomenologically by the dispersion relation
\begin{equation}
\Omega^{2} + 2i\Gamma(P^{2})\Omega = \tilde{m}^{2} + P^{2} + \tilde{f}(P^{2}) \,. 
\label{eq:dispersion_relation}
\end{equation}
Equation (\ref{eq:dispersion_relation}) gives the effective (``dressed'') dispersion relation of the field, once the environment degrees of freedom have been traced over.
The functions $\Gamma(P^{2})$ and $\tilde{f}(P^{2})$ describe the deviations from the Lorentz invariant case, corresponding, respectively, to dissipative and dispersive effects.
For simplicity, we shall set the effective mass and dispersion to zero, i.e. $\tilde{m}^{2}=0$ and $\tilde{f}(P^{2})=0$.
That leaves only $\Gamma(P^{2})$, the dissipative rate measured in the preferred frame.
We shall principally study the case in which $\Gamma(P^{2}) = \gamma P^{2}$ for some constant $\gamma>0$, for this accords with approximate Lorentz invariance when $\Omega$ and $P$ are sufficiently small, i.e. when $\Omega \sim P \ll 1/\gamma$ where $1/\gamma$ defines the dissipative UV scale $\Lambda_{\mathrm{diss}}$.
However, another case of interest is that with $P$-independent $\Gamma$, relevant to systems such as polaritons \cite{Gerace-Carusotto-2012}; this is briefly studied in Appendix \ref{app:Gamma_constant}.


\subsection{Action}

The system as a whole contains two subsystems: the field of interest, $\phi$, and the environment field, $\Psi$.
For simplicity, we assume these to be scalar fields.
The action of the system takes the form
\begin{equation}
S = S_{\phi} + S_{\Psi} + S_{\mathrm{int}} \,.
\label{eq:action_decomp}
\end{equation}
$S_{\phi}$ is the action of the free $\phi$ field.
It has the form
\begin{equation}
S_{\phi} = \frac{1}{2} \int dt \, dx \, \left\{ \left[ \left(\partial_{t} + v\partial_{x}\right)\phi \right]^{2} - \left[ \partial_{x}\phi \right]^{2} -m^{2}\phi^{2} -\phi \, f \left(-\partial_{x}^{2}\right)\phi \right\} \,.
\label{eq:phi_action}
\end{equation}
This is just the standard $1+1$-dimensional relativistic action for a scalar field in the metric (\ref{eq:PG_metric}), except for the last term in $f(-\partial_{x}^{2})$ which accounts for dispersion.
The mass $m$ and dispersion $f(P^{2})$ are ``bare'', i.e. what they would be in the absence of interactions, and as such appear here without tildes.

\begin{figure}
\includegraphics[width=0.6\columnwidth]{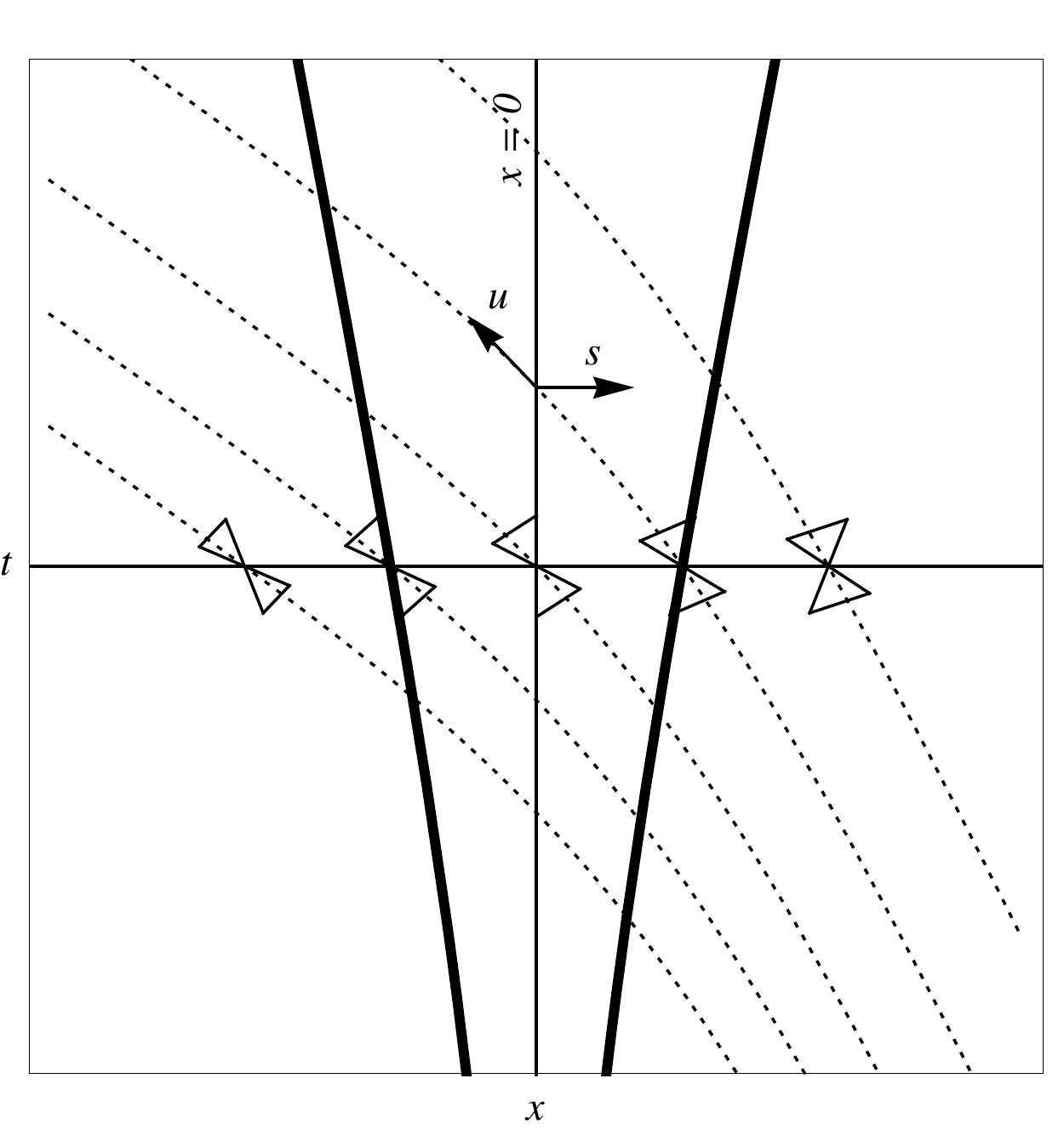}
\caption{\textsc{Environment and field characteristics in space-time}: 
The dotted curves are the freely-falling trajectories, integral curves of the timelike unit vector field $u$; in the moving medium analogy, these are at rest with respect to the medium.  
They are equally spaced in position at time $t=0$, and null cones have been shown for each.
The two solid curves are outgoing null geodesics, which propagate to the right in the freely-falling frame.
Note that these tilt over with $x$, becoming parallel to the Killing vector $\partial_{t}$ at $x=0$; this is the Killing horizon.
On its left, the flow is supersonic ($|v|>  c$), and the outgoing geodesic is moving towards the left in the stationary frame.
\label{fig:freefall}}
\end{figure}

$S_{\Psi}$ is the action of the free environment field, and takes the form
\begin{equation}
S_{\Psi} = \frac{1}{2} \int dt \, dx \int dq \, \left\{ \Big[ \tilde{\partial}_{\tau}\Psi_{q} \Big]^{2} - \Big[ \pi \, q\, \Psi_{q} \Big]^{2} \right\} \,,
\label{eq:Psi_action}
\end{equation}
where we have defined 
\begin{equation}
\tilde{\partial}_{\tau} \equiv \frac{1}{\sqrt{\left|v\right|}} \left( \partial_{t}+v\,\partial_{x} \right) \sqrt{\left|v\right|} = \frac{1}{2}\left(\partial_{t} + \partial_{x}v\right) + \frac{1}{2}\left(\partial_{t} + v\partial_{x}\right)\,.
\label{eq:dtau_defn}
\end{equation}
The environment is thus modeled as a continuous collection of ``independent'' (in the non-interacting case) oscillators, labeled by the continuous parameter $q$ which is proportional to their frequency.
(This parameter can be thought of as the wavevector along an extra dimension, as, for example, in the case of a radiating atom \cite{Unruh-Zurek-1989}.)
A dense set of environment modes is necessary to get a truly dissipative equation for $\phi$; by contrast, a discrete set of environment modes would allow energy to oscillate back and forth between the $\phi$ field and the environment.
The kinetic term in (\ref{eq:Psi_action}), which symmetrises the placement of $v$ and $\partial_{x}$, has been chosen so as to yield a local field equation for $\phi$ when the flow $v$ is not homogeneous (see \cite{Parentani-2007,Adamek-Busch-Parentani-2013}).
It can be seen that the oscillators of the environment are freely-falling, at rest with respect to the moving medium (since they carry no spatial momentum $P$). 
The freely-falling trajectories along which they propagate are illustrated in Figure \ref{fig:freefall}. 

Finally, $S_{\mathrm{int}}$ describes the interaction between $\phi$ and the environment:
\begin{equation}
S_{\mathrm{int}} = \int dt \, dx \, g \, \partial_{x}^{n}\phi \int dq \, \tilde{\partial}_{\tau}\Psi_{q} \,.
\label{eq:int_action}
\end{equation}
Again, $\tilde{\partial}_{\tau}$ of Eq. (\ref{eq:dtau_defn}) appears acting on $\Psi_{q}$ in order to yield a local field equation for $\phi$.
$g$ is a coupling constant, which could depend on space and time.
The non-negative integer $n$ describes how the interaction depends on the proper momentum $P$ of the $\phi$ field: the dissipation rate $\Gamma$ is found to be proportional to $P^{2n}$.
In the main body of this paper, we shall principally consider the case $n=1$; we focus briefly on the case $n=0$ in Appendix \ref{app:Gamma_constant}.


\subsection{Field equations
\label{sub:Field_eqns}}

The action described by Eqs. (\ref{eq:action_decomp})-(\ref{eq:int_action}) leads to the following coupled field equations:
\begin{subequations}\begin{eqnarray}
\left[ \left(\partial_{t}+\partial_{x}v\right)\left(\partial_{t}+v\partial_{x}\right) - \partial_{x}^2
+ m^{2} + f\left(-\partial_{x}^{2}\right) \right] \phi & = & \left(-\partial_{x}\right)^{n} \left( g \int dq \, \tilde{\partial}_{\tau}\Psi_{q} \right) \,, \label{eq:phi_eqn} \\
\left[ \tilde{\partial}_{\tau}^{2} + \left(\pi q\right)^{2} \right] \Psi_{q} & = & - \tilde{\partial}_{\tau} \left( g \partial_{x}^{n}\phi \right) \,. \label{eq:Psi_eqn}
\end{eqnarray}\label{eq:field_eqns}\end{subequations}
We commence their solution by writing a general solution for $\Psi$: on multiplication by $\sqrt{\left|v\right|}$, Eq. (\ref{eq:Psi_eqn}) can be written $\left[\partial_{\tau}^{2}+\left(\pi q\right)^{2}\right]\left(\sqrt{\left|v\right|}\Psi_{q}\right) = -\partial_{\tau}\left(\sqrt{\left|v\right|}g\partial_{x}^{n}\phi\right)$, from which it emerges that, at each $x$, $\sqrt{\left|v\right|}\Psi_{q}$ acts as a driven harmonic oscillator.
Its solution can be straightforwardly written as 
\begin{equation}
\Psi_{q} = \Psi_{q}^{0} + \frac{1}{\sqrt{\left|v\right|}} \int_{-\infty}^{\tau} d\tau^{\prime} \, \frac{\mathrm{sin}\left[\pi \, q \, (\tau-\tau^{\prime})\right]}{\pi \, q} \, \left[-\partial_{\tau}\left(\sqrt{\left|v\right|} \, g \, \partial_{x}^{n}\phi\right) \right]_{\tau^{\prime}} \,,
\label{eq:Psi_soln}
\end{equation}
where $\Psi_{q}^{0}$ is the undriven part, i.e. a solution of the homogeneous form of Eq. (\ref{eq:Psi_eqn}).
When substituting the solution (\ref{eq:Psi_soln}) into the right-hand side of Eq. (\ref{eq:phi_eqn}), the resulting equation is greatly simplified by the identity
\begin{equation}
\partial_{\tau} \left( \int dq \, \frac{\mathrm{sin}\left[\pi \, q \, \left(\tau-\tau^{\prime}\right)\right]}{\pi \, q} \right) = \delta(\tau-\tau^{\prime}) \,.
\label{eq:delta_identity}
\end{equation}
It is precisely this identity that motivates the appearance of $\tilde{\partial}_{\tau}\Psi_{q}$ in $S_{\Psi}$ and $S_{\mathrm{int}}$ of Eqs. (\ref{eq:Psi_action}) and (\ref{eq:int_action}), for it leads to a field equation for $\phi$ which is local in time \footnote{Had we not made this simplifying choice, the wave equation (\ref{eq:final_phi_eqn}) would have been an integralo-differential equation, containing a non-local kernel encoding the damping, and not a standard PDE.  In homogeneous backgrounds, this can still be handled in Fourier space, but it becomes very messy in inhomogeneous spaces.}.
Still, the resulting equation can be further simplified by taking $n=0$ or $1$, assuming the constancy of $g$, and adjusting the bare mass $m^2$ and dispersion $f(P^2)$.
The derivation appears in Appendix \ref{app:Wave_eqn_derivation}; here we simply state the final result:
\begin{equation}
\Box_{t,x}\phi \equiv \left[\partial_{t} + \frac{g^{2}}{2} \left(-i\partial_{x}\right)^{2n} + \partial_{x}\left(v + 1\right)\right] \left[ \partial_{t} + \frac{g^{2}}{2} \left(-i\partial_{x}\right)^{2n} + (v - 1)\partial_{x}\right] \phi = \left(-\partial_{x}\right)^{n} J \,,
\label{eq:final_phi_eqn}
\end{equation}
where we have defined $J \equiv g\int dq \, \tilde{\partial}_{\tau}\Psi_{q}^{0}$.

As shown in Appendix \ref{app:Wave_eqn_derivation}, there are two choices for the bare mass and dispersion yielding a factorized wave equation of the form (\ref{eq:final_phi_eqn}), which differ in the ordering of $v+1$ and $v-1$.
(The particular ordering in Eq. (\ref{eq:final_phi_eqn}) is chosen for later convenience when $g$ is $x$-dependent; this is particularly relevant for the case $n=0$ (see App. \ref{app:Gamma_constant}), where it is useful to ``switch off'' the coupling asymptotically.)
However, during our main focus on the $n=1$ case, we shall assume that $g$ is constant; then Eq. (\ref{eq:final_phi_eqn}) becomes independent of the ordering of $v+1$ and $v-1$. 
The homogeneous solutions are then linear combinations of solutions of the two lower-order equations
\begin{equation}
\left[ \partial_{t} + \frac{g^{2}}{2} \left(-i\partial_{x}\right)^{2n} + \left(v \pm 1\right)\partial_{x} \right] \phi = 0\,.  
\label{eq:source-free_uv-mode_eqns}
\end{equation}
When $g = 0$, these describe waves whose group velocity in the $(t,x)$ coordinate system is $v \pm 1$; that is, since $v$ is the flow of the medium, the waves propagate at speed $1$ with respect to the medium in one of two directions.  
Those moving to the right (left) {\it with respect to the medium} will be referred to here as $u$- ($v$-) modes, respectively.
Since we assume $v<0$, only $v+1$ can vanish at some point, and hence it is principally the $u$-modes that experience the effects of the black hole horizon, which splits them into two disconnected sectors (they correspond to the outgoing null geodesics of Fig. \ref{fig:freefall}).
We shall thus focus mainly on the $u$-modes.
It is true, however, that the presence of the source term on the right-hand side of Eq. (\ref{eq:final_phi_eqn}) induces some higher-order coupling between the $u$- and $v$-modes; we briefly examine the strength of this coupling in Appendix \ref{uv-corr}.


\section{Dissipative field modes
\label{sec:Modes}}

In this section we aim to construct the retarded Green function for the $\phi$ field.
This will allow us, in turn, to solve the sourced wave equation (\ref{eq:final_phi_eqn}) and go on to characterize the state of the $\phi$ field for a given state of the environment.
The treatment here is quite general; the interested reader will find explicit expressions for the field modes and the Green function in a simple black hole profile in Appendix \ref{simplec}. 


\subsection{Retarded Green function
\label{sub:Gret}}

The retarded Green function describes the response of the field to a delta impulse.  
It obeys
\begin{equation}
\Box_{t,x} G_{\mathrm{ret}}\left(t,x;t^{\prime},x^{\prime}\right) = \delta(t-t^{\prime})\delta(x-x^{\prime}) \,,
\label{eq:Gret_defn}
\end{equation}
where $\Box_{t,x}$ is the differential operator of \eq{eq:final_phi_eqn} acting on unprimed coordinates.  
Since this operator is $t$-independent, $G_{\mathrm{ret}}(t,x;t^{\prime},x^{\prime})$ can only depend on the time difference $t-t^{\prime}$. 
It is then convenient to take the Fourier transform:
\begin{equation}
G_{\mathrm{ret}}^{\omega}(x,x^{\prime}) \equiv \int_{-\infty}^{+\infty} G_{\mathrm{ret}}(t,x;t^{\prime},x^{\prime}) \, e^{i\omega(t-t^{\prime})} \, d(t-t^{\prime}) \,,
\end{equation}
upon which Eq. (\ref{eq:Gret_defn}) becomes
\begin{equation}
\Box_{x}^{\omega} G_{\mathrm{ret}}^{\omega}(x,x^{\prime}) = \delta(x-x^{\prime}) \,,
\label{eq:Gret_defn_FT}
\end{equation}
where $\Box_{x}^{\omega}$ is obtained from $\Box_{t,x}$ by substituting $-i\omega$ for all occurrences of the time derivative operator $\partial_{t}$. 

Note that, except for singular behavior along the diagonal $x=x^{\prime}$, $G^{\omega}_{\mathrm{ret}}(x,x^{\prime})$ as a function of $x$ obeys the source-free form of the wave equation, and can thus be decomposed into stationary solutions of
\be
\Box_{x}^{\omega} \phi^{\omega}(x) = 0 \, . 
\label{nmodes}
\ee
Asymptotically in $|x|$ -- where we shall assume $v$ approaches constant limiting values for both $x \to \pm \infty$ -- $G^{\omega}_{\mathrm{ret}}(x,x^{\prime})$ must be a linear combination of plane waves which decrease in amplitude towards infinity. 
But the coefficients of these plane waves will generally depend on $x^{\prime}$, and so it is useful to consider how $G^{\omega}_{\mathrm{ret}}(x,x^{\prime})$ behaves as a function of $x^{\prime}$.  
Postulating that it obeys an equation of the same form as Eq. (\ref{eq:Gret_defn_FT}):
\begin{equation}
\widetilde{\Box}^{-\omega}_{x^{\prime}} G^{\omega}_{\mathrm{ret}}(x,x^{\prime}) = \delta(x-x^{\prime}) \,,
\label{eq:Gret_dual_eqn}
\end{equation}
where the sign of $\omega$ on the operator is chosen for convenience and reflects the fact that the Green function decomposes into products of functions in $x$ and $x^{\prime}$,
we necessarily have
\begin{equation}
\int dx_{1} \left( \widetilde{\Box}^{-\omega}_{x_{1}} G^{\omega}_{\mathrm{ret}}(x,x_{1}) \cdot G^{\omega}_{\mathrm{ret}}(x_{1},x^{\prime}) - G^{\omega}_{\mathrm{ret}}(x,x_{1}) \cdot \Box^{\omega}_{x_{1}} G^{\omega}_{\mathrm{ret}}(x_{1},x^{\prime}) \right) = 0 \,,
\end{equation}
and hence that $\widetilde{\Box}^{-\omega}_{x}$ is the ``dual'' operator defined by
\begin{equation}
\int_{-\infty}^{+\infty} \widetilde{\phi}_{1}^{-\omega} \cdot \Box^{\omega}_{x} \phi_{2}^{\omega} \, dx = \int_{-\infty}^{+\infty} \widetilde{\Box}^{-\omega}_{x} \widetilde{\phi}_{1}^{-\omega} \cdot \phi_{2}^{\omega} \, dx \, .
\end{equation}
The dual operator $\widetilde{\Box}^{\omega}_{x}$ is obtained from $\Box^{\omega}_{x}$ by flipping the sign of $\gamma$, so $\widetilde{\Box}^{-\omega}_{x}$ is obtained by flipping the sign of the entire complex frequency $\omega-i\Gamma$.
Except at $x=x^{\prime}$, $G^{\omega}_{\mathrm{ret}}(x,x^{\prime})$ can, as a function of $x^{\prime}$, be decomposed into solutions of 
\be
\widetilde{\Box}^{-\omega}_{x^{\prime}} \widetilde{\phi}^{-\omega}(x^{\prime}) = 0\, . 
\label{dualm}
\ee 
Note that when $\gamma = 0$, $\widetilde{\Box}^{\omega}_{x} = \Box^{\omega}_{x}$, so that the notion of the ``dual'' solutions becomes redundant and the functions of $x^{\prime}$ in Eq. (\ref{dualm}) are simply solutions of Eq. (\ref{nmodes}) with reversed frequency (i.e. they are complex conjugate solutions).

Overall, $G^{\omega}_{\mathrm{ret}}(x,x^{\prime})$ can be written as a sum of products $\phi^{\omega}(x) \, \widetilde{\phi}^{-\omega}(x^{\prime})$,  where the coefficient of each product depends only on the ordering of $x$ and $x^{\prime}$.
Moreover, when the system is homogeneous in space (i.e. $v$ and $\gamma$ are independent of $x$), the entire Green function can only depend on the interval $x-x^{\prime}$.
The general case is treated in Appendix \ref{app:Gret}.
Here we only consider Eq. (\ref{eq:final_phi_eqn}) with $n = 1$ and $g$ constant.
Defining $\gamma \equiv g^{2}/2$, the dissipative rate is $\Gamma = \gamma \, P^{2}$.
In this case, the Fourier transform obeys 
\begin{equation}
\left[-i\omega + \gamma \left(-i\partial_{x}\right)^{2} + \partial_{x} \left(v+1\right) \right] \left[ -i\omega + \gamma \left(-i\partial_{x}\right)^{2} + \left(v-1\right)\partial_{x}\right]G^{\omega}_{\mathrm{ret}}\left(x,x^{\prime}\right) = \delta(x-x^{\prime}) \,, 
\label{eq:Gret_om}
\end{equation}
and for a homogeneous background, we can Fourier transform also in space to get
\begin{eqnarray}
 \left[-i\omega +\gamma k^{2} + ik(v+1) \right] \left[ -i\omega + \gamma k^{2} + i k (v-1) \right] G^{\omega}_{\mathrm{ret}}(k)  & = & 1\, . 
\label{eq:Gret_FT_hom}
\end{eqnarray}
The roots of the first factor in (\ref{eq:Gret_FT_hom}) are solutions of the $u$-mode differential operator, and those of the second factor are solutions of the $v$-mode equation.
We can use this property to label the roots.
Hence, we can write 
\begin{eqnarray}
G^{\omega}_{\mathrm{ret}}(k) = \frac{1}{\gamma^{2} \left(k-k_{uR}\right) \left(k-k_{uL}\right) \left(k-k_{vR}\right) \left(k-k_{vL}\right)}  \,.
\end{eqnarray}
The additional indices $R$ and $L$ indicate the sign of the imaginary part of the wave vector: it is $R$ ($L$) when $\Im k > 0$ ($\Im k < 0$), so that $k_R$ ($k_L$) modes decay towards the right (left) side.

By inverting the Fourier transform in $k$, the $R$ roots contribute when $x > x'$, and vice versa for the $L$ roots; so we have
\begin{eqnarray}
G_{\mathrm{ret,hom}}^{\omega}(x,x^{\prime}) & = & i\theta(x-x^{\prime}) \left\{ \frac{e^{ik_{uR}(x-x^{\prime})}}{2(\omega-i\gamma k_{uR}^{2})} - \frac{e^{ik_{vR}(x-x^{\prime})}}{2(\omega-i\gamma k_{vR}^{2})} \right\} \nonumber \\
& & - i\theta(x^{\prime}-x) \left\{ \frac{e^{ik_{uL}(x-x^{\prime})}}{2(\omega-i\gamma k_{uL}^{2})} - \frac{e^{ik_{vL}(x-x^{\prime})}}{2(\omega-i\gamma k_{vL}^{2})} \right\} \,. 
\end{eqnarray} 
From this exact equation, we can read out the appropriate boundary conditions that also prevail when working in an inhomogeneous background flow which is asymptotically constant on both sides. 
The key property which should be implemented is that $G_{\mathrm{ret,hom}}^{\omega}(x,x^{\prime})$ is asymptotically bounded for $x \to \pm \infty$ at fixed $x^{\prime}$, and for $x^{\prime} \to \pm \infty$ at fixed $x$.
As a result, the Green function can be written in terms of 8 well-defined modes as
\begin{eqnarray}
G_{\mathrm{ret}}^{\omega}(x,x^{\prime}) & = & i\theta(x-x^{\prime}) \left\{ \frac{\phi_{uR}^{\omega}(x) \widetilde{\phi}^{-\omega}_{uR}(x^{\prime})}{2(\omega-i\gamma k_{uR}^{2})} - \frac{\phi_{vR}^{\omega}(x) \widetilde{\phi}^{-\omega}_{vR}(x^{\prime})}{2(\omega-i\gamma k_{vR}^{2})} \right\} \nonumber \\
& & - i\theta(x^{\prime}-x) \left\{ \frac{\phi_{uL}^{\omega}(x) \widetilde{\phi}^{-\omega}_{uL}(x^{\prime})}{2(\omega-i\gamma k_{uL}^{2})} - \frac{\phi^{\omega}_{vL}(x) \widetilde{\phi}^{-\omega}_{vL}(x^{\prime})}{2(\omega-i\gamma k_{vL}^{2})} \right\} \, , 
\label{eq:Gret_Psq_general}
\end{eqnarray} 
where $k_{uR}$ and $k_{vR}$ in the denominator should be evaluated on the asymptotic right side, and similarly for the $L$ roots on the left.
The 4 solutions of \eq{nmodes} are easy to characterize. 
As an example, the dissipated mode $\phi^{\omega}_{uR}(x)$ is the unique solution which asymptotes on the right (for $x \to \infty$) to the exponential $e^{i k_{uR} x }$ with wave number $k_{uR}$ and a unit amplitude.  
Instead the characterization of the 4 dual modes, solutions of \eq{dualm}, is more subtle (a fuller derivation can be found in Appendix \ref{app:Gret}).
For instance $\widetilde \phi^{-\omega}_{uR}(x')$ is the unique solution which asymptotically grows as $\exp({- i k_{uR} x' })$ on the right side, but which decays on the left side.
Hence it contains on the right side a uniquely determined superposition of two decaying exponentials. 
We could have normalized these modes, as done for relativistic and dispersive fields, in such a way that the denominators $2(\om - i \gamma k^2)$ would not appear in \eq{eq:Gret_Psq_general}.
Since the modes have no sense by themselves due to the coupling to the environment, this does not matter here, and we can work with modes of unit asymptotic amplitude.
Note that, in the limit $\gamma \rightarrow 0$, the (conjugated) dual modes $\widetilde{\phi}^{\omega}(x^{\prime})$ become standard ingoing modes, with the single asymptotic growing mode becoming the single incident wave. 

It is interesting to note that, since Eq. (\ref{eq:final_phi_eqn}) has been fine-tuned so as to give no $u$-$v$ coupling, $u$-modes (dual $u$-modes) are composed only of $u$-modes (dual $u$-modes) in both asymptotic regions.
Therefore, Eq. (\ref{eq:Gret_Psq_general}) can be seen to split exactly into a sum of two terms, one containing only $u$-modes, the other only $v$-modes.
We shall often focus our attention on the $u$-part of $G_{\mathrm{ret}}^{\omega}(x,x^{\prime})$,
\begin{equation}
G_{\mathrm{ret},u}^{\omega}(x,x^{\prime}) = i\theta(x-x^{\prime}) \frac{\phi^{\omega}_{uR}(x) \widetilde{\phi}^{-\omega}_{uR}(x^{\prime})}{2(\omega-i\gamma k_{uR}^{2})} - i\theta(x^{\prime}-x) \frac{\phi^{\omega}_{uL}(x) \widetilde{\phi}^{-\omega}_{uL}(x^{\prime})}{2(\omega-i\gamma k_{uL}^{2})} \,. 
\end{equation}

\begin{figure}
\includegraphics[width=0.8\columnwidth]{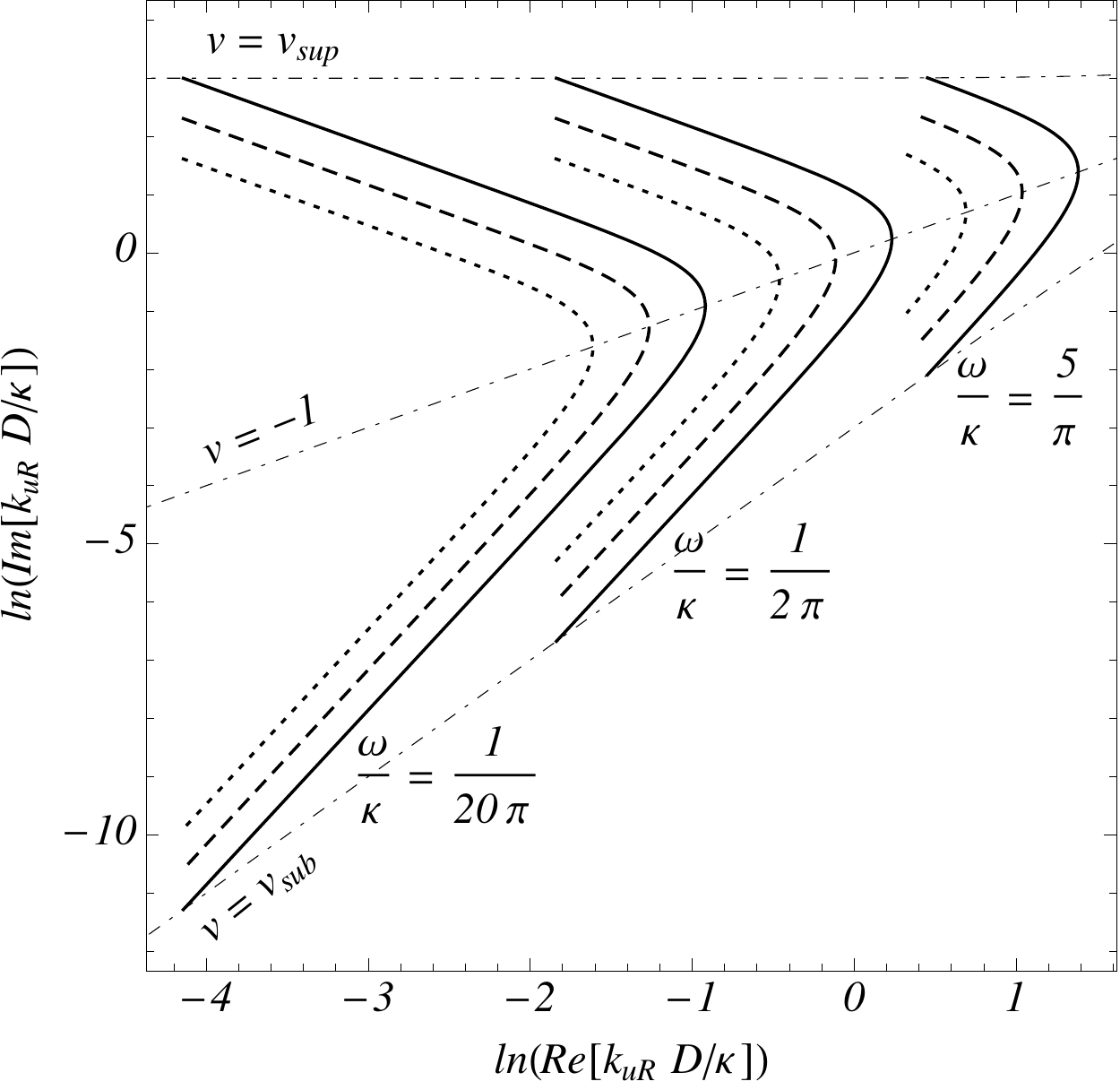}
\caption{\textsc{Roots of dispersion relation of \eq{uDR}}: 
Here is shown, on the complex $k_{uR}$-plane (and in logarithmic scale), the continuous variation of the wave vector $k_{uR}$ as $v+1$ varies between $D$ and $-D$, i.e. as the position is varied from the subsonic to the supersonic region.
$k_{uR}$ has positive real and imaginary parts, and is thus outwardly propagating to the right.
In the subsonic region, as $x\rightarrow \infty$, its imaginary part takes its smallest value, and this corresponds to the lower end of each curve.
As $x$ is varied back towards the horizon, its real part gradually increases, turning around at $x=0$.
As $x$ is further varied into the left-hand (supersonic) region, the real part of $k_{uR}$ decreases again while its imaginary part continues to increase, so that the top end of each curve corresponds to $x\rightarrow -\infty$. 
The various styles of curve correspond to different values of $\lambda^{2}=\gamma\kappa/D^{2}$: $0.05$ (solid), $0.1$ (dashed) and $0.2$ (dotted).
The three different groups correspond to different values of $\omega/\kappa$.  
The three light dot-dashed curves plot the roots at fixed $v+1$ for varying $\omega$, with $\lambda^{2} = 0.05$.
Note that the $(v+1)_{\mathrm{sub}} = D$ curve follows the expected pattern with $\mathrm{Im}(k) \propto \left(\mathrm{Re}(k)\right)^{2}$; the $(v+1)_{\mathrm{sup}}= - D$ curve has $\mathrm{Im}(k)$ independent of $\mathrm{Re}(k)$; and the $v+1 = 0$ curve (i.e., at the horizon) has $\mathrm{Im}(k) = \mathrm{Re}(k)$, and is the same regardless of the value of $\lambda^{2}$.
\label{fig:roots}}
\end{figure}

When considering a black hole flow, there is an interesting switch amongst $\phi^{\omega}_{uR}$ and $\phi^{\omega}_{uL}$.
When the flow vanishes or is subsonic, the $u_R$-mode is the dissipative version of the standard $u$-mode, whereas the additional root $k_{uL}$ is brought in by the dissipative term in $k^4$ in the dispersion relation.
In fact, its imaginary part diverges as $-(v+1)/\gamma$ when $\gamma\to 0$. 
When the flow is supersonic and to the left (i.e. $v+1< 0$), it is now the $u_{L}$-mode that is the dissipative version of the standard $u$-mode, while $k_{uR}$ is the additional root with large imaginary part.
This can be understood from the fact that, in supersonic flows to the left, the group velocity of $u$-modes in the lab frame is to the left, see Fig.~\ref{fig:freefall}. 
This switch can be studied by considering the roots of the $u$-dispersion relation, see \eq{eq:source-free_uv-mode_eqns},
\be
\left[ -i\omega + \gamma k_u^{2} + i k_u (v+1) \right] = 0 \, .
\label{uDR}
\ee
One sees that the standard root in one region varies continuously to become the additional root in the other region. 
The migration of the $k_{uR}$ root is shown in Figure \ref{fig:roots} as $v+1$ varies from $D$ (subsonic region) to $-D$.
(When considering the $v$-modes, no such switch arises as $v-1$ does not flip sign in a black hole flow to the left.)

\subsection{Adimensional units}

When focusing on $u$-modes, we shall see that the dissipative wave equation (\ref{eq:source-free_uv-mode_eqns}) defines an adimensional parameter which depends both on the coupling strength $\gamma = g^2/2$ and the properties of the background flow. 
Recalling that $c=1$, notice that $\gamma$ has dimensions of time, as it is related to the dissipative rate by $\Gamma = \gamma P^2 = P^2/\Lambda_{\rm diss}$. 

Since $u$-modes are sensitive only to the sum $v+1$ we can write
\begin{equation}
v(x)+1 = D \, h\left(\frac{\kappa\,x}{D}\right) \,,
\label{eq:u-mode_profile_adimensional}
\end{equation}
where $\kappa$ and $D$ are parameters while $h$ varies monotonically between two asymptotic limiting values, with $h(0)=0$ and $h^{\prime}(0)=1$.
Then there is a horizon at $x=0$, with ``surface gravity'' $\left|v^{\prime}(0)\right| = \kappa$.
For a given profile, the parameter $D$ allows us to vary the total difference in $v$ between the asymptotic regions, or, equivalently, the size of the near-horizon region (NHR) where $v(x)+1 \approx \kappa \, x$.
In the limit $D \rightarrow \infty$, $v(x)+1$ becomes exactly $\kappa \,x$, which describes the Poincar\'e patch of de Sitter space in the stationary picture \cite{Adamek-Busch-Parentani-2013}. 

We can adimensionalize the $u$-mode wave equation of (\ref{eq:source-free_uv-mode_eqns}) by defining dimensionless time and space variables:
\begin{alignat}{2}
T = \frac{t}{t_{0}} \,, & \qquad X = \frac{x}{x_{0}} \,.
\end{alignat}
By direct substitution, we get 
\begin{equation}
\left[ \frac{1}{t_{0}} \partial_{T} + \frac{\gamma}{x_{0}^{2}} \left(-i\partial_{X}\right)^{2} + \frac{D}{x_{0}} h\left(\frac{\kappa x_{0}}{D} X\right) \partial_{X} \right] \phi = 0 \,.
\label{eq:adimensional_u_eqn}
\end{equation}
It is natural to select $t_{0} = 1/\kappa$, independent of the dissipative rate.
However, there are two natural choices for the length scale $x_{0}$: there is a ``geometrical'' length $x_{0}^{\mathrm{geom}}$ adapted to the geometry described by $h$; and there is a ``dissipative'' length $x_{0}^{\mathrm{diss}}$ adapted to the effective rate of dissipation in space:
\begin{alignat}{2}
x_{0}^{\mathrm{geom}} = \frac{D}{\kappa} \,, & \qquad x_{0}^{\mathrm{diss}} = \sqrt{\frac{\gamma}{\kappa}} \,.
\label{eq:length_scales}
\end{alignat}
Calling $X = x/x_{0}^{\mathrm{geom}}$ and $\chi = x/x_{0}^{\mathrm{diss}}$, Eq. (\ref{eq:adimensional_u_eqn}) can therefore be written in the equivalent forms
\begin{subequations}\begin{eqnarray}
\left[ \partial_{T} + \lambda^{2} \left(-i \partial_{X}\right)^{2} + h\left(X\right) \, \partial_{X} \right] \phi & = & 0 \,, \label{eq:u_eqn_Xgeom} \\
\left[ \partial_{T} + \left(-i\partial_{\chi}\right)^{2} + \frac{h(\lambda \, \chi)}{\lambda} \, \partial_{\chi} \right] \phi & = & 0 \,, \label{eq:u_eqn_Xdiss}
\end{eqnarray}\label{eq:u_eqn_X}\end{subequations}
where we have defined the dimensionless parameter
\begin{equation}
\lambda^{2} \equiv \left( \frac{x_{0}^{\mathrm{diss}}}{x_{0}^{\mathrm{geom}}} \right)^{2} = \frac{\gamma\,\kappa}{D^{2}}\,.
\label{eq:lambda_defn}
\end{equation}
Equations (\ref{eq:u_eqn_X}) are both useful in different regimes: they show how position should be scaled in order to remove any significant dependence on $\lambda^2$.
So when $\lambda^{2}P^{2} \ll h P$, as is most probable in the asymptotic regions, the dissipative term in Eq. (\ref{eq:u_eqn_Xgeom}) can be treated as a small perturbation, and the field modes do not depend significantly on the particular value of $\lambda^2$.
This assumption breaks down near the horizon where $P$ becomes very large, but there Eq. (\ref{eq:u_eqn_Xdiss}) becomes independent of $\lambda^2$ since $h(\lambda \chi)/\lambda \approx \chi$ when $\lambda \chi$ is sufficiently small.

Note also the different limits probed by Eqs. (\ref{eq:u_eqn_Xgeom}) and (\ref{eq:u_eqn_Xdiss}) when $\lambda^2 \rightarrow 0$:  Eq. (\ref{eq:u_eqn_Xgeom}), treating the ``geometrical'' length as the fundamental length unit, sends the dissipative rate to zero so that we recover the standard relativistic wave equation; on the other hand, Eq. (\ref{eq:u_eqn_Xdiss}), which treats the ``dissipative'' length as fundamental, effectively sends $D$ to $\infty$ as $\lambda^2$ as decreased.
In this limit, one is considering a dissipative field in de Sitter space~\cite{Adamek-Busch-Parentani-2013}, since $v+ 1 = \kappa x$ for all $x$. 

\begin{figure}
\subfloat{\includegraphics[width=0.45\columnwidth]{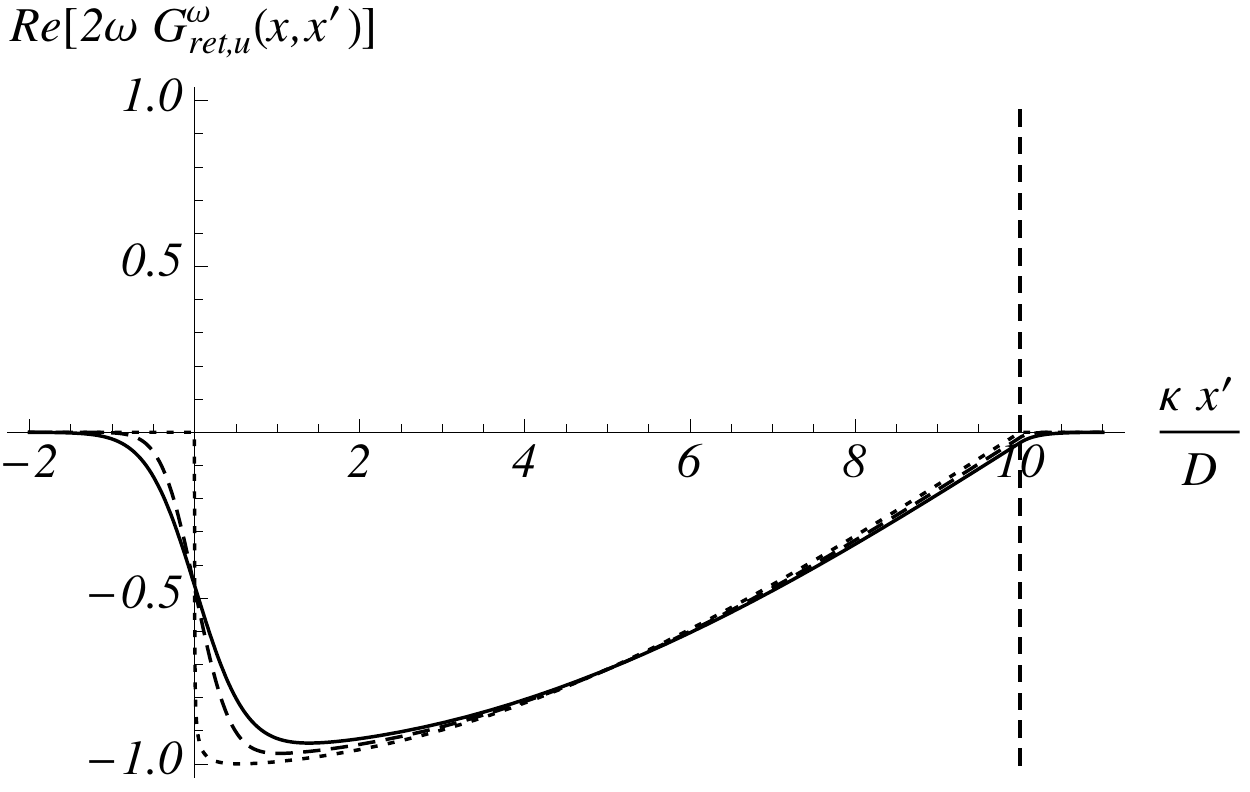}} \qquad \subfloat{\includegraphics[width=0.45\columnwidth]{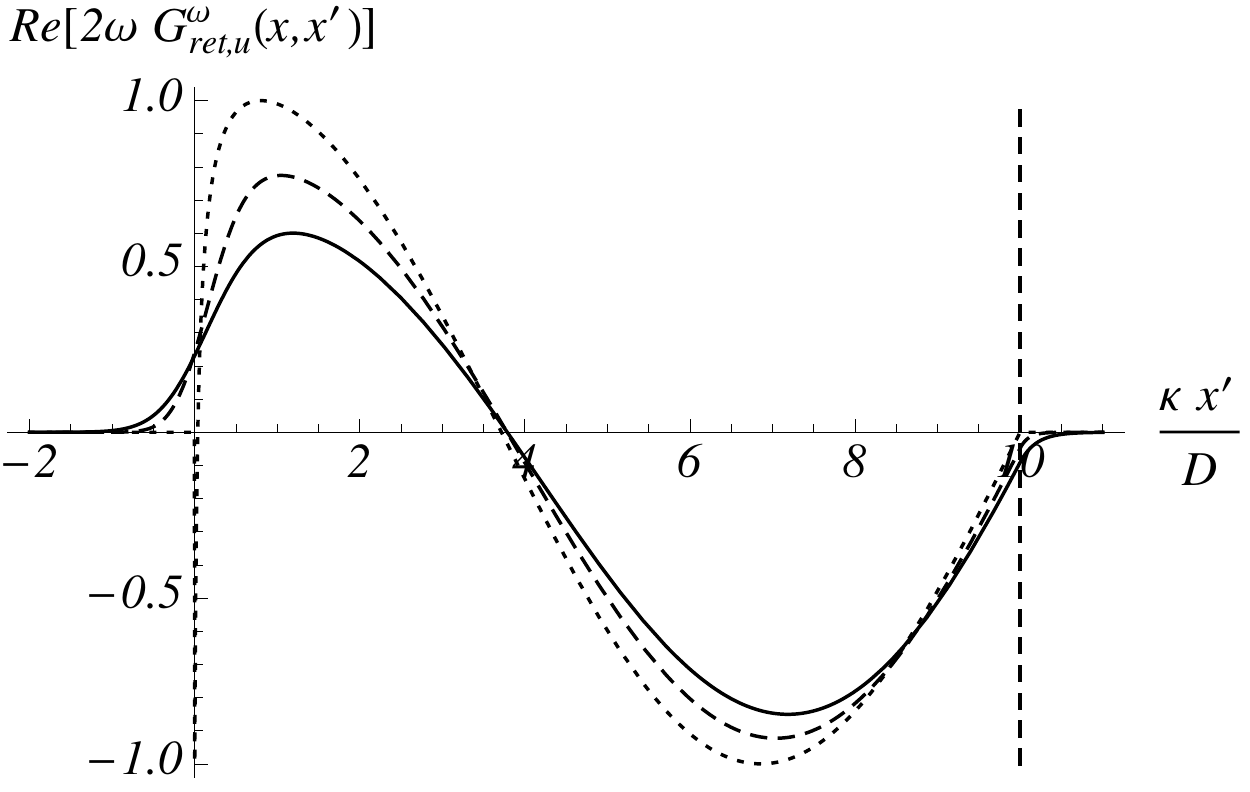}} \\
\subfloat{\includegraphics[width=0.45\columnwidth]{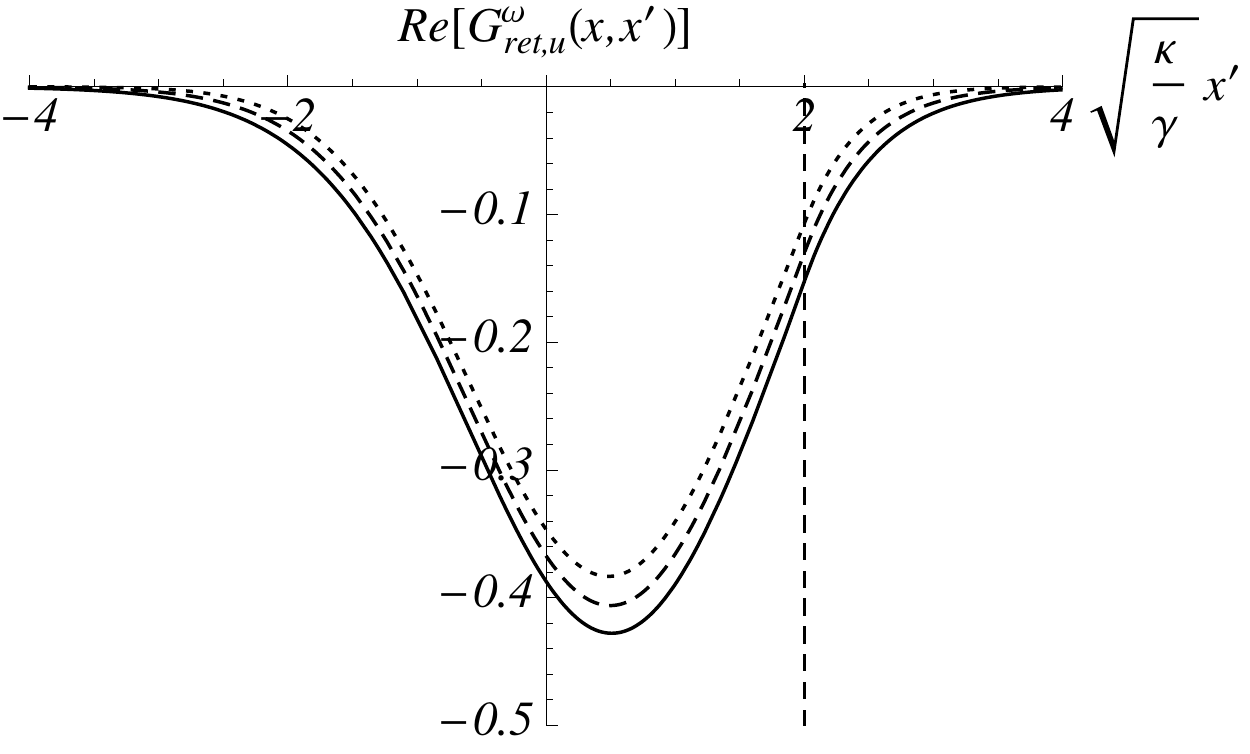}} \qquad \subfloat{\includegraphics[width=0.45\columnwidth]{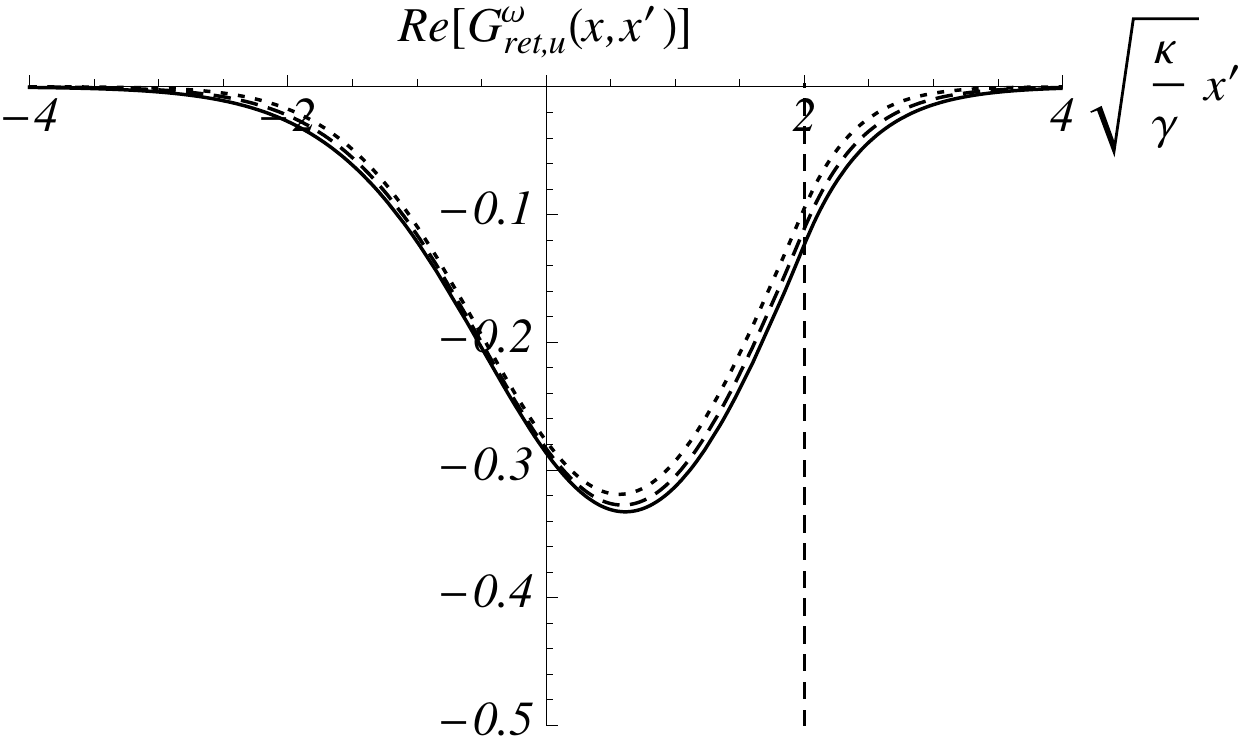}}
\caption{\textsc{Green function}: Here is plotted the $u$ part of the retarded Green function, $G_{\mathrm{ret},u}^{\omega}(x,x^{\prime})$ as a function of the source coordinate $x^{\prime}$ for fixed image coordinate $x$.
The various curves correspond to different values of $\lambda^{2}=\gamma \kappa/D^{2}$: $0$ (dotted curve), $0.1$ (dashed curve) and $0.2$ (solid curve).
The columns correspond to different frequencies, with $\omega/\kappa = 1/2\pi$ on the left and $1/2$ on the right.
In the top row, the position coordinates have been adimensionalised by the ``geometrical'' length (the first of Eqs. (\ref{eq:length_scales})) and the Green function has been multiplied by $2\omega$ so as to see the effects of dissipation through a reduction with respect to a unit amplitude. 
In the bottom row $x$ has been adimensionalised by the ``dissipative'' length (the second of Eqs. (\ref{eq:length_scales})).  
In this representation, the limiting case $\lambda^{2} = 0$ corresponds to working with a dissipative field in a de Sitter geometry.
The small differences with the cases $\lambda^{2}=0.1$ and $0.2$ illustrate that the near-horizon behavior of $G_{\mathrm{ret},u}^{\omega}(x,x^{\prime})$ in a black hole flow is essentially the same as that in de Sitter.
The vertical dashed lines show the fixed position of the image coordinate: $\kappa x/D = 10$ (in the flat asymptotic region) in the top row, and $x \sqrt{\kappa/\gamma} = 2$ (well inside the near-horizon region) in the bottom row.
\label{fig:Gret}}
\end{figure}

In Figure \ref{fig:Gret} are plotted examples of (the $u$ part of) the retarded Green function, for various dissipative rates, in both the ``geometrical'' coordinate $X$ and the ``dissipative'' coordinate $\chi$.
Note the effect of dissipation according to these two viewpoints.
For fixed $X$ with varying $X^{\prime}$, the largest changes induced by dissipation occur around the horizon, where the infinitely many oscillations of the relativistic model are smoothed into an exponential tail; in the asymptotic region, dissipation causes a gradual decrease in the amplitude, but we can see from the position of the node that the phase of the function is hardly affected.
On the other hand, when $\chi$ is fixed inside the near-linear de Sitter region and $\chi^{\prime}$ is varied, the behavior around the horizon possesses a well-defined (and nontrivial) limit as $\lambda^{2} \to 0$. 
Whenever $\lambda^{2} \ll 1$, this establishes that the behaviour in the near-horizon region is indeed very close to that found in de Sitter space (the deviations vanish as $\lambda^{2}$ for $\lambda^{2} \to 0$).
The plots of Fig. \ref{fig:Gret} have been obtained in a particular flow where the decaying modes can be explicitly computed. 
This simple case is studied in Appendix~\ref{simplec}.


\section{State of the field
\label{sec:State}}

In this section we include the effects of the source term $(-\partial_{x})^{n}J$ in Eq. (\ref{eq:final_phi_eqn}) in order to calculate various observables characterizing the state of the $\phi$ field.  
The statistical properties of the operator $\hat J$ are fixed by the initial state of the environment, which is assumed to be thermal, completely characterized by its temperature $T_{\Psi}$.
The entire information is thus contained in the anticommutator $\left\langle \left\{ \hat{J}, \hat J \right\} \right\rangle $, which is called the noise kernel~\cite{Gardiner-QuantumNoise}.
Because the field equations (\ref{eq:field_eqns}) are linear and because the background fields $g_{\mu\nu}$ and $u^{\mu}$ are stationary, the state of the $\phi$ field will also be Gaussian and stationary.  
As a result, each $\omega$-sector can be studied separately, and its state is completely determined by the $\omega$-components of the anticommutator,
\begin{alignat}{3}
G_{\mathrm{ac}}^{\omega}(x,x^{\prime}) = \left\langle \left\{ \hat{\phi}^{\omega}(x), \hat{\phi}^{-\omega}(x^{\prime}) \right\} \right\rangle \,,
\label{eq:Gac_defn}
\end{alignat}
where $\omega$ is real and positive, and where $\hat{\phi}^{-\omega}(x) = (\hat{\phi}^{\omega}(x))^\dagger$ since the field is hermitian.


\subsection{Interpretation of the anticommutator
\label{sub:Gac_interpretation}}

To interpret the form of $G_{\mathrm{ac}}^{\omega}$, let us first consider the $\omega$-component of the field operator in the non-dissipative model (using notation adopted from \cite{Busch-Parentani-2014}):
\begin{equation}
\hat{\phi}^{\omega}(x) = \hat{a}_{\omega} \, \varphi^{\omega}_{uR}(x) + \left(\hat{a}_{-\omega}\right)^{\dagger} \, \left(\varphi^{-\omega}_{uL}(x)\right)^{\star} + \hat{a}_{\omega}^{v} \, \varphi^{\omega}_{v}(x) \,.
\label{eq:phi_decomp_nondiss}
\end{equation}
The amplitude operators obey the usual bosonic commutation relations, and the relativistic modes $\varphi_\omega$ are normalized according to the standard convention ($\sim e^{-i\omega t}/\sqrt{2\omega}$).
For $\omega>0$, the $uL$-mode has negative norm while the others have positive norm, and this is indicated by writing the $uL$-mode contribution to the field operator as the hermitian conjugate of the same term with frequency $-\omega$, which has positive norm.
This allows us to suppress the $u$ label on the amplitude operators and distinguish the left and right amplitude operators by the sign in front of $\omega$.

By direct substitution of Eq. (\ref{eq:phi_decomp_nondiss}) in Eq. (\ref{eq:Gac_defn}), in any stationary state, we find
\begin{multline}
G_{\mathrm{ac}}^{\omega}(x,x^{\prime}) = \left( 2\,n_{\omega} + 1 \right) \, \varphi^{\omega}_{uR}(x) \left(\varphi^{\omega}_{uR}(x^{\prime})\right)^{\star} + \left( 2\,n_{-\omega} + 1 \right) \, \left(\varphi^{-\omega}_{uL}(x)\right)^{\star} \varphi^{-\omega}_{uL}(x^{\prime}) + \left( 2\,n_{\omega}^{v} + 1\right) \, \varphi^{\omega}_{v}(x) \left(\varphi^{\omega}_{v}(x^{\prime})\right)^{\star} \\
+ 2\,c_{\omega} \, \varphi^{\omega}_{uR}(x) \varphi^{-\omega}_{uL}(x^{\prime}) + 2\,c^{v}_{\omega} \, \varphi^{\omega}_{v}(x) \varphi^{-\omega}_{uL}(x^{\prime}) + 2\,d^{v}_{\omega} \, \varphi^{\omega}_{v}(x) \left(\varphi^{\omega}_{uR}(x^{\prime})\right)^{\star} + \left(x \leftrightarrow x^{\prime}\right)^{\star} \,,\label{eq:Gac_decomp_nondiss}
\end{multline}
where the additional contribution $\left(x\leftrightarrow x^{\prime}\right)^{\star}$ only applies to the second line.
We have used the bosonic commutation relations that exist between the amplitude operators in order to replace the anticommutators with standard observables:
\begin{alignat}{5}
n_{\omega} & = \Big\langle \hat{a}^{\dagger}_{\omega} \hat{a}_{\omega} \Big\rangle \,, & \qquad n_{-\omega} & = \Big\langle \hat{a}^{\dagger}_{-\omega} \hat{a}_{-\omega} \Big\rangle \,, & \qquad n_{\omega}^{v} & = \Big\langle \left(\hat{a}^{v}_{\omega}\right)^{\dagger} \hat{a}^{v}_{\omega} \Big\rangle \,, \nonumber \\
c_{\omega} & = \Big\langle \hat{a}_{\omega} \hat{a}_{-\omega} \Big\rangle \,, & \qquad c_{\omega}^{v} & = \Big\langle \hat{a}^{v}_{\omega} \hat{a}_{-\omega} \Big\rangle \,, & \qquad d_{\omega}^{v} & = \Big\langle \hat{a}^{\dagger}_{\omega} \hat{a}^{v}_{\omega} \Big\rangle \,.
\label{eq:nandc}
\end{alignat}
These observables, then, are directly related to the amplitudes of various terms appearing in the decomposition of the anticommutator.
In turn, following \cite{Busch-Parentani-2014} we can construct from them the following quantities:
\begin{alignat}{3}
\Delta_{\omega} = n_{\omega} n_{-\omega} - \left|c_{\omega}\right|^{2} \,, & \qquad \Delta^{v}_{\omega} = n_{\omega}^{v} n_{-\omega} - \left| c_{\omega}^{v}\right|^{2} \,, & \qquad D_{\omega}^{v} = n_{\omega} n_{\omega}^{v} - \left|d_{\omega}^{v}\right|^{2} \,.
\label{eq:nonseparability}
\end{alignat}
If the state of the system can be described classically, all of the quantities in Eqs. (\ref{eq:nonseparability}) must be strictly nonnegative.
Quantum mechanics, however, allows $\Delta_{\omega}$ and $\Delta_{\omega}^{v}$ to be negative, bounded from below by the corresponding values of $-n$. 
($D_{\omega}^{v}$, examining correlations between quasiparticles of the same norm, remains strictly nonnegative even in the quantum description~\cite{Zapata}.) 
States for which either $\Delta_{\omega}$ or $\Delta_{\omega}^{v}$ is negative are said to be {\it nonseparable}: the strength of the correlations is so strong that they cannot be described by a classical ensemble. 

We shall principally be interested in $\Delta_{\omega}$, which measures the ``quantumness'' of the pairs of $u$-quasiparticles which propagate on either side of the horizon.
To extract $\Delta_{\omega}$, we shall use the fact that $G_{\mathrm{ac}}^{\omega}$ is a sum of three terms:
\begin{equation}
G_{\mathrm{ac}}^{\omega}(x,x^{\prime}) = G_{\mathrm{ac},u}^{\omega}(x,x^{\prime}) + G_{\mathrm{ac},v}^{\omega}(x,x^{\prime}) + G_{\mathrm{ac},u/v}^{\omega}(x,x^{\prime})\,.
\label{eq:Gac_decomp_diss}
\end{equation}
The first term governs the sector of interest containing the two $u$-modes, the second one the sector containing the single $v$-mode, and the last one contains the $u/v$-correlations governed by $c_{\omega}^{v} $ and $d_{\omega}^{v}$.
In the present model, the latter do not vanish even though the retarded Green function is a sum of a $u$ and a $v$ contribution, for both sectors are sourced by the same environment.
However the strength of the $u/v$ correlations are very small (see Appendix \ref{uv-corr}). 


\subsection{Behavior of the anticommutator
\label{sub:Gac_calculation}}

From the wave equation (\ref{eq:final_phi_eqn}), taking the Fourier transform in time, we obtain
\begin{equation}
\Box_{x}^{\omega} \Box_{x^{\prime}}^{-\omega} G_{\mathrm{ac}}^{\omega}(x,x^{\prime}) = \partial_{x}\partial_{x^{\prime}} \left\langle \left\{ \hat{J}^{\omega}(x), \hat{J}^{-\omega}(x^{\prime}) \right\} \right\rangle \equiv \partial_{x}\partial_{x^{\prime}} N^{\omega}(x,x^{\prime}) \,.
\label{eq:Gac_om_wave_eqn}
\end{equation}
Note that the operator $\Box^{-\omega}_{x^{\prime}}$ appearing in Eq. (\ref{eq:Gac_om_wave_eqn}) is {\it not} a dual operator of \eq{dualm}.
The solution can be built from the retarded Green function:
\begin{eqnarray}
G^{\omega}_{\mathrm{ac}}(x,x^{\prime}) & = & \int dx_{1} \int dx_{2} \, G^{\omega}_{\mathrm{ret}}(x, x_{1}) \, G^{-\omega}_{\mathrm{ret}}(x^{\prime},x_{2}) \, \partial_{x_{1}}\partial_{x_{2}} N^{\omega}(x_{1},x_{2}) \nonumber \\
& = & \int dx_{1} \int dx_{2} \, \partial_{x_{1}}G^{\omega}_{\mathrm{ret}}(x, x_{1}) \, \partial_{x_{2}}G^{-\omega}_{\mathrm{ret}}(x^{\prime},x_{2}) \, N^{\omega}(x_{1},x_{2}) \,,
\label{eq:Gac_from_Gret}
\end{eqnarray}
where in the second line we have used integration by parts to move the derivatives from the noise kernel to the Green functions.
The anticommutator is thus determined from a knowledge of the retarded Green function (calculated in \S\ref{sub:Gret}) and of the noise kernel (calculated in Appendix \ref{app:Noise} -- see Eq. (\ref{eq:noise_derived})):
\begin{eqnarray}
N^{\omega}(x_{1},x_{2}) & = & \frac{\gamma}{\pi} \frac{e^{i\omega\left(\tau_{0}(x_{1})-\tau_{0}(x_{2})\right)}}{\sqrt{v(x_{1})v(x_{2})}} \int_{-\infty}^{+\infty} d\omega_{q} \, \omega_{q} \, \mathrm{coth}\left(\frac{\omega_{q}}{2T_{\Psi}}\right) \, e^{-i\omega_{q}\left(\tau_{0}(x_{1})-\tau_{0}(x_{2})\right)} \nonumber \\
& = & \frac{\gamma}{2\pi} \frac{e^{i\omega\left(\tau_{0}(x_{1})-\tau_{0}(x_{2})\right)}}{\sqrt{v(x_{1})v(x_{2})}} \int_{-\infty}^{+\infty} d\omega_{q} \, \mathrm{coth}\left(\frac{\omega_{q}}{2T_{\Psi}}\right) \, \left(iv(x_{1})\partial_{x_{1}}-iv(x_{2})\partial_{x_{2}}\right) e^{-i\omega_{q}\left(\tau_{0}(x_{1})-\tau_{0}(x_{2})\right)}\,,
\label{eq:noise}
\end{eqnarray}
where in the second line we have rewritten $\omega_{q}$ in terms of derivatives of the exponential, and where
\begin{equation}
\tau_{0}(x) \equiv \int_{0}^{x} \frac{dx_{1}}{v(x_{1})} 
\end{equation}
is the proper time along a freely-falling geodesic between positions $0$ (the horizon) and $x$. 
(Since $v < 0$, $\tau_{0}(x)$ decreases when increasing $x$.) 
Plugging the expression for the noise kernel back into the expression for the anticommutator, we can integrate $x_{1}$ and $x_{2}$ by parts and integrate over $\omega_{q}$
-- using the fact that the Fourier transform of the $\mathrm{coth}$ function is another $\mathrm{coth}$ function of inverse width, as can be shown by expanding it into a sum of poles in the complex plane -- to obtain
\begin{multline}
G^{\omega}_{ac}(x,x^{\prime}) = \frac{\gamma}{\pi}\int_{-\infty}^{+\infty}dx_{1}\int_{-\infty}^{+\infty}dx_{2} \left\{ \frac{e^{i\omega \tau_{0}(x_{1})}}{\sqrt{\left|v(x_{1})\right|}}\partial_{x_{1}}G^{\omega}_{\mathrm{ret}}(x,x_{1}) \cdot \partial_{x_{2}}\left[ \frac{e^{-i\omega \tau_{0}(x_{2})}}{\sqrt{\left|v(x_{2})\right|}} v(x_{2})\partial_{x_{2}}G^{-\omega}_{\mathrm{ret}}(x^{\prime},x_{2}) \right] \right. \\
\left. -\partial_{x_{1}}\left[ \frac{e^{i\omega \tau_{0}(x_{1})}}{\sqrt{\left|v(x_{1})\right|}} v(x_{1})\partial_{x_{1}}G^{\omega}_{\mathrm{ret}}(x,x_{1})\right] \cdot \frac{e^{-i\omega \tau_{0}(x_{2})}}{\sqrt{\left|v(x_{2})\right|}} \partial_{x_{2}}G^{-\omega}_{\mathrm{ret}}(x^{\prime},x_{2}) \right\}
\times \pi T_{\Psi} \, \mathrm{coth}\left(\pi T_{\Psi} (\tau_{0}(x_{1})-\tau_{0}(x_{2}))\right) \,.
\label{eq:Gac_Nsub}
\end{multline}
This is a useful expression in that the singularity at $x_{1}=x_{2}$ is of the form of the simple pole $\left(x_{1}-x_{2}\right)^{-1}$, and can thus be integrated by taking the Cauchy principal value.  
An effective way of doing this is to switch the dummy variables $x_{1}$ and $x_{2}$, and take the mean of the two integrands; since $\mathrm{coth}(z)$ is an odd function, this has the effect of removing the singularity altogether, so that numerical integration can proceed.

\begin{figure}
\subfloat{\includegraphics[width=0.45\columnwidth]{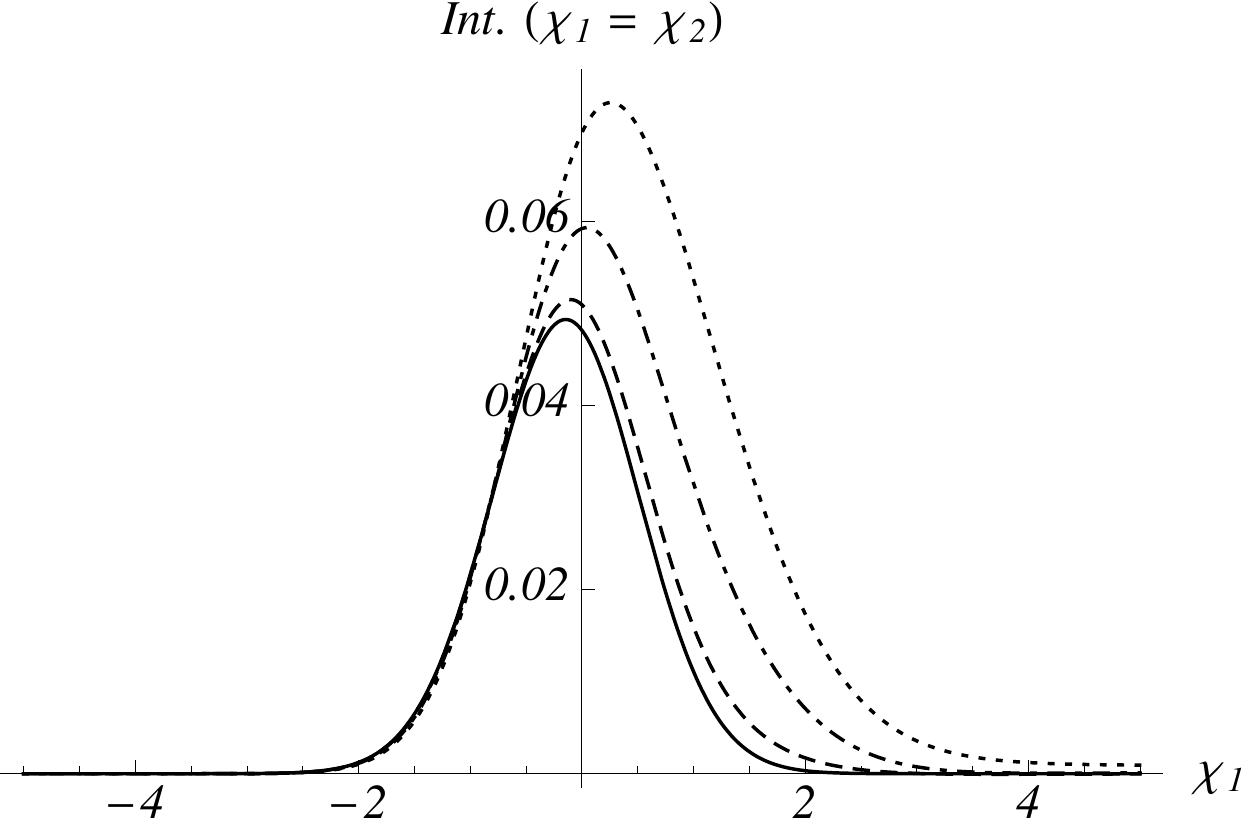}} \qquad \subfloat{\includegraphics[width=0.45\columnwidth]{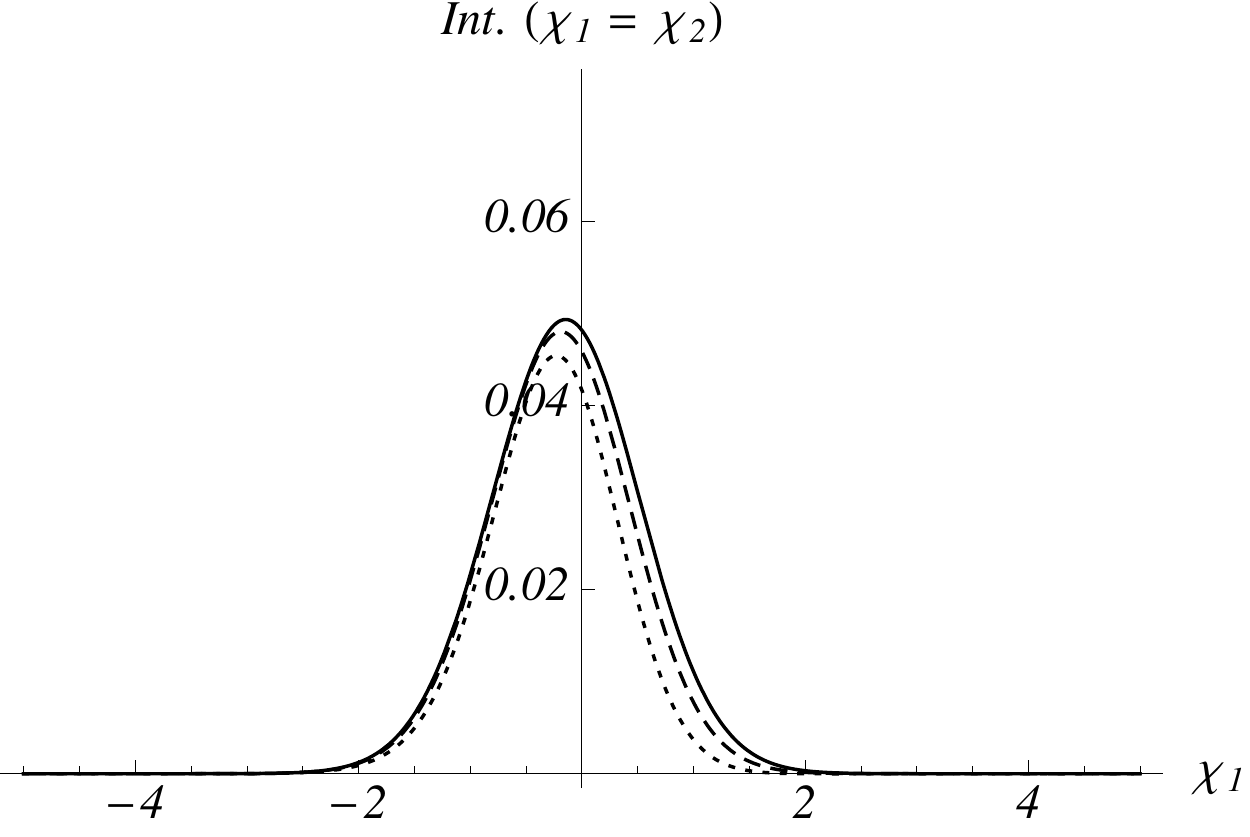}}
\caption{\textsc{Integrand for anticommutator}: Here is plotted the integrand of Eq. (\ref{eq:Gac_Nsub}) re-expressed in adimensional ``dissipative'' length units $\chi = x \sqrt{\kappa/\gamma}$.
We plot the integrand along the diagonal $\chi_{1}=\chi_{2}$, having sent $x$ and $x^{\prime}$ into the right-hand asymptotic region and factored out the product of $uR$-modes $\phi^{\omega}_{uR}(x) \phi^{-\omega}_{uR}(x^{\prime})/(2\omega)^{2}$. 
In the left plot, $\lambda^{2}=0.1$ while the various curves correspond to different values of $\omega/\kappa$: $1/20\pi$ (solid), $1/2\pi$ (dashed), $1/3$ (dot-dashed) and $1/2$ (dotted).
In the right plot, $\omega/\kappa = 1/2\pi$, while the various curves correspond to different values of $\lambda^{2}$: $0.1$ (solid), $0.2$ (dashed) and $0.4$ (dotted). 
Throughout, the geometry is fixed at $D=0.5$.
We clearly see that the integrand is peaked in the near-horizon region with a half-width of about one dissipative length $x_{0}^{\mathrm{diss}}$ (see Eqs. (\ref{eq:length_scales})).
This remains true in the limit $\lambda^2 \to 0$.
\label{fig:integrand}}
\end{figure}

In Figure \ref{fig:integrand} we illustrate the integrand of Eq. (\ref{eq:Gac_Nsub}) having re-expressed the integral using the adimensional lengths $\chi_{1}$ and $\chi_{2}$ (see Eqs. (\ref{eq:length_scales})) as dummy variables.
Using these variables, the limit $\gamma \to 0$ is finite.
The plots show the integrand along the diagonal $\chi_{1}=\chi_{2}$, having sent $x$ and $x^{\prime}$ into the right-hand asymptotic region and extracted the product of two $uR$-modes $\phi^{\omega}_{uR}(x) \phi^{-\omega}_{uR}(x^{\prime})/(2\omega)^{2}$.  
We clearly see that $\chi$ is the appropriate integration variable, with variations in $\lambda^{2}$ and $\omega/\kappa$ only slightly deforming the shape of the integrand in these units. 
Note that, as well as the modes in $x$ and $x^{\prime}$, two factors of $1/2\omega$ have been extracted so that, as can be seen in the left plot of Fig. \ref{fig:integrand}, the remaining integrand approaches a well-defined limit as $\omega \to 0$.
One of these factors ``normalizes'' the asymptotic plane waves, so that the coefficient of $\phi_{uR}^{\omega}(x) \phi_{uR}^{-\omega}(x^{\prime})/2\omega$ corresponds to $2n_{\omega}+1$.
The second factor of $1/2\omega$ shows that, as $\omega \to 0$, $n_{\omega}+1/2 \sim 1/\omega$, which corresponds to the infrared divergence of the Planck spectrum.
On the right plot, we can verify that the integrand possesses a well-defined and non-trivial limit when $\lambda^{2} \to 0$, guaranteeing that the anticommutator is a well-defined function of $x$ and $x'$ in the relativistic limit.

\begin{figure}
\subfloat{\includegraphics[width=0.45\columnwidth]{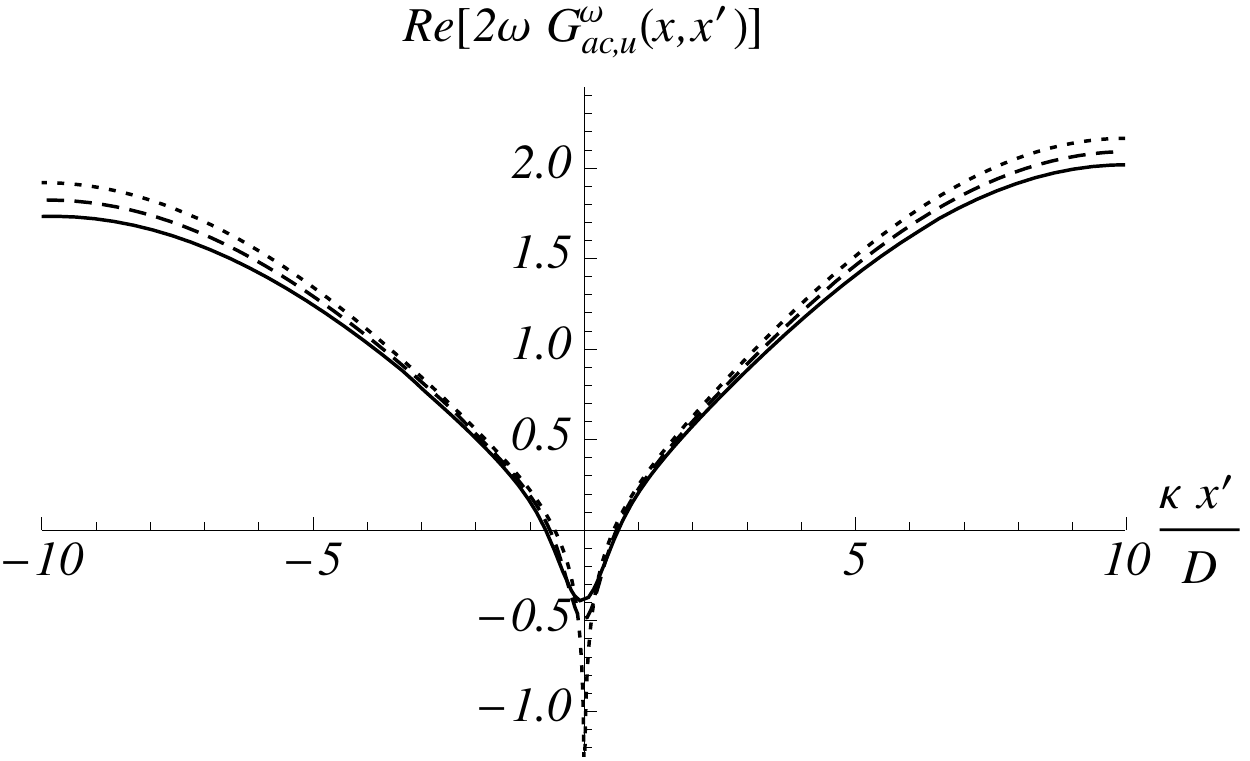}} \qquad \subfloat{\includegraphics[width=0.45\columnwidth]{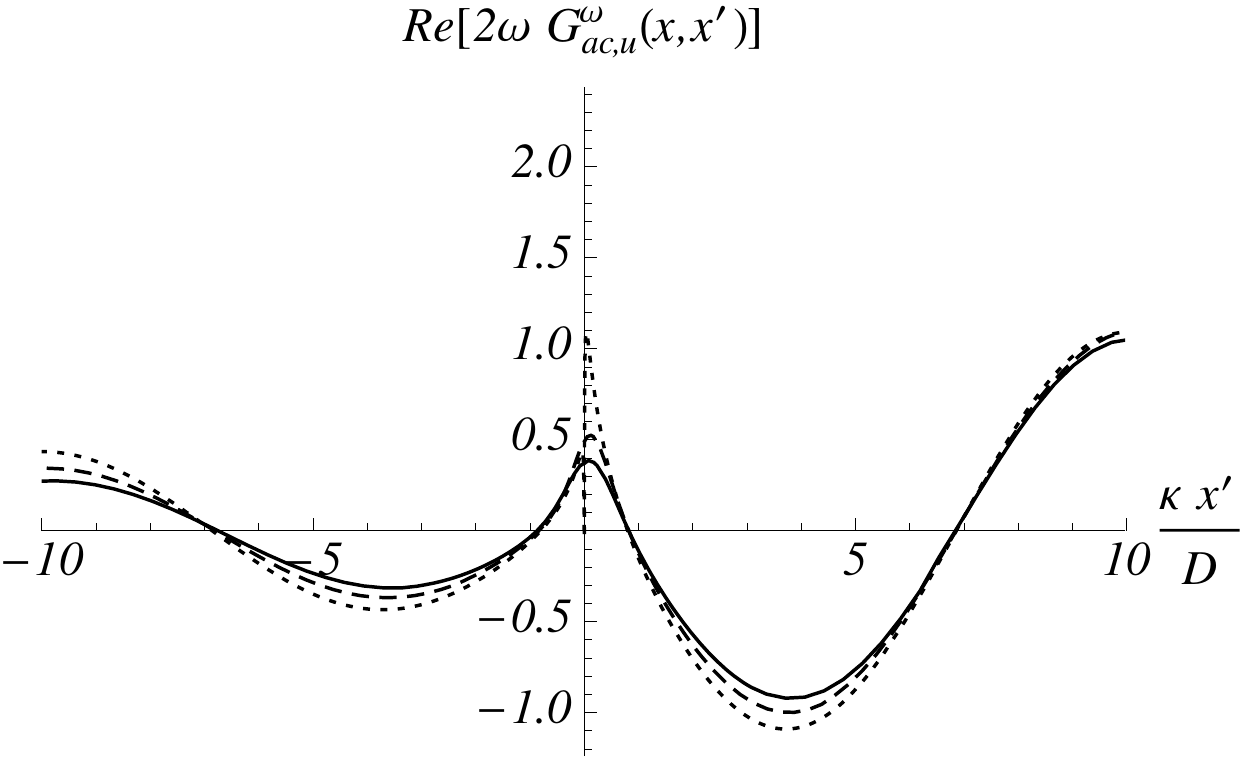}} \\
\subfloat{\includegraphics[width=0.45\columnwidth]{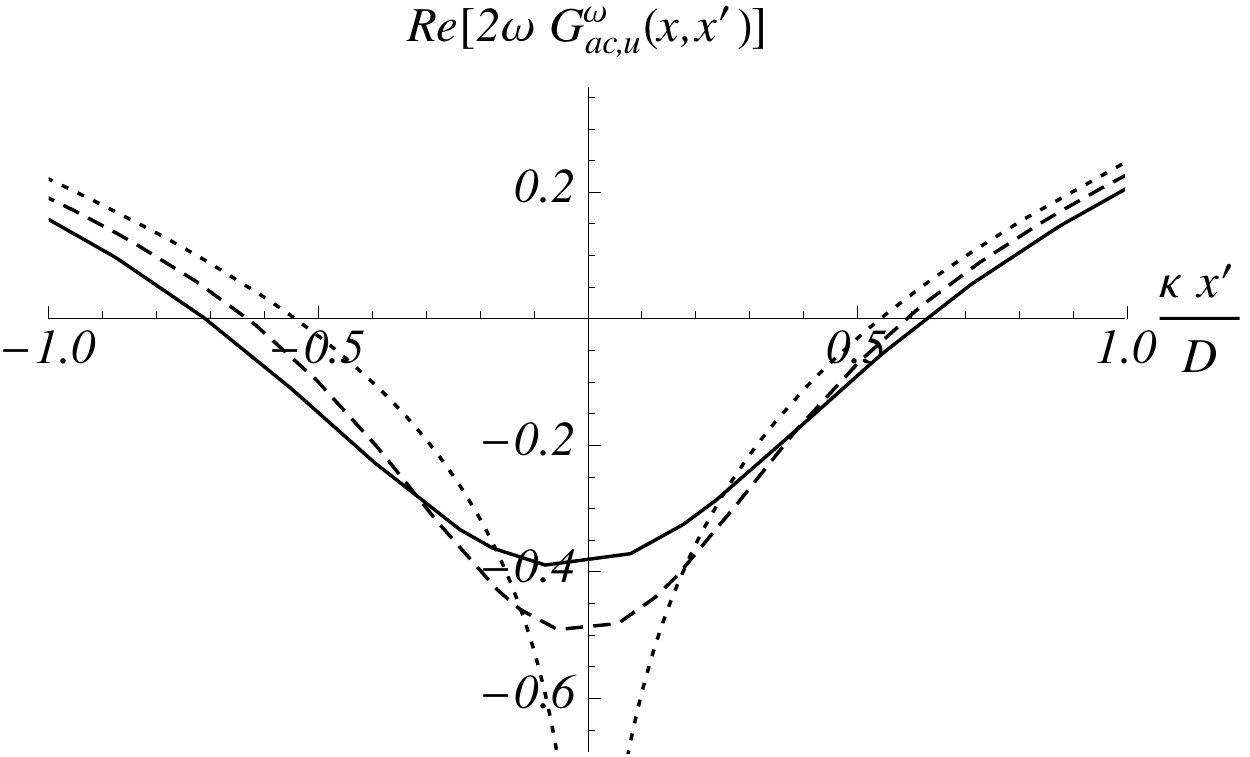}} \qquad \subfloat{\includegraphics[width=0.45\columnwidth]{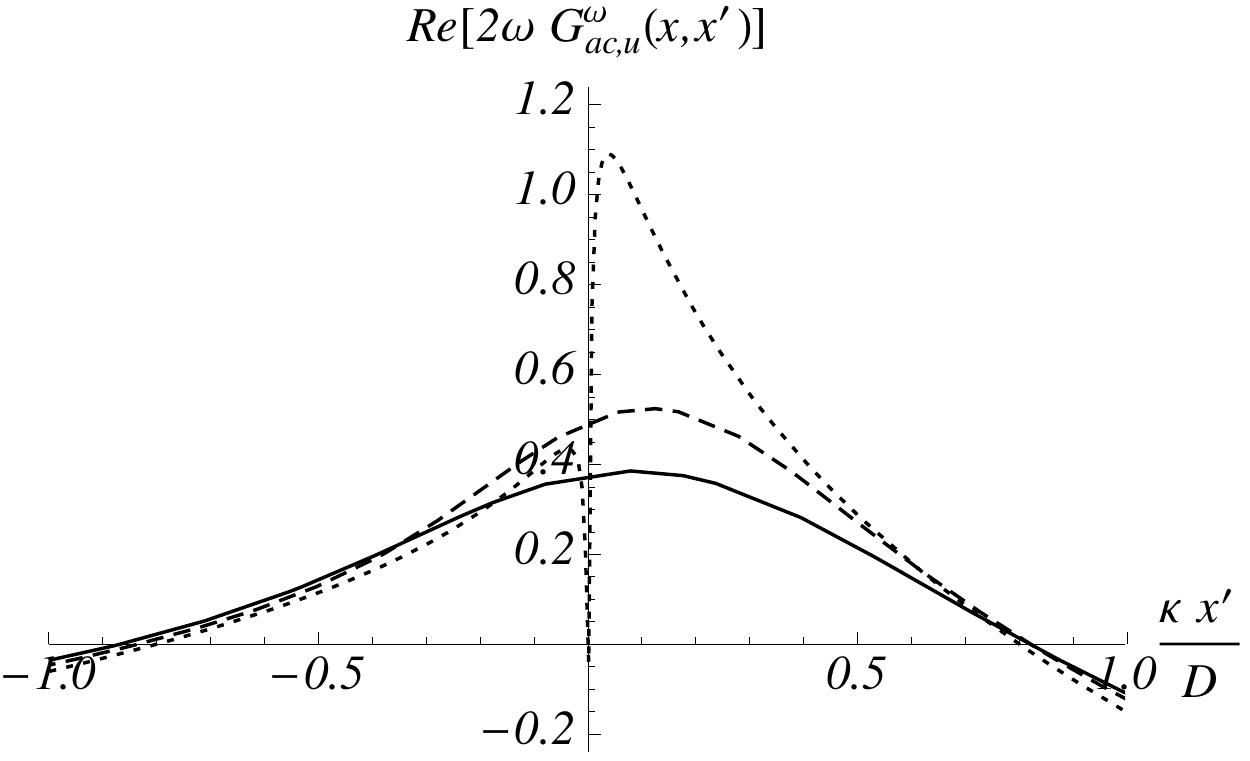}}
\caption{\textsc{Anticommutator}: We plot here the real part of the $u$-mode contribution to the anticommutator, $G_{\mathrm{ac,u}}^{\omega}(x,x^{\prime})$, as a function of $x^{\prime}$, with $x$ fixed at $\kappa x/D=10$, the geometry fixed at $D=0.5$ and the environment temperature fixed at $T_{\Psi}/\kappa = 10^{-2}$.
The dotted curves plot the non-dissipative limit (note that $G_{\mathrm{ac,u}}^{\rm relat.}$ has a phase singularity as $x^{\prime} \to 0$); the dashed and solid curves have $\lambda^{2} = 0.05$ and $0.1$, respectively.
The two columns correspond to different frequencies: $\omega/\kappa = 1/2\pi$ (left column) and $1/2$ (right column).
The second row shows the same information as the first row, but zoomed in to the near-horizon region, so that effects of dissipation can be better observed.
As explained in the text, up to (small) damping factors discussed in the next section, on the asymptotic right side, the value of $(2 \omega) \times G_{\mathrm{ac,u}}$ gives $2n_{\omega}+1$ whereas when evaluated on opposite sides it gives the real part of $ 2 c_{\omega} e^{i \theta(x,x^{\prime})}$ where the phase $\theta(x,x')$ accounts for the propagation from $x$ to $x'$.
We clearly see that dissipation in the UV removes the singular behavior as $x' \to 0$ of the relativistic case while preserving the asymptotic properties of $G_{\mathrm{ac},u}$ on both sides of the horizon. 
\label{fig:Gac}}
\end{figure}

In Figure \ref{fig:Gac} are plotted examples of the $u$ part of the anticommutator $G_{\mathrm{ac},u}^{\omega}(x,x^{\prime})$ as a function of the source coordinate $x^{\prime}$, $x$ being fixed at $\kappa x/D = 10$, far in the subsonic asymptotic region; the geometry is fixed at $D=0.5$, and the environment temperature is very low: $T_{\Psi}/\kappa = 10^{-2}$. 
(This case is very close to the zero temperature limit since $T_{\Psi}/T_H \sim 1/15$, where $T_H = \kappa/2\pi$ is the Hawking temperature.) 
More precisely, we represent the first term of \eq{eq:Gac_decomp_diss} (found by replacing $G^{\omega}_{\mathrm{ret}}$ with its $u$ part $G^{\omega}_{\mathrm{ret},u}$ in Eq. (\ref{eq:Gac_Nsub})) multiplied by $2 \omega$ so as to be able to read off the values of the power spectrum $(= 2n_\omega + 1)$ for $x' = x$, and of the strength of the correlation $2 c_\omega$ for $x' = -x$, see \eq{eq:Gac_decomp_nondiss}.
Various values of $\lambda^2$ have been used, including $\lambda^2=0$.
In this relativistic case, the state of the field has been taken to be the standard Unruh vacuum~\cite{Brout-et-al-Primer}. 
The two columns correspond to different frequencies, with $\omega/\kappa = 1/2\pi$ in the left column and $\omega/\kappa = 1/2$ in the right column. 

Three important observations should be made.
Firstly, on the right side of the two upper plots, by comparing the two curves ($\lambda^2 = 0.05$ and $0.1$) with the relativistic case, we can see with the naked eye that dissipation does not have a significant effect on the mean number of quasiparticles emitted to the right.
This means that the low temperature fluctuations of the environment feed the radiation field (in a vicinity of the horizon approximately given by one dissipative length $x_{0}^{\mathrm{diss}}$ of Eq. (\ref{eq:length_scales})) in such a manner as to bring its state to the Unruh vacuum in the limit $\lambda^{2} \to 0$.
The smallness of the deviations for $\lambda^2 \ll 1$ confirms that the asymptotic spectrum is robust against introducing dissipation in the ultra-violet sector. 
(In appendix~\ref{hightemp}, we shall examine to what extent this remains true when increasing the temperature of the environment above the Hawking temperature.)
On the right upper plot, notice that the asymptotic value $\sim 1$ means that we are close to the vacuum for this value of the Killing frequency $\omega = \kappa/2$, i.e., about three times the Hawking temperature $T_H = \kappa/2\pi$.

Secondly, as for the retarded Green function, see Fig.~\ref{fig:Gret}, dissipation smooths out the logarithmic phase singularity (in $|x'|^{i \omega/\kappa}$ as $|x'| \to 0$) that occurs at the horizon in the relativistic case; this is most clearly seen in the second row, which zooms in on the near-horizon region.
In this sense, the ``trans-Planckian'' behavior in the vicinity of the horizon is completely erased in our dissipative model with $\Gamma = \gamma P^2$.

Thirdly, and most interestingly, unlike for the retarded Green function and despite the large deviations caused by dissipation in the near-horizon region, it is seen that on the {\it other} side of the horizon the anticommutator redevelops so as to smoothly join to the oscillatory behavior that occurs in the absence of dissipation (and dispersion). 
While the amplitude is slightly decreased with increasing dissipation, the nodes are seen to occur at the same places.
As a result, the {\it relative phase} between the two asymptotic regions, which is encoded in the phase of the $c_\omega$ coefficient of \eq{eq:nandc}, is preserved. 
In this we generalize the observation that $n_\omega$ {\it and} $c_\omega$ are both robust when introducing high-frequency dispersion~\cite{Brout-et-al-1995}.
In other words, the asymptotic state of the $u$-modes is robust against UV dissipation, i.e., the differences of $n_\omega$ {\it and} $c_\omega$ with respect to the relativistic values linearly vanish with $\lambda^2$ when $\lambda^2 \ll 1$. 

Yet, in order to determine the domain where the state is nonseparable, which is that where $\Delta_\omega$ of \eq{eq:nonseparability} is negative, we need to extract the occupation numbers and the correlation term with precision. 


\subsection{Spectrum and entanglement for low environment temperature
\label{lowtemp}}

\begin{figure}
\subfloat{\includegraphics[width=0.45\columnwidth]{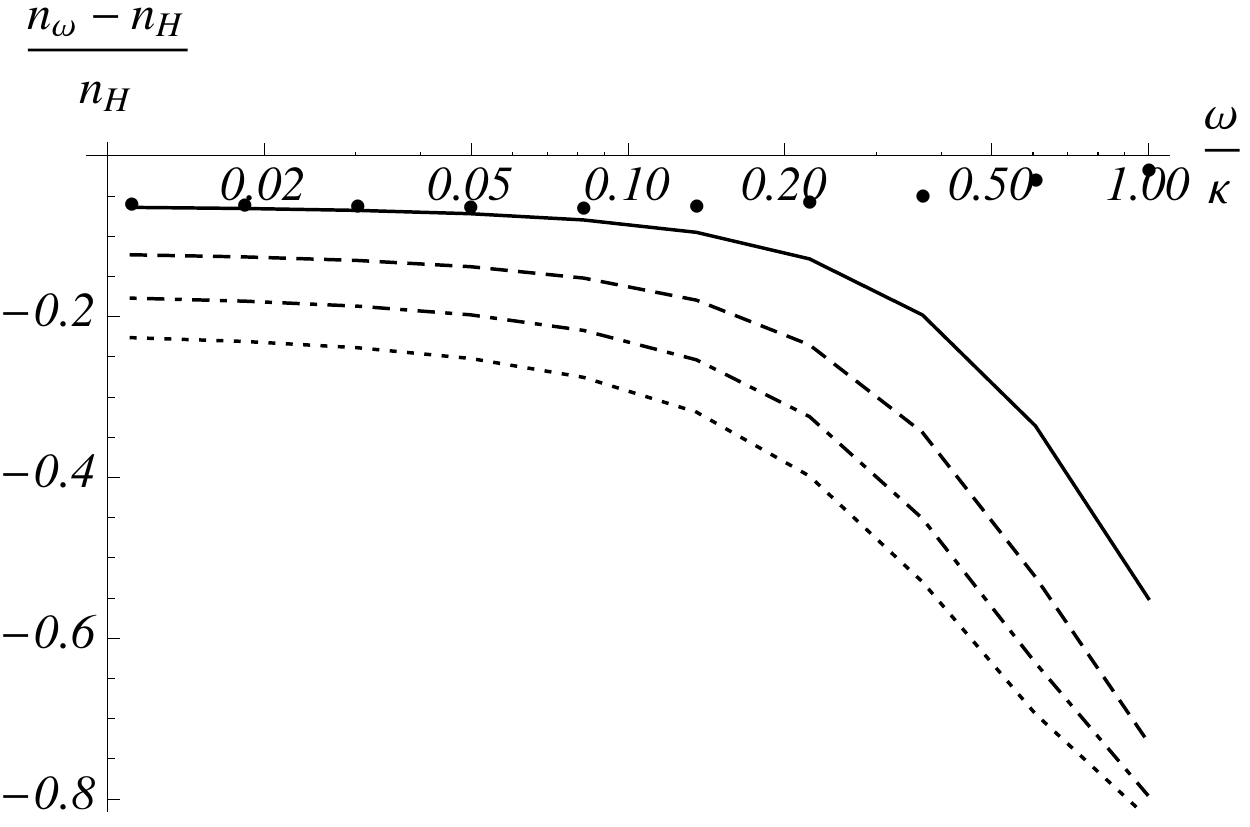}} \qquad \subfloat{\includegraphics[width=0.45\columnwidth]{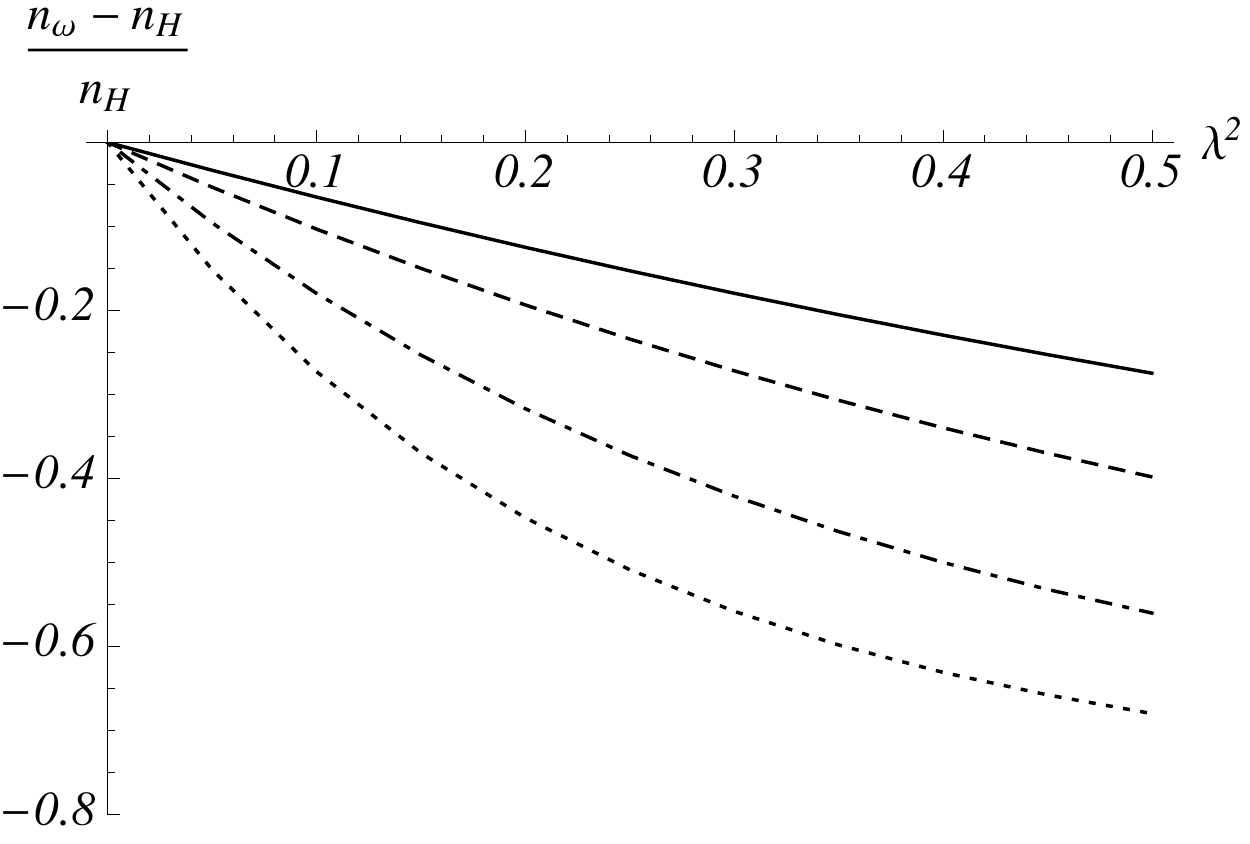}}
\caption{\textsc{Power spectrum at low temperature}: Shown here is the relative change in the number of quasiparticles from the Hawking thermal prediction, in the right (subsonic) asymptotic region. 
The temperature is fixed at $T_{\Psi}/\kappa = 10^{-2}$, with measurements made at a fixed distance from the horizon of $\kappa \left|x\right|/D = 5$.
The geometry itself is fixed at $D=0.5$.  
In the left column, the variable is $\omega/\kappa$, and the various curves correspond to different values of $\lambda^2$: $0.1$ (solid curve), $0.2$ (dashed curve), $0.3$ (dot-dashed curve) and $0.4$ (dotted curve).
The dots show the data points of the solid curve corrected for the residual damping outside of the near-horizon region, which is most significant at high frequencies. 
In the right column, the variable is instead the adimensional parameter $\lambda^2$ of \eq{eq:lambda_defn}, while the various curves correspond to different values of $\omega/\kappa$: $1/20\pi$ (solid curve), $1/2\pi$ (dashed curve), $1/3$ (dot-dashed curve) and $1/2$ (dotted curve).  
On both plots, we clearly see the linearity of the deviations in $\lambda^2$ when $\lambda^2\ll 1$.
\label{fig:LowTemp}}
\end{figure}

Because of the residual coupling to the environment, the modes are still slightly damped in the asymptotic regions.
Hence, the extraction from the anticommutator of the values of $n_\omega$ and $c_\omega$ must either be done at a finite distance from the horizon, or one should explicitly switch off the interactions between the radiation field and its environment. 
As a result, the identification of  $n_\omega$ and $c_\omega$ is inevitably slightly imprecise. 
The importance of the errors is governed by $\Gamma_{\rm as}/\omega$, where $\Gamma_{\rm as}$ is the asymptotic residual decay rate.
For low values of $\lambda^2$ of \eq{eq:lambda_defn}, this is not a serious problem in our model because $\Gamma_{\rm as}/\omega$ is equal to $\lambda^2 \omega/\kappa$.
Similar errors have been discussed in homogeneous time dependent settings in Sec. IV.B.2 of~\cite{Busch-Parentani-2013}. 
To minimize the total absorption during outwards propagation from the black hole, while also having a locally flat geometry, we work at $\kappa |x| = 5 D$. 

In Figure \ref{fig:LowTemp} we plot the spectral deviations on the subsonic side when the environment is again at a low value of $T_{\Psi}/\kappa = 10^{-2}$, the velocity profile fixed at $D=0.5$ and the positions of the hypothetical particle detectors fixed at $\kappa \left|x\right|/D = 5$.
The limit $\lambda^2 \rightarrow 0$ thus corresponds to the relativistic limit. 
(We do not represent the spectrum of negative energy partners in the left (supersonic) asymptotic region because it behaves very similarly.)
From the left plot, we see that for low frequencies, the relative difference in the quasiparticle numbers from their relativistic values tends to a constant, which translates into an effective change (specifically a reduction, since $n_\omega$ decreases) in the black hole temperature. 
At higher frequencies, we notice that the relative difference is greater, meaning that the spectrum falls off more rapidly than it would for a purely Planckian spectrum. 

To see to what extent this reduction is due to damping of the modes as they propagate from the horizon to the detector, we multiplied $n_{\omega}$ by the correction factor $\exp(2 \int^{x^{\rm det.}}_{x^{\rm em.}} dx' \Im k_\om (x'))$ which accounts for the damping from $x^{\rm em.}$ to $x^{\rm det.}$.
If the final value $x^{\rm det.}$ is clear, and given here by $\kappa x/D = 5$, some rule should be adopted for determining the locus of emission $x^{\rm em.}$. 
We adopt the location where the diagonal of the integrand of \eq{eq:Gac_Nsub} is equal to $1/5$ its maximum value, see Figure~\ref{fig:integrand}. 
The result is represented by the dots in Figure \ref{fig:LowTemp}. 
From this we conclude that the higher damping observed at high frequency is largely due to the mode propagation, and not intrinsic to the black hole emission. 

From the right plot, we see that, for a fixed frequency $\omega$, the relative difference in occupation number linearly increases with the parameter $\lambda^2$ when it is smaller than $0.1$.
For $\omega/\kappa \sim 1$, the coefficient of proportionality is of order 1.
The fact that the relative difference in $n_\omega$ due to dissipation is proportional to $\lambda^2$ was anticipated in \cite{Adamek-Busch-Parentani-2013} on the basis that, firstly, the near-horizon black hole physics is similar to that occurring in de Sitter space, and secondly, that the relative difference in $n_\omega$ due to dissipation in de Sitter was found to be extremely small.
(The subdominant deviation of the spectrum due to $D$ at fixed $\lambda^2$ is studied in appendix~\ref{changeD}.) 
It is therefore the departure from the de Sitter physics that fixes the leading spectral deviations of the black hole spectrum. 
It should be noticed that similar observations apply to dispersion effects~\cite{Coutant-RP-SF}, and also to the local description of the correlations (encoded in the anticommutator): as long as one probes the $G_{\rm ac}$ in the de Sitter-like near-horizon region, there is no departure from the vacuum structure (both for relativistic fields, and for weakly dispersive/dissipative fields). 
The typical long distance behavior~\cite{MassarRP1996,Carusotto-et-al-2008} is gradually obtained as one {\it leaves} the de Sitter-like region~\cite{parentani2010}. 

\begin{figure}
\subfloat{\includegraphics[width=0.45\columnwidth]{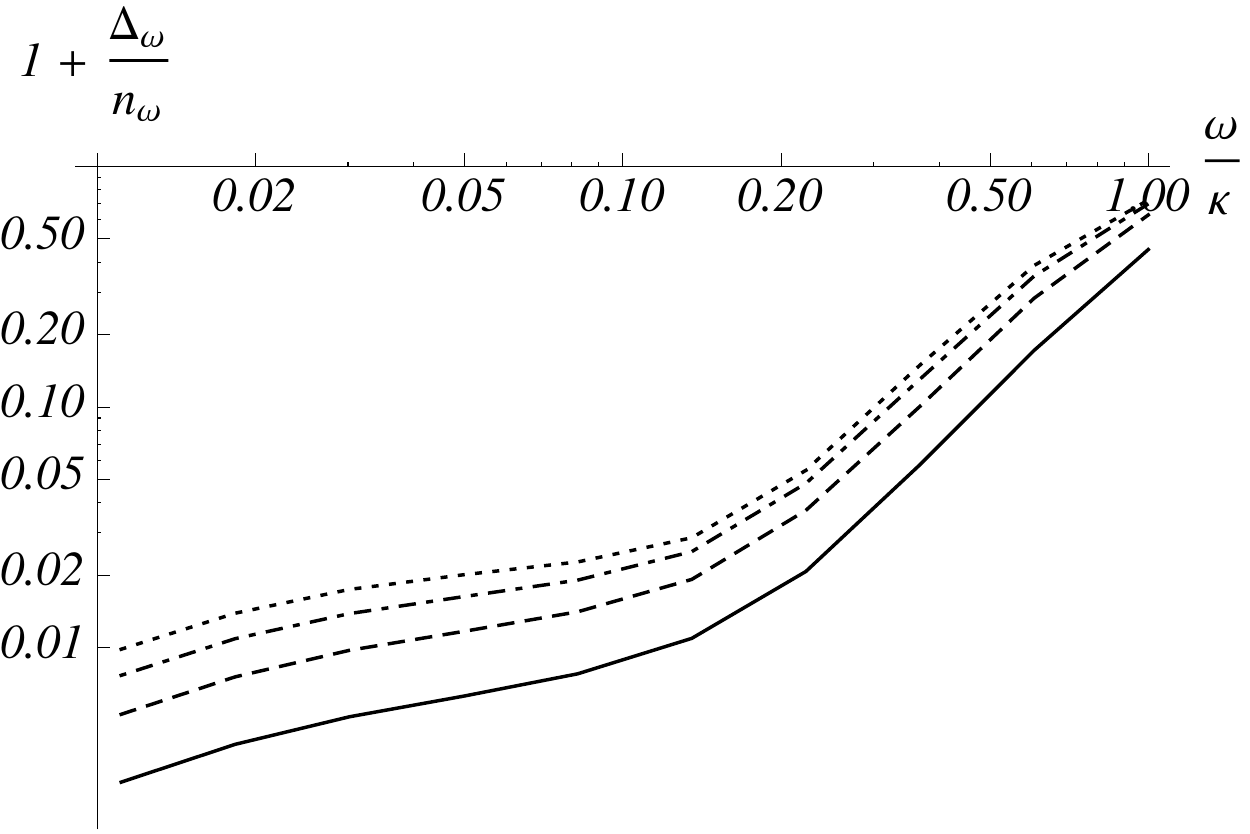}} \qquad \subfloat{\includegraphics[width=0.45\columnwidth]{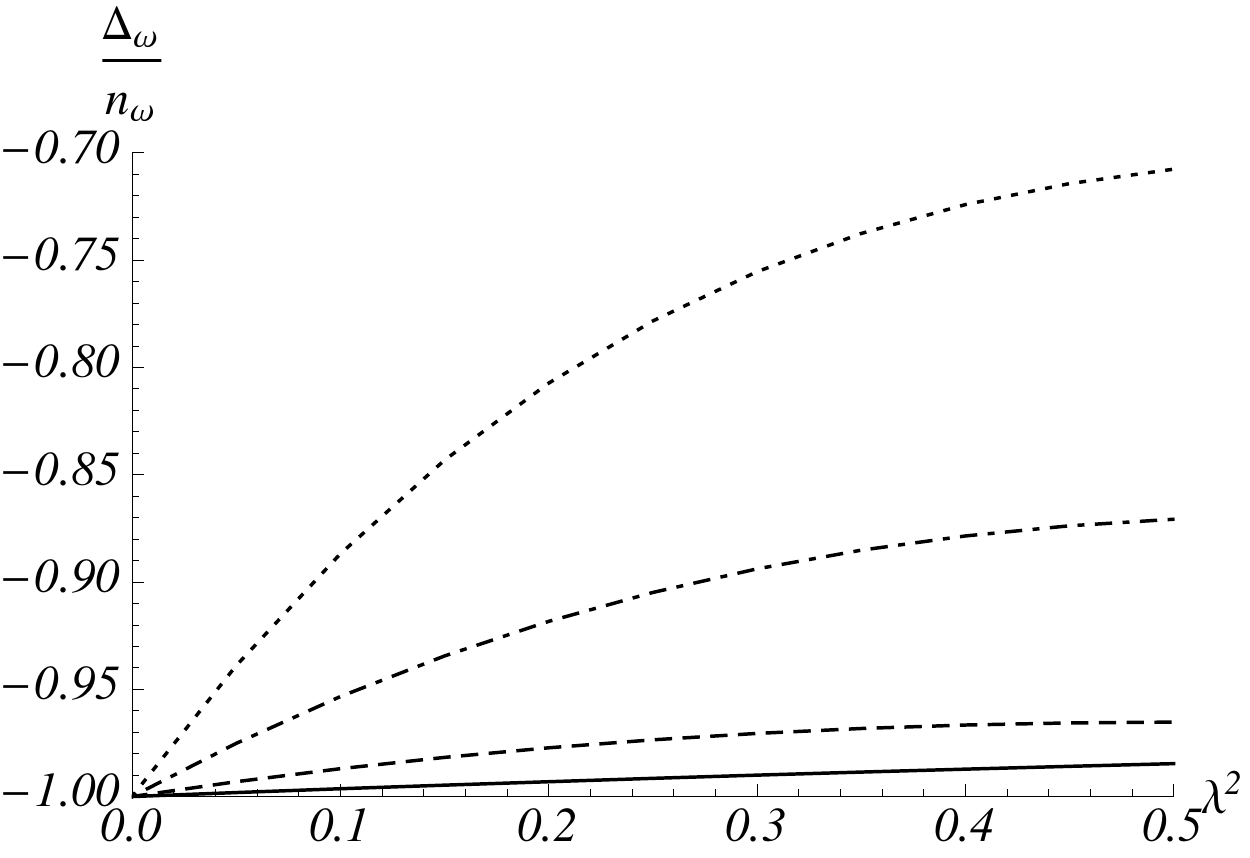}}
\caption{\textsc{Nonseparability at low temperature}: Here is plotted the rescaled nonseparability parameter, $\Delta_{\omega}/n_{\omega}$ of \eq{eq:nonseparability}, which is bounded from below by $-1$; to make the differences more visible, $1+\Delta_{\omega}/n_{\omega}$ has been plotted in logarithmic scale in the left plot.
As in Fig. \ref{fig:LowTemp}, we work with $T_{\Psi}/\kappa = 10^{-2}$, and $\kappa \left|x\right|/D = 5$, where $D=0.5$.
In the left column, the variable is $\omega/\kappa$, and the various curves correspond to $\lambda^2$: $0.1$ (solid curve), $0.2$ (dashed curve), $0.3$ (dot-dashed curve) and $0.4$ (dotted curve).
In the right column, the variable is instead $\lambda^2$ of \eq{eq:lambda_defn}, while the various curves correspond to $\omega/\kappa$: $1/20\pi$ (solid curve), $1/2\pi$ (dashed curve), $1/3$ (dot-dashed curve) and $1/2$ (dotted curve).
Even though the departures from the relativistc case $\lambda^2 = 0$ are clear, quite remarkably, in all represented cases, the two-mode states of $u$-quanta remain nonseparable. 
\label{fig:DeltaLowTemp}}
\end{figure}

In Figure \ref{fig:DeltaLowTemp} we plot the parameter $\Delta_\omega$ of \eq{eq:nonseparability}, the sign of which determines if the final state is entangled or not.
We observe a strong robustness of the nonseparability of the emitted quasiparticles: no loss of nonseparability is observed for the values considered, though it appears that it would be lost when we consider higher frequencies, $\omega/\kappa \gg 1$, which are subject to dissipation outside the near-horizon region.
The strong robustness can be understood as a new manifestation of the fact that the mode mixing responsible for the Hawking effect effectively occurs at low momentum $P\sim \kappa$. 
In effect, strong dissipation occurs at high momenta $P$, before vacuum configurations are converted into on shell particles, as discussed above. 
Therefore dissipation does not significantly reduce the entanglement. 
When studying dissipative effects in cosmological settings, the importance of the time ordering of dissipation and pair creation was clearly established~\cite{adamek-campo-niemeyer}.  
As a result, when working in de Sitter space~\cite{Adamek-Busch-Parentani-2013}, for a given field with dissipative effects in the deep UV sector, it was found that nonseparability can be lost for the cosmological pairs with opposite momenta $(k,-k)$ (because they are produced when dissipation is still important), whereas it can be preserved for the $u$-pairs $(\omega,-\omega)$. 


\section{Conclusions
\label{sec:Conclusions}}

We have studied the effects of dissipation on the properties of the quasiparticle pairs emitted by an analogue black hole.
Since dissipation implies the absence of well-defined particle modes and their corresponding annihilation and creation operators, we found the properties of the radiation field through examination of its anticommutator, which for our Gaussian model completely determines the asymptotic state of the field.
We worked with $1+1$-dimensional stationary models.

To preserve unitarity of the system as a whole, we began with an action that included a dense environment, which was later traced over to obtain a driven-dissipative equation for the radiation field. 
The kinematical term of the environmental degrees of freedom, and their coupling to the radiation field were fine-tuned so as to yield a simple, local dissipative equation. 
In particular, the decoupling of $u$- and $v$-modes (which is obtained for massless relativistic fields in two dimensions) was preserved for the dissipated modes in a model where the decay rate grows like the square of the momentum measured in a freely falling frame ($\Gamma \propto P^{2}$).  
This allowed relatively easy computation of the retarded Green function, which was found to depend on only two dimensionless parameters: the frequency $\omega/\kappa$ and $\lambda^{2} = \gamma\kappa/D^{2}$, which combines the dissipative frequency $\Lambda_{\rm diss} = 1/\gamma$, the black hole surface gravity $\kappa$, and $D$, the extension of the near-horizon region which can be mapped onto a stationary portion of de Sitter space. 
As a result, the limit $\lambda^{2}\to 0$ can be conceived in two different ways: as giving the relativistic limit in a fixed metric, or as working in de Sitter space with a fixed dissipative frequency $\Lambda_{\rm diss}$. 
At fixed Killing frequency, we showed that the retarded Green function can be canonically expressed in a double mode basis containing, on the one hand, the dissipated modes, and on the other hand, dual modes obeying well-defined conditions on asymptotic spatial infinities.
The origin of this writing, which should apply to all unitary models as indicated in Appendix~\ref{app:Gret}, comes from the fact that the retarded Green function itself is the unique solution which is asymptotically bounded. 
We then studied its near-horizon properties in the model $\Gamma \propto P^{2}$, and saw that the singular behavior of the relativistic theory in $\theta(x) \, x^{i \omega/\kappa}$ as $x \to 0$ was fully regularized by dissipation.

The Green function was then used to calculate the anticommutator, the calculation of the double integral being done numerically.
Focusing on the $\Gamma \propto P^{2}$ model, the asymptotic spectra and long-distance correlations of the emitted pairs were calculated along with the local properties of the anticommutator.
Two apparently contradictory results were found. 
On the one hand, when sending one point into the near-horizon region while keeping the other fixed far away, we recovered the radical departure from the relativistic case (already found when studying the retarded Green function) by which the singular behavior at the horizon is smoothed out.  
On the other hand, despite this radical departure close to the horizon, we find that when $\lambda^2 \ll 1$, the anticommutator far from the horizon in both directions behaves essentially like that of a relativistic field in the Unruh vacuum.
In other words, the strong interactions occurring at high momenta $P^2 \sim \kappa \Lambda_{\rm diss}$ in the near-horizon region are such that the fluctuations of the environment bring the state of the radiation field to its regular ground state. 
The origin of this fact can be understood by analyzing the behavior of the integrand of the double integral defining the anticommutator: one finds that it possesses a scaling property which guarantees that the decoupling limit $\lambda^2 \to 0$, i.e., $\Lambda_{\rm diss}\to \infty$, is non-trivial and implements the fluctuation-dissipation theorem, namely that the state of the radiation field is the same as that of its environment. 
We then computed the leading spectral deviations due to dissipation.
For a fixed value of $\omega/\kappa$, we found that the deviations principally (and linearly) depend on $\lambda^{2}$. 
This confirmed that the spatial extension of the de Sitter-like near-horizon region plays a crucial role in determining the deviations.

In Appendix \ref{hightemp} we verified that the asymptotic spectrum and the correlations remain as in the Unruh vacuum when the temperature of the environment is higher than the Hawking temperature, provided dissipation only occurs above a sufficiently high energy scale. 
The robustness stems from the large redshift which effectively suppresses the stimulated emission.
Hence, when $\Lambda_{\rm diss}$ is sufficiently high, only the spontaneous contribution due to vacuum fluctuations is responsible for the asymptotic properties. 
This was explicitly verified by calculating the parameter $\Delta_\omega$ which measures the degree of the nonseparability of the emitted pairs: we observed that the final state remains entangled over a wide range of environment temperatures and dissipative rates.

In conclusion, even though we worked with fined-tuned dissipative models, mainly characterized by an absence of dispersion in the limit $\lambda^2 \to 0$ and decoupling of the $u$- and $v$-sectors at the level of the field modes, we claim that our findings are generic in character. 
Our principal findings concern, on the one hand, the mode structure of the retarded Green function at fixed $\omega$, and how it determines both the local and the asymptotic properties of the anticommutator; and on the other hand, the robustness of the spectrum and the correlation strength when $\lambda^2 \ll 1$, and the properties of their leading deviations. 
These results should also apply to more complicated models which describe analogue black hole flows in water waves, in acoustics, and in polariton systems.
Note however that the latter case is more sensitive to dissipation as the rate $\Gamma$ is independent of the momentum; see Appendix \ref{app:Gamma_constant}.


\acknowledgements{We thank Florent Michel for a careful reading of the manuscript and useful comments.
We also thank Xavier Busch for helpful discussions on the technical aspects of our model.
This work was supported by the French National Research Agency under the Program Investing in the Future Grant No. ANR-11-IDEX-0003-02 associated with the project QEAGE (Quantum Effects in Analogue Gravity Experiments).} 


\appendix
\appendixpage
\numberwithin{equation}{section}


\section{Derivation of the driven-dissipative wave equation
\label{app:Wave_eqn_derivation}}

In this appendix, we show how Eq. (\ref{eq:final_phi_eqn}) is derived from Eqs. (\ref{eq:field_eqns})-(\ref{eq:delta_identity}) plus some additional simplifying assumptions.

Substituting Eq. (\ref{eq:Psi_soln}) in Eq. (\ref{eq:phi_eqn}) and using Eq. (\ref{eq:delta_identity}) yields the following:
\begin{multline}
\left[ \left( \partial_{t}+\partial_{x}v\right)\left(\partial_{t}+v\partial_{x}\right) - \partial_{x}^{2} + m^{2} + f\left(-\partial_{x}^{2}\right) + (-1)^{n} \partial_{x}^{n} g \tilde{\partial}_{\tau} g \partial_{x}^{n} \right] \phi \\
= \left[ \left( \partial_{t}+\partial_{x}v\right)\left(\partial_{t}+v\partial_{x}\right) - \partial_{x}^{2} + m^{2} + f\left(-\partial_{x}^{2}\right) +\frac{(-1)^{n}}{2} \partial_{x}^{n} g \left(\partial_{t}+\partial_{x}v\right)g\partial_{x}^{n} + \frac{(-1)^{n}}{2} \partial_{x}^{n}g\left(\partial_{t}+v\partial_{x}\right)g\partial_{x}^{n} \right] \phi \\
= (-1)^{n}\partial_{x}^{n} g \int dq \, \tilde{\partial}_{\tau} \Psi_{q}^{0} \,,
\label{eq:phi_diss-driven}\end{multline}
where in the second line we have used Eq. (\ref{eq:dtau_defn}) to expand $\tilde{\partial}_{\tau}$.
It is convenient to combine the last two terms of the second line with the first term, but to do this we must first change the order of some of the differential operators.
We require the following operator identities:
\begin{subequations}\begin{eqnarray}
\partial_{x}^{n}g\left(\partial_{t}+\partial_{x}v\right)g\partial_{x}^{n} & = & \left(\partial_{t}+\partial_{x}v\right)\partial_{x}^{n}g^{2}\partial_{x}^{n} + \left(\partial_{x}^{n}g\partial_{x}v-\partial_{x}v\partial_{x}^{n}g\right)g\partial_{x}^{n} \nonumber \\
& = & \left(\partial_{t}+\partial_{x}v\right)\partial_{x}^{n}g^{2}\partial_{x}^{n} + \left( \partial_{x}^{n+1}g v - \partial_{x}^{n} g^{\prime} v - \partial_{x}v\partial_{x}^{n}g\right)g\partial_{x}^{n} \nonumber \\
& = & 2 \left(\partial_{t}+\partial_{x}v\right)\partial_{x}^{n}\gamma\partial_{x}^{n} + 2 \partial_{x} \left( \partial_{x}^{n}v-v\partial_{x}^{n} \right) \gamma \partial_{x}^{n} - \partial_{x}^{n} v\gamma^{\prime} \partial_{x}^{n} \,, \\
\partial_{x}^{n}g\left(\partial_{t}+v\partial_{x}\right)g\partial_{x}^{n} & = & \partial_{x}^{n}g^{2}\partial_{x}^{n}\left(\partial_{t}+v\partial_{x}\right) + \partial_{x}^{n}g\left(v\partial_{x}g\partial_{x}^{n}-g\partial_{x}^{n}v\partial_{x}\right) \nonumber \\
& = & \partial_{x}^{n}g^{2}\partial_{x}^{n}\left(\partial_{t}+v\partial_{x}\right) + \partial_{x}^{n}g\left(vg\partial_{x}^{n+1} + vg^{\prime}\partial_{x}^{n}-g\partial_{x}^{n}v\partial_{x}\right) \nonumber \\
& = & 2\partial_{x}^{n}\gamma\partial_{x}^{n}\left(\partial_{t}+v\partial_{x}\right) - 2\partial_{x}^{n}\gamma\left(\partial_{x}^{n}v-v\partial_{x}^{n}\right)\partial_{x} + \partial_{x}^{n}v\gamma^{\prime}\partial_{x}^{n} \,,
\end{eqnarray}\label{eq:diff_ops_rearranged}\end{subequations}
where in the last lines we have defined $\gamma \equiv g^{2}/2$, and a prime denotes the $x$-derivative.
Using Eqs. (\ref{eq:diff_ops_rearranged}) to replace the last two terms of Eq. (\ref{eq:phi_diss-driven}) and rearranging, we find
\begin{multline}
\left[ \left(\partial_{t}+\partial_{x}v + (-1)^{n}\partial_{x}^{n}\gamma\partial_{x}^{n}\right)\left(\partial_{t}+v\partial_{x} + (-1)^{n}\partial_{x}^{n}\gamma\partial_{x}^{n}\right) - \partial_{x}^{2} \right. \\
\left. + \, m^{2} + f\left(-\partial_{x}^{2}\right) - \left(\partial_{x}^{n}\gamma\partial_{x}^{n}\right)^{2} + (-1)^{n} \partial_{x}\left(\partial_{x}^{n}v-v\partial_{x}^{n}\right)\gamma\partial_{x}^{n} - (-1)^{n} \partial_{x}^{n}\gamma\left(\partial_{x}^{n}v-v\partial_{x}^{n}\right)\partial_{x} \right] \phi = \left(-\partial_{x}\right)^{n} J(t,x) \,,
\label{eq:phi_eqn_rearranged}
\end{multline}
where $J(t,x) \equiv g \int dq \, \tilde{\partial}_{\tau}\Psi_{q}^{0}$.

Equation (\ref{eq:phi_eqn_rearranged}) is quite general: $v$ and $\gamma$ are arbitrary functions of $x$, $n$ is an arbitrary (nonnegative) integer, and $f(\partial_{x}^{2})$ remains unspecified. 
Interestingly, the equation becomes much simpler in certain cases.
For example, if we restrict $n$ to be $0$ or $1$, then the last two terms on the right-hand side automatically vanish: when $n=0$ the operator $\partial_{x}^{n}v-v\partial_{x}^{n}$ vanishes, and when $n=1$ it is simply $v^{\prime}$ and the two terms cancel out.
We shall thus restrict ourselves to these two cases to exploit this simplification.
(We note in passing that the last two terms on the right-hand side of (\ref{eq:phi_eqn_rearranged}) only contain dervatives of $v$ of at least second order; therefore, in de Sitter space where $v$ is proportional to $x$, they vanish identically for {\it all} values of $n$.)

It is also useful to choose the parameters of Eq. (\ref{eq:phi_eqn_rearranged}) such that the differential operator on the right-hand side factorizes (thus leading to the $u$-$v$ decoupling described in \S\ref{sub:Field_eqns}).
This is done on the first line, splitting the $\partial_{x}^{2}$ into two occurrences of $\partial_{x}$ which are absorbed into the terms in brackets:
\begin{multline}
\left[ \left(\partial_{t} + \partial_{x}(v \pm 1) +(-i\partial_{x})^{n}\gamma(-i\partial_{x})^{n}\right) \left(\partial_{t} + (v \mp 1)\partial_{x}+(-i\partial_{x})^{n}\gamma(-i\partial_{x})^{n}\right) \right. \\
\left. \pm (-1)^{n}\partial_{x}^{n}\gamma\partial_{x}^{n+1} \mp (-1)^{n}\partial_{x}^{n+1}\gamma\partial_{x}^{n} + m^{2} + f\left(-\partial_{x}^{2}\right) - \left(\partial_{x}^{n}\gamma\partial_{x}^{n}\right)^{2} \right] \phi = \left(-\partial_{x}\right)^{n} J(t,x)
\label{eq:phi_eqn_nearly_factorized}
\end{multline}
Note the sign ambiguity, which depends on how we choose to place $\pm \partial_{x}$ in the two brackets of the first line.
The idea is to make Eq. (\ref{eq:phi_eqn_nearly_factorized}) exactly factorizable by choosing parameters such that the second line vanishes.
The first two terms of the second line give
\begin{equation}
\pm (-1)^{n} \left( \partial_{x}^{n}\gamma\partial_{x}^{n+1} - \partial_{x}^{n+1}\gamma\partial_{x}^{n} \right) = \pm (-1)^{n} \partial_{x}^{n} \left( \gamma\partial_{x} - \partial_{x}\gamma \right) \partial_{x}^{n} = \mp (-1)^{n} \partial_{x}^{n} \gamma^{\prime} \partial_{x}^{n} \,.
\label{eq:gamma_prime_term}
\end{equation}
This vanishes when $\gamma^{\prime}=0$, in which case the two possible orderings of $(v+1)$ and $(v-1)$ in the first line become equivalent.
On the other hand, when $\gamma^{\prime} \neq 0$, it is still possible to factorize the equation by choosing $m^{2}$ and/or $f(P^{2})$ appropriately so as to cancel out the second line of (\ref{eq:phi_eqn_nearly_factorized}); but since this choice depends on the sign of the term in (\ref{eq:gamma_prime_term}), the ordering of $(v+1)$ and $(v-1)$ will matter.

The simplest way to factorize Eq. (\ref{eq:phi_eqn_nearly_factorized}) is to take $n=0$ or $1$, and, if $n=1$, take $\gamma$ to be constant.  
Then, to cancel out the second line of (\ref{eq:phi_eqn_nearly_factorized}), for $n=0$ one should take the bare mass function to be $m^2 = \gamma^{2} \pm \gamma^{\prime}$ and the dispersive term $f^2 = 0$, while for $n= 1$ one should take $m^2 = 0$ and $f(P^2) =  \gamma^{2}P^4$. 
(Note that the dimensions of $\gamma$ are different for the two values of $n$; in fact, when $n=0$, $\gamma$ is simply the dissipation rate and is usually called $\Gamma$, as in App. \ref{app:Gamma_constant}.) 
Finally, in both cases, we have 
\begin{equation}
\left[ \partial_{t} + \gamma\left(-i\partial_{x}\right)^{2n} + \partial_{x}\left(v\pm 1\right) \right] \left[ \partial_{t} + \gamma\left(-i\partial_{x}\right)^{2n} + \left(v\mp 1\right)\partial_{x} \right] \phi = \left(-\partial_{x}\right)^{n} J(t,x) \, ,
\label{eq:phi_eqn_factorized}
\end{equation}
which, with a choice of sign, is exactly Eq. (\ref{eq:final_phi_eqn}).


\section{Retarded Green function
\label{app:Gret}}

Here we generalize the considerations of \S\ref{sub:Gret}, finding the form of the Green function for a general wave equation with polynomial dispersion relation, for a system which is asymptotically homogeneous.
This can be considered as a generalization of the Jost construction \cite{Newton-Scattering}.

Firstly, we consider the purely homogeneous case.
Then the Fourier transformed (in both space and time) Green function is found straightforwardly from the Fourier transformed wave equation, and is of the form
\begin{equation}
G^{\omega}_{\mathrm{ret,hom}}(k) = \frac{1}{G_{0} \, \prod_{i} \left(k-k_{i}^{R}\right) \, \prod_{j} \left(k-k_{j}^{L}\right)} \,,
\end{equation}
where the superscript $R$ ($L$) corresponds to a positive (negative) imaginary part of $k$, so that the mode decays to the right (left).
Inverting the spatial Fourier transform, and using the Cauchy residue theorem, we then have
\begin{equation}
G^{\omega}_{\mathrm{ret,hom}}(x,x^{\prime}) = i \theta(x-x^{\prime}) \sum_{i} \mathcal{N}^{R}_{i} e^{ik^{R}_{i}(x-x^{\prime})} + i \theta(x^{\prime}-x) \sum_{j} \mathcal{N}^{L}_{j} e^{ik^{L}_{j}(x-x^{\prime})} \,,
\label{eq:Gret_hom}
\end{equation}
where we have defined
\begin{alignat}{2}
\mathcal{N}^{R}_{k} = \frac{1}{G_{0} \, \prod_{i \neq k} \left(k_{k}^{R}-k_{i}^{R}\right) \, \prod_{j} \left(k_{k}^{R}-k_{j}^{L}\right)} \,, & \qquad \mathcal{N}^{L}_{k} = -\frac{1}{G_{0} \, \prod_{i} \left(k_{k}^{L}-k_{i}^{R}\right) \, \prod_{j \neq k} \left(k_{k}^{L}-k_{j}^{L}\right)} \,.
\end{alignat}
The inverse coefficients $1/\mathcal{N}^{R/L}_{k}$ can be thought of as generalized squared ``norms'' of the unit plane waves.
Note that they can be written as derivatives of $1/G_{\mathrm{ret,hom}}^{\omega}$, which itself is just the dispersion relation.
In the relativistic and dispersive cases, this means that they can be written as $\pm 2\left(\omega-v\,k\right) v_{g}(k)$, where $v_{g}(k)=d\omega/dk$ is the group velocity, see~\cite{Coutant-RP-SF}.

Now consider an inhomogeneous system which becomes homogeneous asymptotically.
In the diagonal asymptotic regions $x,x^{\prime} \rightarrow \pm\infty$, the Green function will be $G^{\omega}_{\mathrm{ret,hom}}(x,x^{\prime})$ of Eq. (\ref{eq:Gret_hom}) plus a solution of the source-free form of the wave equation, i.e. a solution which is just a sum of products of plane waves containing no Heaviside step functions.
This additional term must obey appropriate boundary conditions; in particular, it must vanish asymptotically, in both $x$ and $x^{\prime}$.
So, in the right-hand asymptotic region $x,x^{\prime} \rightarrow \infty$, we can add to (\ref{eq:Gret_hom}) a term of the form
\begin{equation}
i\sum_{i}\sum_{j} c_{i,j} e^{ik_{i}^{R}x} e^{-ik_{j}^{L}x^{\prime}} \nonumber
\end{equation}
since this is the only source-free solution that converges as $x,x^{\prime} \rightarrow \infty$.
Therefore, the general form of the Green function in the right-hand asymptotic region is
\begin{equation}
G^{\omega}_{\mathrm{ret}}(x,x^{\prime}) = i\theta(x-x^{\prime}) \sum_{i} e^{ik_{i}^{R}x} \left[ \mathcal{N}_{i}^{R} e^{-ik_{i}^{R}x^{\prime}} + \sum_{j} c_{i,j} \, e^{-ik_{j}^{L}x^{\prime}} \right] + i\theta(x^{\prime}-x) \sum_{j} \left[ \mathcal{N}_{j}^{L} e^{ik_{j}^{L}x} + \sum_{i} c_{i,j} \, e^{ik_{i}^{R}x} \right] e^{-ik_{j}^{L}x^{\prime}} \,.
\label{eq:Gret_inhom_RHS}
\end{equation}
Let us emphasise that Eq. (\ref{eq:Gret_inhom_RHS}) applies only in the right-hand asymptotic region where the system becomes homogeneous.
In particular, the wave vectors $k_{k}^{R/L}$ are solutions of the dispersion relation in the right-hand asymptotic region, and the ``normalization'' factors $\mathcal{N}_{k}^{R/L}$ come from the expansion of $G^{\omega}_{\mathrm{ret,hom}}(x,x^{\prime})$ using the background values in this region.
Similarly, the coefficients $c_{i,j}$ will be particular to the right-hand side, and are fixed by the requirement that, when this asymptotic solution is extended throughout the space, the resulting $G^{\omega}_{\mathrm{ret}}(x,x^{\prime})$ must vanish asymptotically.
The functions of $x$ and $x^{\prime}$ appearing in Eq. (\ref{eq:Gret_inhom_RHS}) -- which are solutions of the wave equation and its dual, respectively -- can then be used as a set of basis modes.
However, we make another choice here: since the interpretation of the anticommutator in \S\ref{sub:Gac_interpretation} relies on the modes $\phi^{\omega}(x)$ containing a single outgoing wave in one of the asymptotic regions, we impose that this be the case.
Those modes multiplying $\theta(x-x^{\prime})$ in the first term of (\ref{eq:Gret_inhom_RHS}) are exactly of this form, and so we define $\phi^{\omega,R}_{i}(x)$ and $\widetilde{\phi}^{\omega,R}_{j}(x^{\prime})$ such that, as $x,x^{\prime} \rightarrow \infty$,
\begin{alignat}{2}
\phi^{\omega,R}_{i}(x) \rightarrow e^{ik_{i}^{R}x} \,, & \qquad \widetilde{\phi}^{\omega,R}_{i}(x^{\prime}) \rightarrow e^{-ik_{i}^{R}x^{\prime}} + \frac{1}{\mathcal{N}^{R}_{i}} \sum_{j} c_{i,j} \, e^{-ik_{j}^{L}x^{\prime}} \,.
\label{eq:modes_RHS}
\end{alignat}
Thus we see that the definition of $\phi_{i}^{\omega,R}(x)$ is straightforward while that of $\widetilde{\phi}_{i}^{\omega,R}(x^{\prime})$ is more subtle: it contains a single growing mode in the right-hand asymptotic region, but must be purely decaying when continued into the left-hand region, and thus contains whatever combination of decaying modes in the right-hand region makes this so.

The second collection of wave functions in (\ref{eq:Gret_inhom_RHS}), which multiply $\theta(x^{\prime}-x)$, are not of the required form, but can be made so by redoing the above analysis in the left-hand asymptotic region.  
Using bars to denote wave vectors and ``normalization'' factors evaluated on the left-hand side, the homogeneous form of the Green function there is
\begin{equation}
G^{\omega}_{\mathrm{ret,hom}}(x,x^{\prime}) = i \theta(x-x^{\prime}) \sum_{i} \bar{\mathcal{N}}^{R}_{i} e^{i\bar{k}^{R}_{i}(x-x^{\prime})} + i \theta(x^{\prime}-x) \sum_{j} \bar{\mathcal{N}}^{L}_{j} e^{i\bar{k}^{L}_{j}(x-x^{\prime})} \,,
\label{eq:Gret_hom_LHS}
\end{equation}
to which we can add a term of the form
\begin{equation}
i\sum_{i}\sum_{j} \bar{c}_{i,j} e^{i\bar{k}_{j}^{L}x} e^{-i\bar{k}_{i}^{R}x^{\prime}} \nonumber
\end{equation}
since this is the only source-free solution that decays as $x,x^{\prime} \rightarrow -\infty$.  
So the asymptotic form of the Green function as $x,x^{\prime} \rightarrow -\infty$ is
\begin{equation}
G^{\omega}_{\mathrm{ret}}(x,x^{\prime}) = i\theta(x-x^{\prime}) \sum_{i} \left[\bar{\mathcal{N}}^{R}_{i} e^{i\bar{k}_{i}^{R}x} + \sum_{j} \bar{c}_{i,j} \,e^{i\bar{k}_{j}^{L}x} \right] e^{-i\bar{k}_{i}^{R}x^{\prime}} + i\theta(x^{\prime}-x) \sum_{j} e^{i\bar{k}_{j}^{L}x} \left[ \bar{\mathcal{N}}^{L}_{j} e^{-i\bar{k}_{j}^{L}x^{\prime}} + \sum_{i} \bar{c}_{i,j} \, e^{-i\bar{k}_{i}^{R}x^{\prime}} \right] \,.
\label{eq:Gret_inhom_LHS}
\end{equation}
The second term is now of the form required, and we define $\phi^{\omega,L}_{j}(x)$ and $\widetilde{\phi}^{\omega,L}_{j}(x^{\prime})$ such that, as $x,x^{\prime} \rightarrow -\infty$,
\begin{alignat}{2}
\phi^{\omega,L}_{j}(x) \rightarrow e^{i\bar{k}_{j}^{L}x} \,, & \qquad \widetilde{\phi}^{\omega,L}_{j}(x^{\prime}) \rightarrow e^{-i\bar{k}_{j}^{L}x^{\prime}} + \frac{1}{\bar{\mathcal{N}}_{j}^{L}} \sum_{i} \bar{c}_{i,j} \, e^{-i\bar{k}_{i}^{R} x^{\prime}} \,,
\label{eq:modes_LHS}
\end{alignat}
where, as before, the coefficients $\bar{c}_{i,j}$ are fixed by the requirement that $\widetilde{\phi}^{\omega,L}_{j}(x^{\prime})$ be purely decaying when continued into the right-hand asymptotic region.

Finally, using the modes defined in (\ref{eq:modes_RHS}) and (\ref{eq:modes_LHS}), the full Green function can be written
\begin{equation}
G^{\omega}_{\mathrm{ret}}(x,x^{\prime}) = i\theta(x-x^{\prime}) \sum_{i} \mathcal{N}^{R}_{i} \, \phi^{\omega,R}_{i}(x) \, \widetilde{\phi}^{-\omega,R}_{i}(x^{\prime}) + i\theta(x^{\prime}-x) \sum_{j} \bar{\mathcal{N}}^{L}_{j} \, \phi^{\omega,L}_{j}(x) \, \widetilde{\phi}^{-\omega,L}_{j}(x^{\prime}) \,.
\end{equation}


\section{Modes and dual modes for a simple black hole profile \label{simplec}}

In this appendix, we give explicit expressions for the flow velocity profile and the modes and dual modes used to construct the figures in Secs. \ref{sec:Modes} and \ref{sec:State}.

The adimensionalized velocity profile described by Eq. (\ref{eq:u-mode_profile_adimensional}) is taken to be
\begin{equation}
h(X) = \mathrm{tanh}\left(X\right) \,.
\label{eq:adimensionalised_velocity}
\end{equation}
This satisfies the conditions given just after Eq. (\ref{eq:u-mode_profile_adimensional}): $h(0)=0$ and $h^{\prime}(0)=1$, and it asymptotes to constant limiting values in the right (subsonic) and left (supersonic) regions, with flow velocities $-1+D$ and $-1-D$, respectively.
We note that the profile is symmetrical in the sense that $\left|v+1\right|$ is symmetrical around the horizon.

\subsection{Modes}

Let us consider stationary modes of the form $\phi(T,X) = \phi^{w}(X) e^{-i w T}$, where $w = \omega/\kappa$ is the adimensionalized Killing frequency.
Then Eq. (\ref{eq:u_eqn_Xgeom}) becomes
\begin{equation}
\left[ -i w - \lambda^{2} \partial_{X}^{2} + \mathrm{tanh}\left(X \right) \, \partial_{X} \right] \phi^{w}(X) = 0 \,.
\label{eq:u_eqn_Xgeom_tanh}
\end{equation}
First of all, we note that since the velocity profile is asymptotically constant, the general solution in the asymptotic regions decomposes into a sum of exponentials, and Eq. (\ref{eq:u_eqn_Xgeom_tanh}) reduces there to a dispersion relation:
\begin{equation}
\lambda^{2} \, K^{2} + i \, h \, K - i w = 0 \,,
\label{eq:asymptotic_dispersion}
\end{equation}
where $h=1$ in the right-hand (subsonic) region and $h=-1$ in the left-hand (supersonic) region.
In the geometrical optics approximation, we can think of the roots as varying continuously with $X$ as $h$ varies between $-1$ and $1$, and it is precisely this migration of the roots that is shown in Fig. \ref{fig:roots}.
Note that what is plotted there is the root with positive real and imaginary parts, i.e. it is propagating to the right in the subsonic region (the other root, which propagates to the left in the supersonic region, follows curves which are simply the negatives of those in Fig. \ref{fig:roots}, due to the invariance of Eq. (\ref{eq:asymptotic_dispersion}) under $h \rightarrow -h$ and $K \rightarrow -K$). 
It is precisely the dissipative version of the standard right-propagating root; for this reason, we label it $K_{uR}$, while the other root, which decays to the left, is $K_{uL}$.
As $h$ (and $X$) are decreased, the real and imaginary parts of $K_{uR}$ both increase, becoming equal at the horizon where $h$ vanishes, and where, according to Eq. (\ref{eq:asymptotic_dispersion}), $K \propto \sqrt{i}$.
As $h$ is decreased further, the real part of $K_{uR}$ decreases again while the imaginary part continues to increase, until at $h=-1$ ($X \rightarrow -\infty$) it settles on a value with the same real part it started with but a much larger imaginary part.
It has thus, in the left-hand region, become the {\it additional} root due to the dissipative $k^{4}$ term in the dispersion relation.

Equation (\ref{eq:u_eqn_Xgeom_tanh}) is exactly solvable using trigonometric and hypergeometric functions.
Firstly, $\phi^{w}(X)$ is written in the form
\begin{equation}
\phi^{w}(X) = \psi^{w}(X) \, \mathrm{exp}\left(\frac{1}{2\lambda^{2}} \int^{X} \mathrm{tanh}(X^{\prime}) dX^{\prime}\right) = \psi^{w}(X) \, \left( 2\,\mathrm{cosh}(X) \right)^{1/2\lambda^{2}} \,,
\end{equation}
with $\psi^{w}(X)$ obeying the transformed differential equation
\begin{equation}
\left[ i\frac{w}{\lambda^{2}} - \frac{1}{4\lambda^{4}} + \frac{1}{2\lambda^{2}}\left(1+\frac{1}{2\lambda^{2}}\right) \mathrm{sech}^{2}(X) \right] \psi^{w}(X) = 0 \,.
\label{eq:transformed_mode_eqn}
\end{equation}
The solution of (\ref{eq:transformed_mode_eqn}) is given in \S 23, Problem 4 of \cite{Landau-Lifshitz-QM}.
One of the two final independent solutions is
\begin{equation}
\phi^{w}_{uR}(X) = \left(2\,\mathrm{cosh}\left(X\right)\right)^{iK_{uR}} \tensor[_2]{F}{_1}\left(-iK_{uR}, 1+iK_{uL}; 1-\frac{i}{2}\left(K_{uR}-K_{uL}\right);\frac{1}{2}\left(1 - \mathrm{tanh}\left(X\right)\right) \right) \,,
\label{eq:stationary_mode}
\end{equation}
where, following the notation of App. \ref{app:Gret}, the absence of overbars indicates that the wave vectors are to be evaluated in the right-hand asymptotic region, i.e. with $h=1$ in Eq. (\ref{eq:asymptotic_dispersion}).
To see that (\ref{eq:stationary_mode}) obeys the correct boundary conditions, recall that the modes $\phi^{\omega}$ are defined as containing a single outgoing (hence decaying) wave of unit amplitude in one of the asymptotic regions.
Taking the limit $X \rightarrow \infty$, the hypergeometric function approaches $1$, so $\phi^{w}_{uR}(X) \rightarrow \mathrm{exp}\left(i\,K_{uR} \, X\right)$ as required.
The other independent solution is obtained via reflection, $X\rightarrow -X$:
\begin{equation}
\phi^{w}_{uL}(X) = \phi^{w}_{uR}(-X) \,,
\end{equation}
which, in the left-hand asymptotic region, is equal to $\mathrm{exp}\left(-i\,K_{uR}\,X\right) = \mathrm{exp}\left(i\,\bar{K}_{uL}\,X\right)$, as required (the overbar on $\bar{K}_{uL}$ indicating that it is to be evaluated in the left-hand asymptotic region, with $h=-1$ in Eq. (\ref{eq:asymptotic_dispersion})).


\subsection{Dual modes}

In constructing the retarded Green function for the $\phi$ field, we need its dependence on both the image coordinate $x$ {\it and} the source coordinate $x^{\prime}$. 
As explained in \S\ref{sub:Gret}, the latter obeys the dual of Eq. (\ref{eq:final_phi_eqn}); at the level of the $u$- and $v$-mode source-free equations, this amounts to switching the sign of $\gamma$
(and as such describes dissipation acting {\it backwards} in time).
Using a tilde to denote the dual mode, it obeys
\begin{equation}
\left[ iw + \lambda^{2} \partial_{X}^{2} + \mathrm{tanh}\left(X\right) \, \partial_{X} \right] \widetilde{\phi}^{-w}(X) = 0 \,.
\label{eq:u_dual-eqn_Xgeom_tanh}
\end{equation}
At the level of the dispersion relation (\ref{eq:asymptotic_dispersion}), this effectively switches the sign of $h$ so that the wave vector solutions for the dual wave equation are simply the negatives of those of the original wave equation.
This means that the propagating solutions now steadily {\it increase} in amplitude towards infinity, whereas the additional complex modes decay rapidly to zero.  

Equation (\ref{eq:u_dual-eqn_Xgeom_tanh}) can be solved exactly in a similar manner to Eq. (\ref{eq:u_eqn_Xgeom_tanh}).
One of the independent dual modes is
\begin{equation}
\widetilde{\phi}^{-w}_{uR}(X) = \widetilde{A}^{-w}(\lambda) \,\, \left(2\,\mathrm{cosh}\left(X\right)\right)^{-iK_{uL}} \tensor[_2]{F}{_1}\left(iK_{uL}, 1-iK_{uR}; 1-\frac{i}{2}\left(K_{uR}-K_{uL}\right);\frac{1}{2}\left(1 + \mathrm{tanh}\left(X\right)\right) \right) \,,
\label{eq:dual_stationary_mode}
\end{equation}
where the amplitude $\widetilde{A}^{-w}(\lambda)$ is given by
\begin{equation}
\widetilde{A}^{-w}(\lambda) = \frac{\Gamma\left(iK_{uL}\right)\Gamma\left(1-iK_{uR}\right)}{\Gamma\left(1-\frac{i}{2}\left(K_{uR}-K_{uL}\right)\right)\Gamma\left(-\frac{i}{2}\left(K_{uR}-K_{uL}\right)\right)} \,.
\label{eq:dual_stationary_mode_amplitude}
\end{equation}
The amplitude is fixed by the boundary conditions imposed on $\widetilde{\phi}^{-w}_{uR}$, which (as mentioned in \S\ref{sub:Gret}) are that it decays to the left and has a growing mode $\mathrm{exp}\left(-i\,K_{uR}\,X\right)$ with unit amplitude on the right.
The decay to the left means that it must be proportional to $\mathrm{exp}\left(-i\,\bar{K}_{uR}\,X\right) = \mathrm{exp}\left(i\,K_{uL}\,X\right)$ in the limit $X\rightarrow -\infty$; examination of Eq. (\ref{eq:dual_stationary_mode}) shows that this is the case.
To ensure that the growing mode on the right has unit amplitude, we use the following transformation law in order to decompose the dual mode into its exponential components on the right-hand side:
\begin{multline}
\tensor[_2]{F}{_1}(a,b\,;c\,;z) = \frac{\Gamma(c)\Gamma(c-a-b)}{\Gamma(c-a)\Gamma(c-b)} \, \tensor[_2]{F}{_1}(a,b\,; a+b+1-c\,; 1-z) \\
+ \frac{\Gamma(c)\Gamma(a+b-c)}{\Gamma(a)\Gamma(b)} (1-z)^{c-a-b} \, \tensor[_2]{F}{_1}(c-a,c-b\,; 1+c-a-b\,; 1-z) \,.
\label{eq:hypergeometric_transformation}
\end{multline}
The amplitude (\ref{eq:dual_stationary_mode_amplitude}) thus ensures that the coefficient of $\mathrm{exp}\left(-i\,K_{uR}\,X\right)$ is $1$.

As before, the other independent solution can be obtained via reflection about the horizon, $X \rightarrow -X$:
\begin{equation}
\widetilde{\phi}^{-w}_{uL}(X) = \widetilde{\phi}^{-w}_{uR}(-X) \,.
\end{equation}
The symmetry properties relating the various wave vectors ensure that $\widetilde{\phi}^{-w}_{uL}(X)$ also obeys appropriate boundary conditions.

Examples of the dual modes can be seen in Fig. \ref{fig:Gret}, since $G_{\mathrm{ret}}^{w}(X,X^{\prime})$ is proportional to them at fixed image coordinate $X$.
In the top row, we can see the evolution of a propagating wave into a strongly damped wave as the horizon is crossed, and that this gradual transition becomes a sharp cutoff in the relativistic limit $\lambda^2 \to 0$.


\section{Noise kernel
\label{app:Noise}}

In this appendix, we derive the form of the noise kernel in Eq. (\ref{eq:noise}) from the action of the free environment field given in Eq. (\ref{eq:Psi_action}).

The noise kernel $N^{\omega}(x,x^{\prime})$ represents the sourcing of the $\phi$ field due to the homogeneous part $\Psi_{q}^{0}$ of the environment field.
We thus need an explicit form of the operator $\hat{\Psi}_{q}^{0}$ in order to find it.
For this purpose, it is convenient to work in a coordinate system $(\tau,z)$ adapted to freely-falling observers, in which the oscillators of the environment are at rest:
\begin{alignat}{2}
\tau = t \,, & \qquad \int_{z}^{x} \frac{dx^{\prime}}{v(x^{\prime})} = t \,.
\end{alignat}
In this coordinate system, the partial derivative $\partial_{\tau} = \partial_{t} + v \, \partial_{x}$, which is exactly the differential operator $\partial_{\tau}$ introduced in Eq. (\ref{eq:dtau_defn}) and the unit timelike vector field $u$ of Eqs. (\ref{eq:preferred_unit_vectors}).
The free action $S_{\Psi}$ of the environment takes the form
\begin{equation}
S_{\Psi} = \frac{1}{2} \int d\tau \, dz \int dq \left\{ \left( \partial_{\tau} \sqrt{\frac{v(x)}{v(z)}} \Psi_{q} \right)^{2} - \left( \omega_{q} \sqrt{\frac{v(x)}{v(z)}} \Psi_{q} \right)^{2} \right\}
\label{eq:Psi_action_comovcoords}
\end{equation}
and the free equation of motion for $\Psi$ becomes
\begin{equation}
\sqrt{\frac{v(z)}{v(x)}} \left( \partial_{\tau}^{2} + \omega_{q}^{2} \right) \sqrt{\frac{v(x)}{v(z)}} \Psi_{q}^{0} = 0 \,.
\end{equation}
Restricting $\Psi_{q}^{0}$ to be real, the general solution is
\begin{equation}
\Psi_{q}^{0}(\tau,z) = \sqrt{\frac{v(z)}{v(x)}} \left( c_{q}(z) \frac{e^{-i\omega_{q}\tau}}{\sqrt{2\omega_{q}}} + c_{q}^{\star}(z) \frac{e^{i\omega_{q}\tau}}{\sqrt{2\omega_{q}}} \right) \,,
\end{equation}
where the factors of $1/\sqrt{2\omega_{q}}$ are included to give normalized plane waves of unit Wronskian.
From the action (\ref{eq:Psi_action_comovcoords}), the canonical momentum is
\begin{equation}
\Pi_{q}^{0} = \sqrt{\frac{v(x)}{v(z)}} \partial_{\tau} \sqrt{\frac{v(x)}{v(z)}} \Psi_{q}^{0} \,,
\end{equation}
and upon quantization and the imposition of the equal time canonical commutator
\begin{equation}
\left[ \hat{\Psi}^{0}_{q}(\tau,z), \hat{\Pi}^{0}_{q^{\prime}}(\tau,z^{\prime}) \right] = i \, \delta(q-q^{\prime}) \, \delta(z-z^{\prime})
\end{equation}
we find that the quantum amplitudes $\hat{c}_{q}(z)$ obey the bosonic commutation relations
\begin{equation}
\left[ \hat{c}_{q}(z) , \hat{c}^{\dagger}_{q^{\prime}}(z^{\prime}) \right] = \delta(q-q^{\prime}) \, \delta(z-z^{\prime}) \,.
\label{eq:noise_bosonic_commutator}
\end{equation}

Considering first the case $n=0$, the source operator $\hat{J}$ is
\begin{eqnarray}
\hat{J} & = & \gamma(x) \sqrt{\frac{v(z)}{v(x)}} \partial_{\tau} \sqrt{\frac{v(x)}{v(z)}} \int dq \Psi_{q}^{0} \nonumber \\
& = & -i \, \gamma(x) \sqrt{\frac{v(z)}{v(x)}} \int dq \omega_{q} \left( \hat{c}_{q}(z) \frac{e^{-i\omega_{q}\tau}}{\sqrt{2\omega_{q}}} - \hat{c}^{\dagger}_{q}(z) \frac{e^{i\omega_{q}\tau}}{\sqrt{2\omega_{q}}} \right) \,.
\label{eq:source_operator}
\end{eqnarray}
Using Eq. (\ref{eq:noise_bosonic_commutator}) and the fluctuation-dissipation theorem in a heat bath at temperature $T_\Psi$,
\begin{eqnarray}
\langle \left\{ \hat{c}_{q}(z) , \hat{c}^{\dagger}_{q^{\prime}}(z^{\prime}) \right\} \rangle & = & \mathrm{coth}\left(\frac{\omega_{q}}{2T_\Psi}\right) \left[ \hat{c}_{q}(z) , \hat{c}^{\dagger}_{q^{\prime}}(z^{\prime}) \right] \nonumber \\
& = & \mathrm{coth}\left(\frac{\omega_{q}}{2T_\Psi}\right) \, \delta(q-q^{\prime}) \, \delta(z-z^{\prime}) \,,
\end{eqnarray}
we find
\begin{eqnarray}
N(t,x;t^{\prime},x^{\prime}) & \equiv & \langle \left\{ \hat{J}(t,x) , \hat{J}(t^{\prime},x^{\prime}) \right\} \rangle \nonumber \\
& = & \delta\left(t-t^{\prime}-\int_{x^{\prime}}^{x} \frac{dx^{\prime\prime}}{v(x^{\prime\prime})} \right) \frac{\gamma(x)\gamma(x^{\prime})}{\sqrt{v(x)v(x^{\prime})}} \int dq \, \omega_{q} \, \mathrm{coth}\left(\frac{\omega_{q}}{2T_\Psi}\right) \, \mathrm{cos}\left(\omega_{q}\int_{x^{\prime}}^{x} \frac{dx^{\prime\prime}}{v(x^{\prime\prime})}\right) \,.
\end{eqnarray}
Because of stationarity, this is seen to be a function only of the time difference $t-t^{\prime}$, and we can Fourier transform in time to get
\begin{eqnarray}
N^{\omega}(x,x^{\prime}) & \equiv & \int_{-\infty}^{+\infty} d(t-t^{\prime}) e^{i\omega(t-t^{\prime})} N(t,x;t^{\prime},x^{\prime}) \nonumber \\
& = & \frac{\gamma(x)\gamma(x^{\prime})}{\sqrt{v(x)v(x^{\prime})}} \mathrm{exp}\left(\int_{x^{\prime}}^{x} \frac{dx^{\prime\prime}}{v(x^{\prime\prime})}\right) \int dq \, \omega_{q} \, \mathrm{coth}\left(\frac{\omega_{q}}{2T_\Psi}\right) \, \mathrm{cos}\left(\omega_{q}\int_{x^{\prime}}^{x} \frac{dx^{\prime\prime}}{v(x^{\prime\prime})}\right) \nonumber \\
& = & \frac{1}{2\pi} \frac{\gamma(x)\gamma(x^{\prime})}{\sqrt{v(x)v(x^{\prime})}} \int d\omega_{q} \, \omega_{q} \, \mathrm{coth}\left(\frac{\omega_{q}}{2T_\Psi}\right) \, \mathrm{exp}\left(i(\omega-\omega_{q})\int_{x^{\prime}}^{x} \frac{dx^{\prime\prime}}{v(x^{\prime\prime})}\right) \,,
\label{eq:noise_derived}
\end{eqnarray}
where in the last line we have effected the change of variable $\omega_{q}=\pi q$.


\section{Subdominant effects}

In the body of the text, we saw that the main parameter governing the deviations from the non-dissipative case is $\lambda^2$ of \eq{eq:lambda_defn}.
In this appendix we study the residual effects of changing $D$ and increasing the environment temperature.
We also study the strength of the correlations between $u$- and $v$-modes, which are here very small since they are induced by the coupling to the environment.


\subsection{Changing the extension of the near-horizon region }
\label{changeD}

\begin{figure}
\subfloat{\includegraphics[width=0.45\columnwidth]{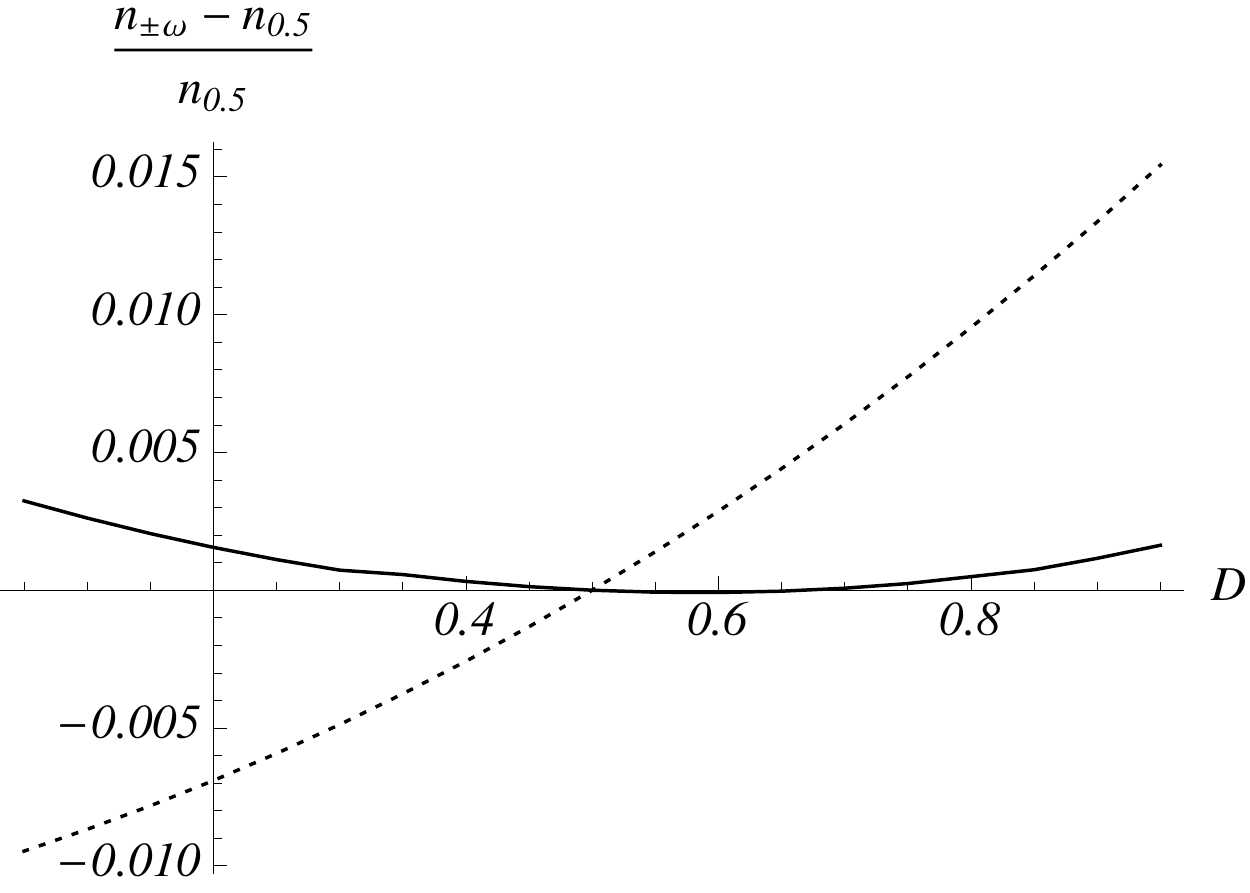}} \qquad \subfloat{\includegraphics[width=0.45\columnwidth]{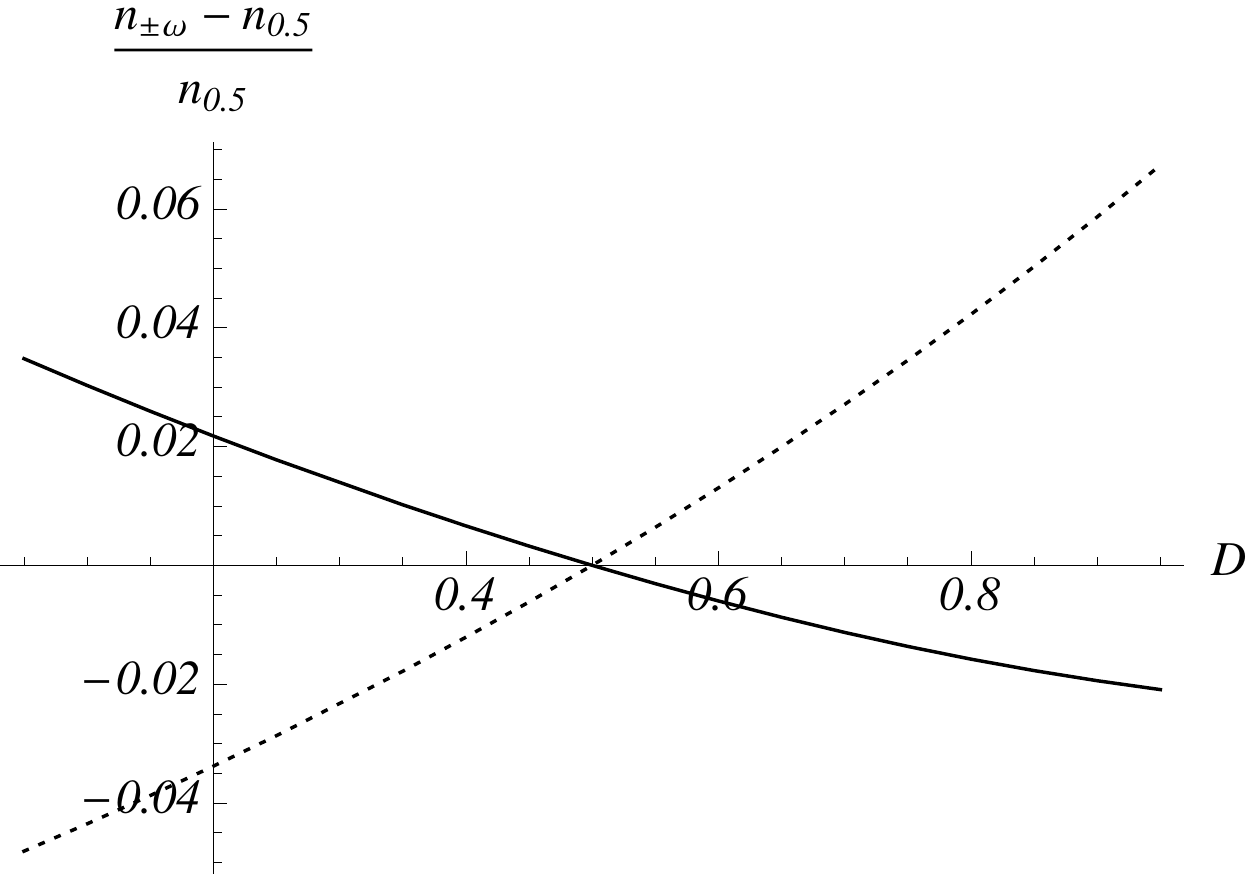}}
\caption{\textsc{Variation with $D$}: Here is plotted the relative difference in $n_{\omega}$ (solid curves) and $n_{-\omega}$ (dotted curves) as a function $D$, the reference values being taken at $D=0.5$ (i.e. the values plotted in Fig. \ref{fig:LowTemp}).
The temperature is fixed at $T_{\Psi}/\kappa=10^{-2}$ and $\lambda^{2} = 10^{-1}$ (so $\gamma$ varies as $D^{2}$).
The left panel has $\omega/\kappa = 1/2\pi$, the lowest of the frequencies plotted in Fig. \ref{fig:LowTemp}, whereas the right panel has $\omega/\kappa = 1/2$, the highest frequency plotted in Fig. \ref{fig:LowTemp}.
\label{fig:VarDLowT}}
\end{figure}

To examine the effects of $D$, in Figure \ref{fig:VarDLowT} we plot the relative change in $n_{\omega}$ and $n_{-\omega}$ from their values at $D=0.5$, at fixed low temperature $T_{\Psi}/\kappa = 10^{-2}$ and dissipative parameter $\lambda^{2} = 10^{-1}$.
There is clearly an increased sensitivity to $D$ at higher frequency.
Also very clear from these plots is the asymmetry between $n_{\omega}$ and $n_{-\omega}$, whose values tend in different directions.
However, the largest total variation seen in any of the observables is around $10\%$, while $D$ varies by almost $100\%$ (i.e. from $0.05$ to $0.95$).
We can thus conclude that it is the combined parameter $\lambda^2= \gamma \kappa/D^{2}$ which is most relevant in determining the values of the observables, while variation of the individual factors (keeping $\lambda^2$ constant) produces only a subdominant variation.


\subsection{High temperature effects}
\label{hightemp}

As it propagates outwards from the horizon, the $\phi$ field will thermalize, and if the environment temperature $T_{\Psi}$ is sufficiently high this will cause the particle number to approach an equilibrium value $n_{\mathrm{eq}}$.  Since the environment temperature applies in the freely-falling frame, $n_{\mathrm{eq}}$ is determined by the ``preferred'' frequency $\Omega$:
\begin{equation}
2n_{\mathrm{eq}}+1 \approx \mathrm{coth}\left|\frac{\Omega}{2T_{\Psi}}\right| \approx \mathrm{coth}\left|\frac{\omega-v_{\rm as}\,k}{2T_{\Psi}}\right| \,,
\label{eq:equilibrium}
\end{equation}
where $v_{\rm as}$ is the asymptotic value of $v$.
Hence, in an asymptotic region, the total value of $n$ as a function of position is thus approximately given by
\begin{equation}
n(x) \approx n_{\mathrm{eq}} + \delta n \, e^{-2\Gamma_{\mathrm{as}} \left|x\right|} \,,
\end{equation}
where $\delta n$ itself is independent of $x$.
(Note that $\delta n$ may be negative, in which case $n$ grows with $x$.)

\begin{figure}
\subfloat{\includegraphics[width=0.45\columnwidth]{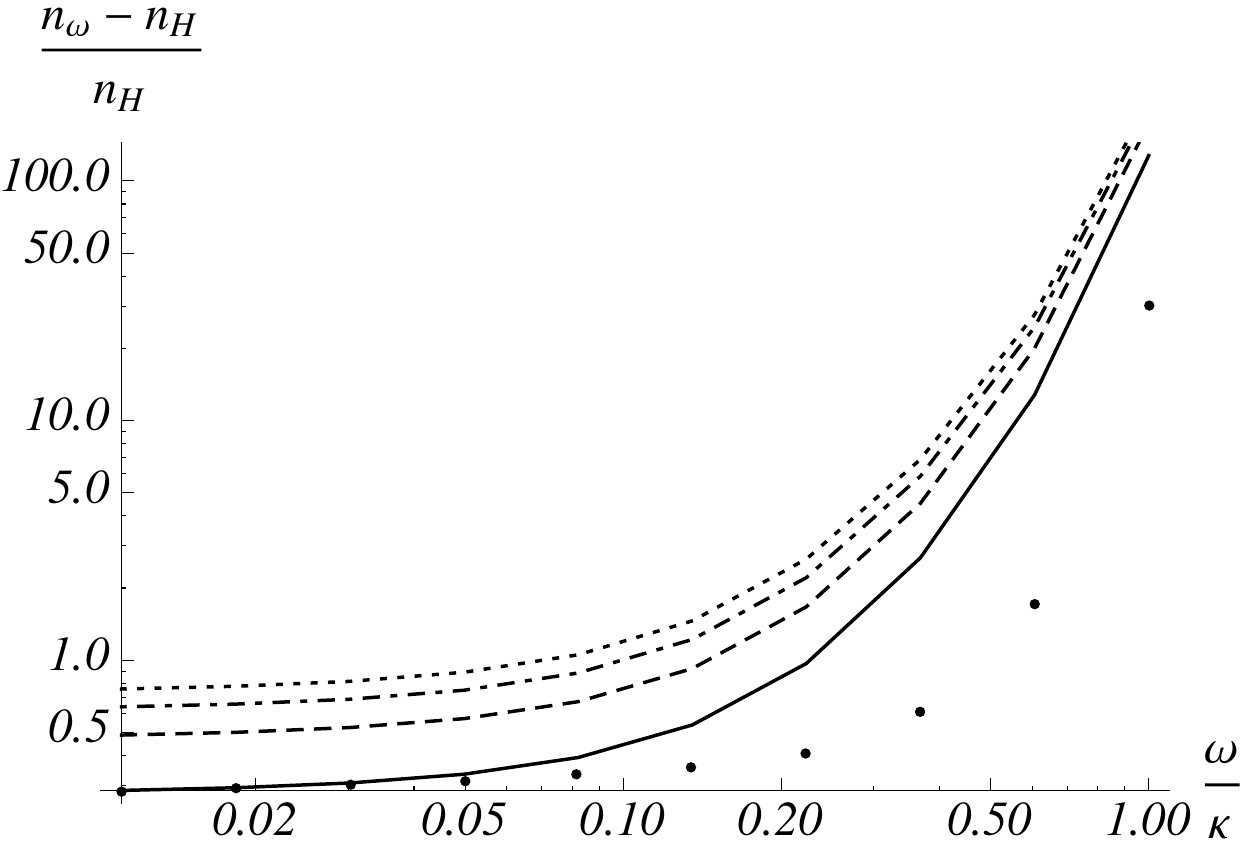}} \qquad \subfloat{\includegraphics[width=0.45\columnwidth]{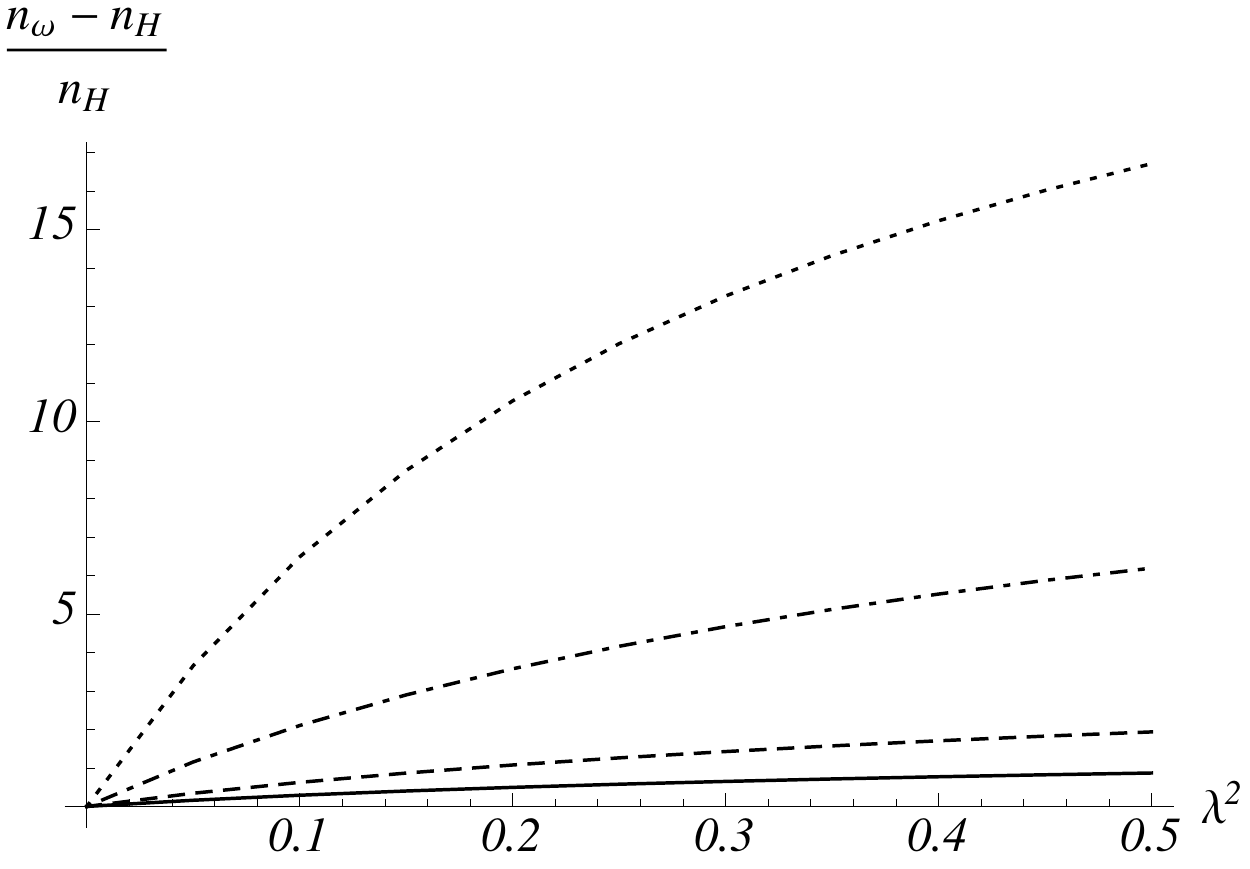}} \\
\subfloat{\includegraphics[width=0.45\columnwidth]{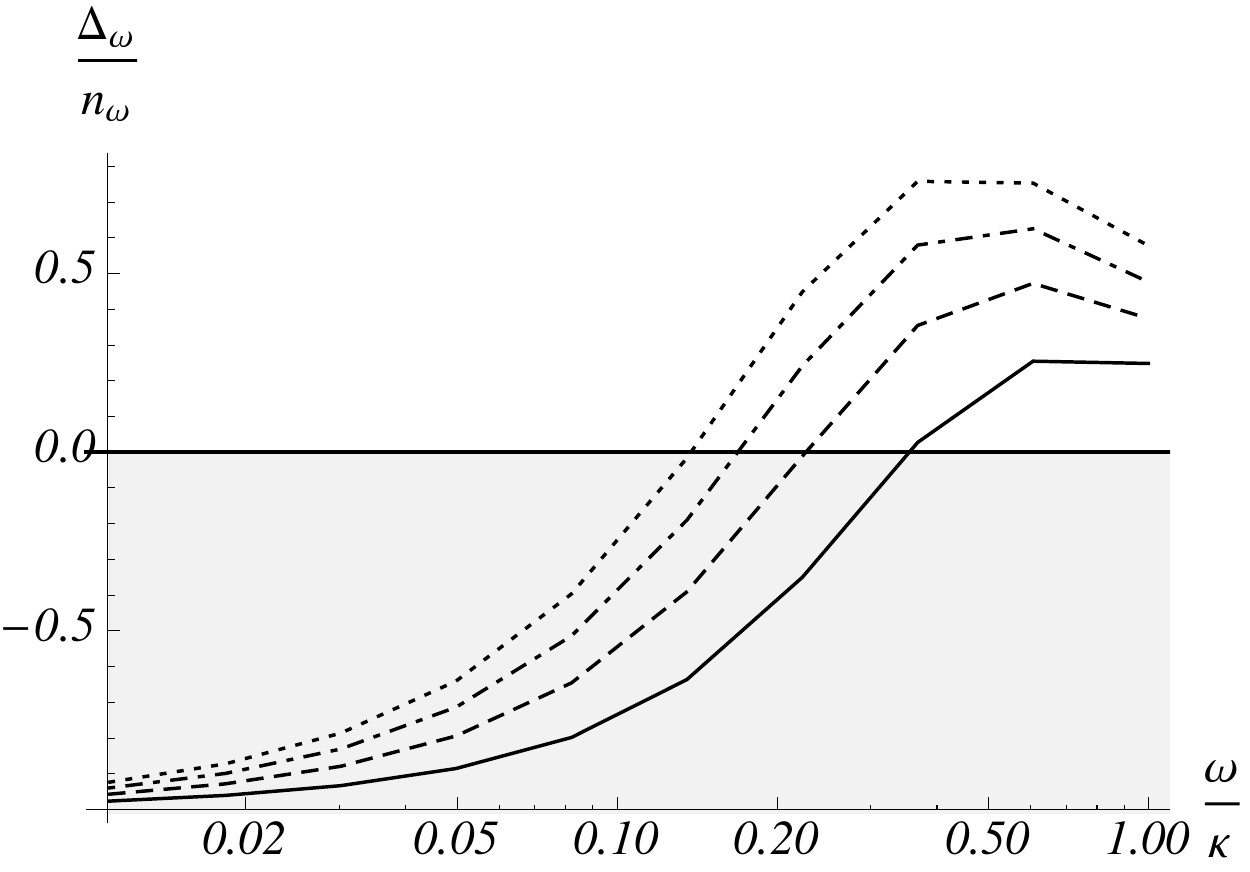}} \qquad \subfloat{\includegraphics[width=0.45\columnwidth]{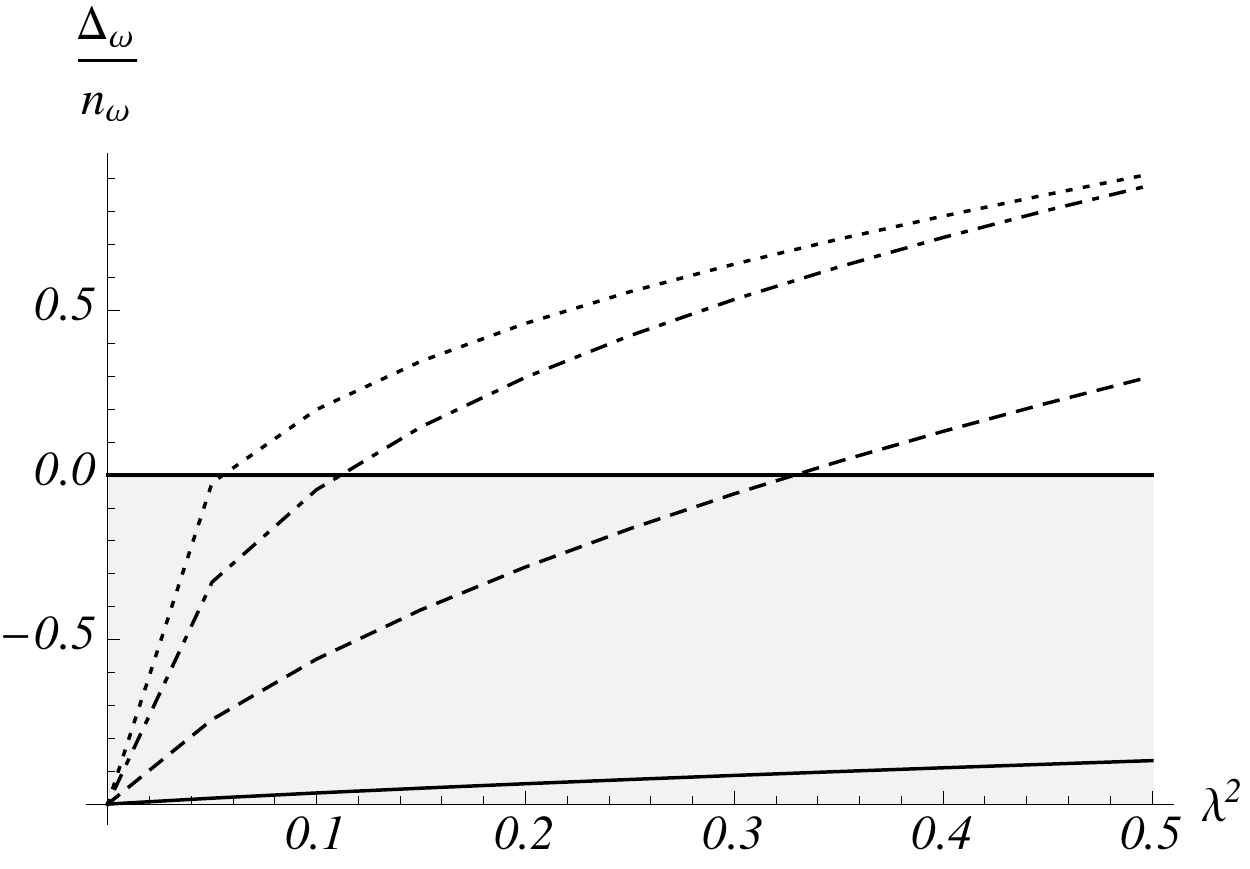}}
\caption{\textsc{Observables at high temperature}: Here are plotted the same observables as in Figs. \ref{fig:LowTemp} and \ref{fig:DeltaLowTemp}, except that the temperature is now fixed at the higher value of $T_{\Psi}/\kappa = \pi/2$ (about $10$ times the Hawking temperature).
$D$ is again fixed at $0.5$, and the measurements are made at fixed distance from the horizon, $\kappa \left|x\right|/D = 5$.
The first row shows the relative difference of the subsonic quasiparticle number from the standard Hawking values (note that, in the upper left plot, the vertical axis uses a logarithmic scale); again, the corresponding value in the supersonic region behaves similarly, and is not shown here.  
The dots in the upper left plot show the data points of the solid curve corrected for the mode propagation from the emission region to the detector.
The second row plots the rescaled nonseparability parameter, with the shaded area corresponding to $\Delta_{\omega}<0$ and hence to a nonseparable state.
In the left column, the variable is $\omega/\kappa$, and the various curves correspond to different values of $\lambda^{2}$: $0.1$ (solid curve), $0.2$ (dashed curve), $0.3$ (dot-dashed curve) and $0.4$ (dotted curve).
In the right column, the variable is instead $\lambda^{2}$, while the various curves correspond to different values of $\omega/\kappa$: $1/20\pi$ (solid curve), $1/2\pi$ (dashed curve), $1/3$ (dot-dashed curve) and $1/2$ (dotted curve).
\label{fig:HighTemp}}
\end{figure}

In Figure \ref{fig:HighTemp}, the same observables as in Figs. \ref{fig:LowTemp} and \ref{fig:DeltaLowTemp} are plotted, but the environment temperature has been fixed at the higher value of $T_{\Psi}/\kappa = \pi/2$ (about $10$ times the Hawking temperature).
We now see a change in the behavior of the spectrum, with the relative difference now increasing both with $\lambda^{2}$ and with $\omega$, and the low-frequency temperature being higher than the standard Hawking value.  
As before, we take our measurement of the particle number at $\kappa x/D = 5$, which is in the flat asymptotic region, but close enough to the horizon so that the coupling to the environment has the least amount of time to act as the quasiparticles propagate outwards from the emission region.
We also perform the same correction described in \S\ref{lowtemp}, multiplying $\delta n$ by $\mathrm{exp}\left(\int_{x^{\mathrm{em.}}}^{x^{\mathrm{det.}}} dx^{\prime} \, \Im k_{\omega}(x^{\prime}) \right)$ to account for the field-environment interaction that occurs between the emission region and the detector.
We adopt the same rule for $x^{\mathrm{em.}}$ as in \S\ref{lowtemp}, taking it as the point where the integrand of Eq. (\ref{eq:Gac_Nsub}) and Fig. \ref{fig:integrand} drops to $1/5$ of its maximum value.
The corrected plot (shown as dots) shows that, for small values of $\lambda^2$, the increase in $n_{\mathrm{sub}}$ at high frequencies is largely due to injection of quasiparticles by the environment during propagation, rather than any increase in emission from the black hole. 
From the right upper plot, when sending $\lambda^2 \to 0$, we see that for all $\omega/\kappa$, the limiting value of the mean occupation number in our dissipative model is that found by Hawking. 
In other words, even if the environment temperature $T_\Psi$ is much higher than the Hawking temperature, if the dissipative scale is sufficiently high (with respect to that temperature) the outgoing spectrum is still characterized to very good approximation by the standard Hawking temperature, as if one were in the Unruh vacuum.

The trends in the nonseparability parameter $\Delta$ remain similar to the low temperature case: it increases with both $\lambda^{2}$ and $\omega$, though we are now able to see a loss of nonseparability where $\Delta = 0$.  However, the main result of Fig. \ref{fig:HighTemp} is that there exists, even at this high temperature, a wide window over which $\Delta<0$ and where the Hawking radiation remains quantum in character. 
The origin of this is to be found in the fact that, for parameters in this window, there is a large redshift in the temperature in the near-horizon region which takes place {\it before} the pairs separate. 
As a result, the state is close to the standard Unruh vacuum when the pair creation takes place.


\subsection{The strength of $u - v$ correlations}
\label{uv-corr}

\begin{figure}
\subfloat{\includegraphics[width=0.45\columnwidth]{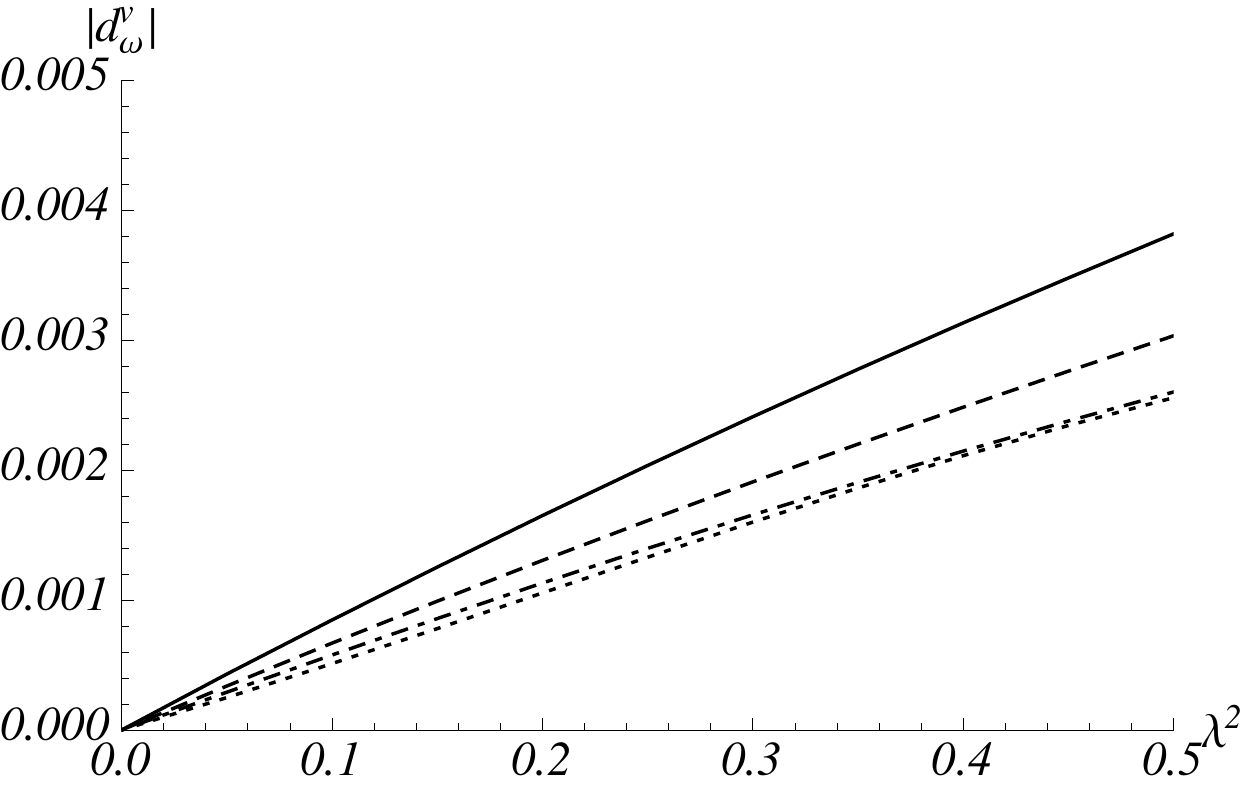}} \qquad \subfloat{\includegraphics[width=0.45\columnwidth]{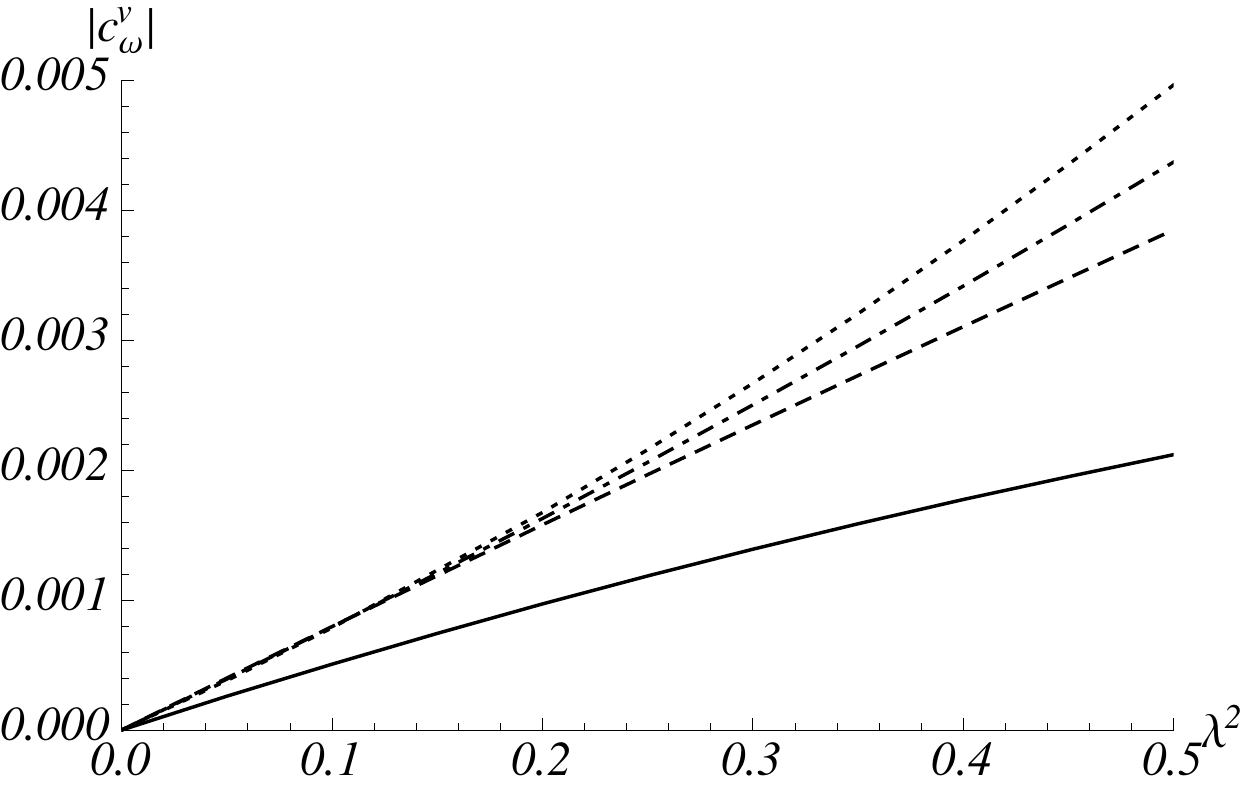}}
\caption{\textsc{$u$-$v$ correlations}: Here are plotted the magnitudes of the two $u$-$v$ correlations, $d^v_\omega$ (left panel) and $c^v_\omega$ (right panel).
The geometry is fixed at $D=0.5$, and the temperature at $T_{\Psi}/\kappa = 10^{-2}$.
The various curves correspond to different frequencies: $\omega/\kappa = 1/20\pi$ (solid), $1/2\pi$ (dashed), $1/3$ (dot-dashed) and $1/2$ (dotted).
These are much smaller than the correlations between the two $u$-modes, which is of order $n_\omega$ as can be deduced from the smallness of $\Delta_\omega/n_{\omega}$ in Figures~\ref{fig:DeltaLowTemp} and \ref{fig:HighTemp}.
\label{fig:uv-correlations}}
\end{figure}

Another subdominant effect concerns the strength of the correlations between outgoing $u$- and $v$-modes.
It is shown in Figure \ref{fig:uv-correlations}.
There are two such correlations, $d_{\omega}^{v}$ and $c_{\omega}^{v}$, corresponding to the amplitudes of the terms $\phi^{\omega}_{uR}(x) \left(\phi^{\omega}_{v}(x^{\prime})\right)^{\star}$ and $\left(\phi^{-\omega}_{uL}(x)\right)^{\star} \left(\phi^{\omega}_{v}(x^{\prime})\right)^{\star}$, respectively, in the full expression for $G^{\omega}_{\mathrm{ac}}(x,x^{\prime})$, see \eq{eq:Gac_decomp_nondiss}. 
The strength of these correlations is always much smaller than that of the $u-u$ correlations encoded in $c_\omega$.
It is seen to increase approximately linearly in $\lambda^{2}$.
They behave in opposite ways regarding their $\omega$-dependence: as $\omega$ increases, $\left|d_{\omega}^{v}\right|$ decreases while $\left|c_{\omega}^{v}\right|$ increases. 
It should be emphasized that the strength of the $u - v$ correlations has a severe impact on the nonseparability of the $u - u$ pairs, see~\cite{Busch-Parentani-2014}. 
We strongly suspect that the robustness of the nonseparability obtained in the present model is to a large extent due to the smallness of the $u - v$ correlations.


\section{The case $\Gamma$ independent of $P$
\label{app:Gamma_constant}}

In this appendix, we consider the case where $n=0$ in Eq. (\ref{eq:final_phi_eqn}), so that the dissipative rate $\Gamma$ is independent of the momentum $P$, and is in fact simply equal to $g^{2}/2$.
Then, after Fourier transforming in time, the mode equation (\ref{nmodes}) becomes
\begin{equation}
\Box_{x}^{\omega} \phi^{\omega} = \left[ -i\left(\omega+i\Gamma\right) + \partial_{x} \left( v+1 \right) \right] \left[ -i\left(\omega+i\Gamma\right) + \left( v-1\right) \partial_{x} \right] \phi^{\omega} = 0 \,,
\label{eq:GammaConst_modes}
\end{equation}
while its dual is
\begin{equation}
\widetilde{\Box}_{x}^{-\omega} \widetilde{\phi}^{-\omega} = \left[ i\left(\omega+i\Gamma\right) + \partial_{x} \left(v-1\right) \right] \left[ i\left(\omega+i\Gamma\right) + \left(v+1\right)\partial_{x} \right] \widetilde{\phi}^{-\omega} = 0 \,.
\label{eq:GammaConst_dual_modes}
\end{equation}
Note that we can remove the first factor in square brackets from each of Eqs. (\ref{eq:GammaConst_modes}) and (\ref{eq:GammaConst_dual_modes}) to get simpler, first order differential equations whose solutions are automatically also solutions of (\ref{eq:GammaConst_modes}) and (\ref{eq:GammaConst_dual_modes}).
Given the sign appearing in the $v\pm 1$ term, these solutions are $v$-modes for (\ref{eq:GammaConst_modes}) and dual $u$-modes for (\ref{eq:GammaConst_dual_modes}).  
As mentioned in \S\ref{sub:Field_eqns}, the ordering of $v+1$ and $v-1$ is irrelevant when $\Gamma$ is independent of $x$, in which case all solutions are solutions of simpler first order equations.  
However, in the general case where $\Gamma$ is $x$-dependent, this is only true of the solutions singled out above.  
So, while we cannot give an explicit general expression for the $u$-mode $\phi^{\omega}_{u}$, we can for its dual:
\begin{equation}
\widetilde{\phi}^{-\omega}_{u}(x) = \mathrm{exp}\left(-i\int^x \frac{\omega+i\Gamma}{v+1} dx' \right) \equiv \mathrm{exp}\left(-i\omega x_{u}\right) \times A_{u}\,, 
\label{eq:GammaConst_dual_u-mode}
\end{equation}
where we have defined
\begin{alignat}{2}
x_{u} = \int^x \frac{dx'}{v+1} \,, & \qquad A_{u} = \mathrm{exp}\left(\int^x \frac{\Gamma(x')}{v+1} dx' \right) \,. 
\label{eq:GammaConst_xu-amplitude}
\end{alignat}
This explains the particular ordering chosen in Eqs. (\ref{eq:GammaConst_modes}) and (\ref{eq:GammaConst_dual_modes}): since it is the dual mode $\widetilde{\phi}^{-\omega}_{u}(x)$ that is integrated over to find the anticommutator, we wish for it to have the simple expression (\ref{eq:GammaConst_dual_u-mode}).  
Although the $u$-mode equation is first order, it does have two independent solutions due to the divergence of the integrand in Eq. (\ref{eq:GammaConst_dual_u-mode}) at the horizon where $v+1=0$; since $\Gamma>0$, $\widetilde{\phi}^{-\omega}_{u}$ must vanish there.  
So, as for the $\Gamma \propto P^{2}$ model studied in the main body of this paper, there are two distinct $u$-modes localized on opposite sides of the horizon, which we label $\phi^{\omega}_{uR}$ and $\phi^{\omega}_{uL}$, and their dual modes $\widetilde{\phi}^{-\omega}_{uR}$ and $\widetilde{\phi}^{-\omega}_{uL}$.  

\begin{figure}
\includegraphics[width=0.8\columnwidth]{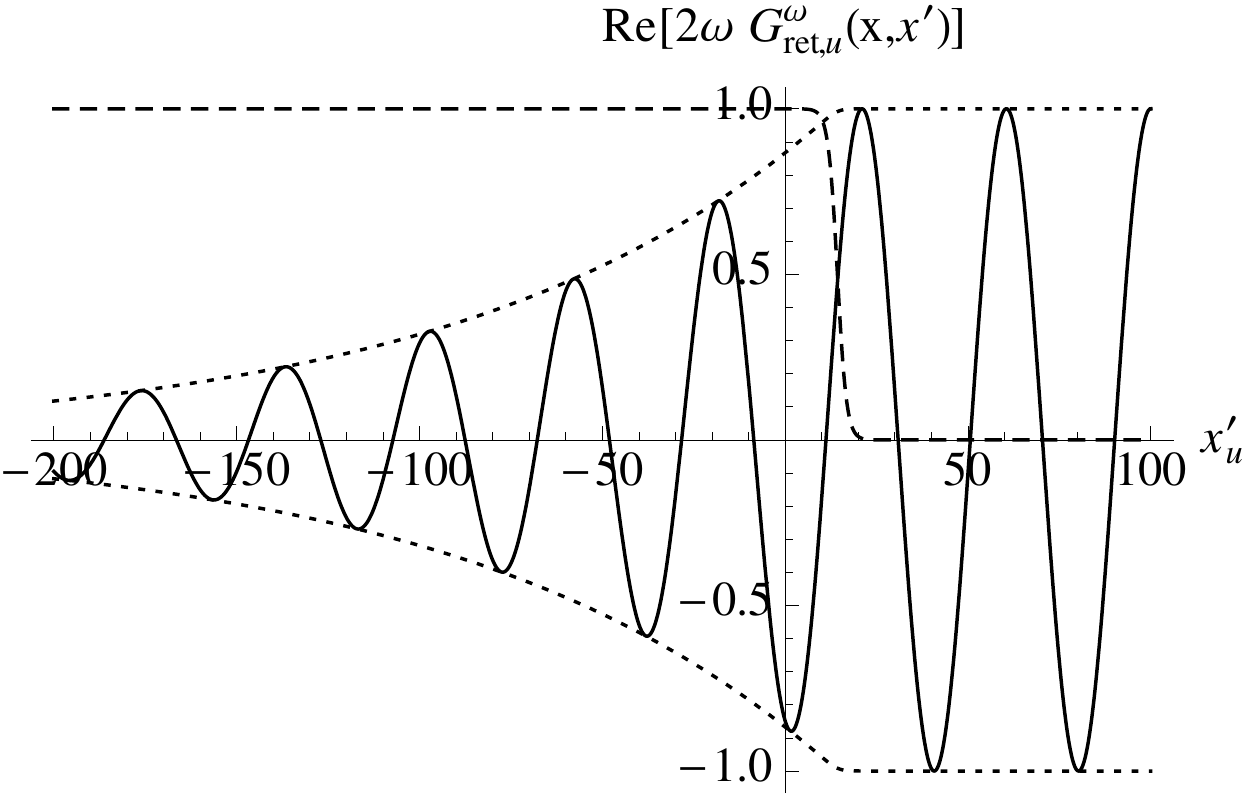}
\caption{\textsc{Green function with $P$-independent $\Gamma$}: Here is plotted, for $\Gamma$ independent of the momentum, the $u$-part of the retarded Green function, $G^{\omega}_{\mathrm{ret,u}}(x,x^{\prime})$ as a function of $x^{\prime}$ with $x$ fixed (solid curve).
For convenience we have used the transformed (tortoise) coordinate $x_{u}$ of Eqs. (\ref{eq:GammaConst_xu-amplitude}), which in the near-horizon region behaves like $x_{u} \approx \mathrm{log}(x)/\kappa$.  
We have taken $\omega/\kappa = 1/2\pi$, $D=0.2$, and $x$ such that $x_{u} = 100$.  
The dashed line plots $\Gamma/\Gamma_{\mathrm{hor}}$, where $\Gamma_{\mathrm{hor}}$ is the value of $\Gamma$ at $x=0$; here, we have taken $\Gamma_{\mathrm{hor}}/\kappa = 10^{-2}$.  
The dotted lines plot $\pm A_{u}$ of Eqs. (\ref{eq:GammaConst_xu-amplitude}), and since this is $\omega$-independent, changing the frequency would change the wavelength in $x_{u}$ but not the envelope, which behaves like $x^{\Gamma_{\mathrm{hor}}/\kappa}=\mathrm{exp}\left(\Gamma_{\mathrm{hor}} x_{u}\right)$ in the near-horizon region.
\label{fig:GretGammaConst}}
\end{figure}

The analysis of the main paper progresses in much the same way, and for the anticommutator, Eq. (\ref{eq:Gac_Nsub}) becomes
\begin{multline}
G^{\omega}_{\mathrm{ac}}(x,x^{\prime}) = \frac{1}{\pi} \int_{-\infty}^{+\infty} dx_{1} \int_{-\infty}^{+\infty} dx_{2} \left\{ \sqrt{\frac{\Gamma(x_{1})}{\left|v(x_{1})\right|}} e^{i\omega\tau_{0}(x_{1})} G^{\omega}_{\mathrm{ret}}(x,x_{1}) \cdot \partial_{x_{2}}\left[ \sqrt{\frac{\Gamma(x_{2})}{\left|v(x_{2})\right|}} v(x_{2}) e^{-i\omega\tau_{0}(x_{2})} G^{-\omega}_{\mathrm{ret}}(x^{\prime},x_{2}) \right] \right. \\
\left. - \partial_{x_{1}}\left[ \sqrt{\frac{\Gamma(x_{1})}{\left|v(x_{1})\right|}} v(x_{1}) e^{i\omega\tau_{0}(x_{1})} G^{\omega}_{\mathrm{ret}}(x,x_{1}) \right] \cdot \sqrt{\frac{\Gamma(x_{2})}{\left|v(x_{2})\right|}} e^{-i\omega\tau_{0}(x_{2})} G^{-\omega}_{\mathrm{ret}}(x^{\prime},x_{2}) \right\} \times \pi T_{\Psi} \mathrm{coth}\left(\pi T_{\Psi} \left(\tau_{0}(x_{1}) - \tau_{0}(x_{2}) \right) \right) \,.
\label{eq:GammaConst_Gac}
\end{multline}
There are two differences with respect to Eq. (\ref{eq:Gac_Nsub}): there are now no derivatives acting on the Green functions, since there are no derivatives in the $n=0$ version of Eq. (\ref{eq:Gac_from_Gret}); and $\gamma$ has been replaced by $\sqrt{\Gamma(x_{1})\Gamma(x_{2})} = g(x_{1})g(x_{2})/2$, since $g$ is not now assumed to be $x$-independent.  
The reason we have allowed $g$, or equivalently $\Gamma$, to depend on $x$ can be appreciated by examining Eq. (\ref{eq:GammaConst_Gac}) and comparing it with Eq. (\ref{eq:Gac_Nsub}).  
We expect the main contribution to the integrand for $G_{\mathrm{ac}}^{\omega}$ to come from the near-horizon region.  
In the $\Gamma \propto P^{2}$ case, Eq. (\ref{eq:Gac_Nsub}), this is achieved via the appearance of the $x$-derivative of $G^{\omega}_{\mathrm{ret}}$, which is largest around $x=0$.  
In Eq. (\ref{eq:GammaConst_Gac}), however, setting $\Gamma$ to be constant everywhere means that all regions contribute equally, and it becomes difficult to distinguish the contribution from the black hole and that from the local environment.  
Another way of saying this is that, while the corrections to the Hawking spectrum should be of order $O\left(\Gamma_{\mathrm{hor}}/\omega\right)$, where $\Gamma_{\mathrm{hor}}$ is the value of $\Gamma$ in the near-horizon region, the quasiparticle number in the asymptotic region is only well-defined up to a correction of order $O\left(\Gamma_{\mathrm{as}}/\omega\right)$, where $\Gamma_{\mathrm{as}}$ is the asymptotic value of $\Gamma$; therefore, if $\Gamma_{\mathrm{hor}}=\Gamma_{\mathrm{as}}$, it is not possible to give a meaningful correction to the Hawking spectrum due to dissipative effects.  
We thus introduce an asymptotic switch-off of $\Gamma$, so that $\Gamma_{\mathrm{as}}=0$; yet, while this means that the asymptotic fluxes are unambiguously defined for any given $v$ and $\Gamma$ profiles, it also introduces extra degrees of freedom in the precise form of $\Gamma$.  
We demand that $\Gamma(x)$ satisfy two criteria: it is constant in the near-horizon region, and it falls off to zero at a rate slower than $\kappa$, so that spectral corrections due to a non-adiabatic switch-off are negligible.

\begin{figure}
\subfloat{\includegraphics[width=0.45\columnwidth]{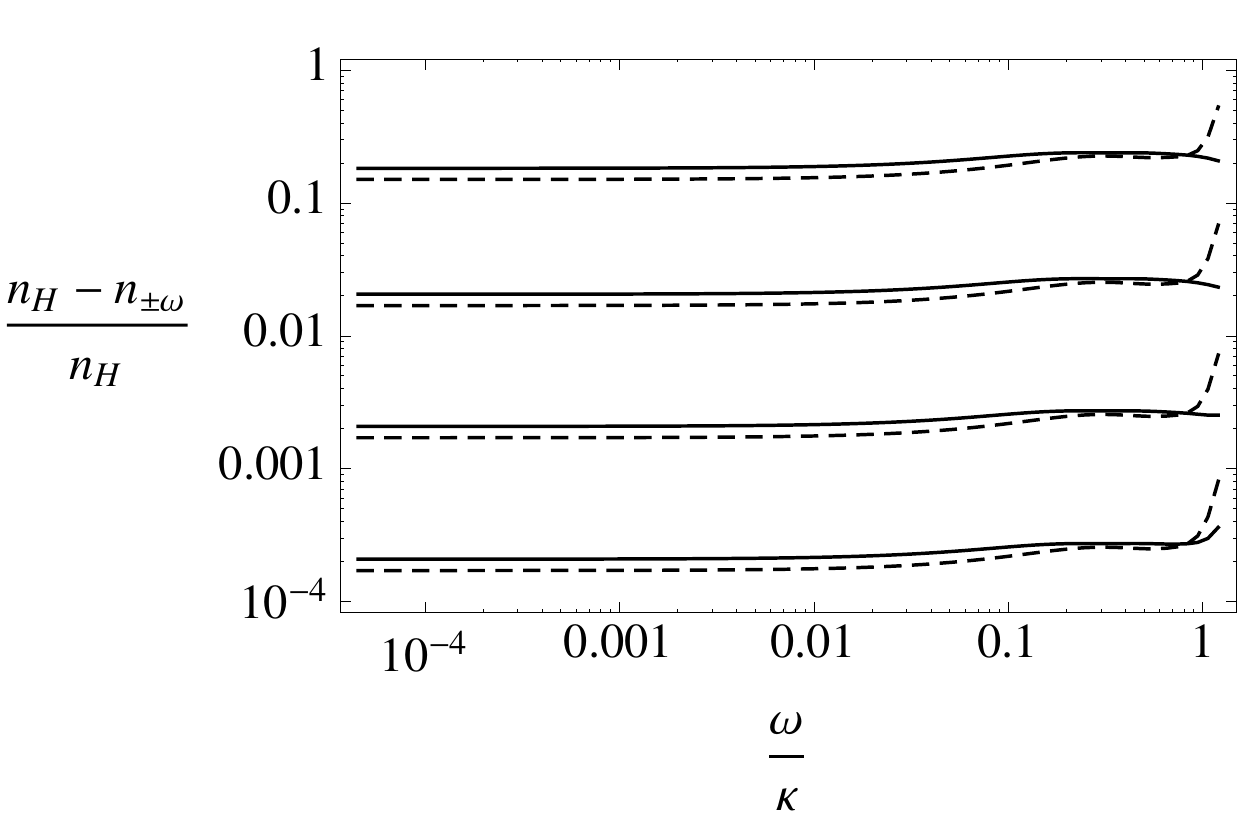}} \qquad \subfloat{\includegraphics[width=0.45\columnwidth]{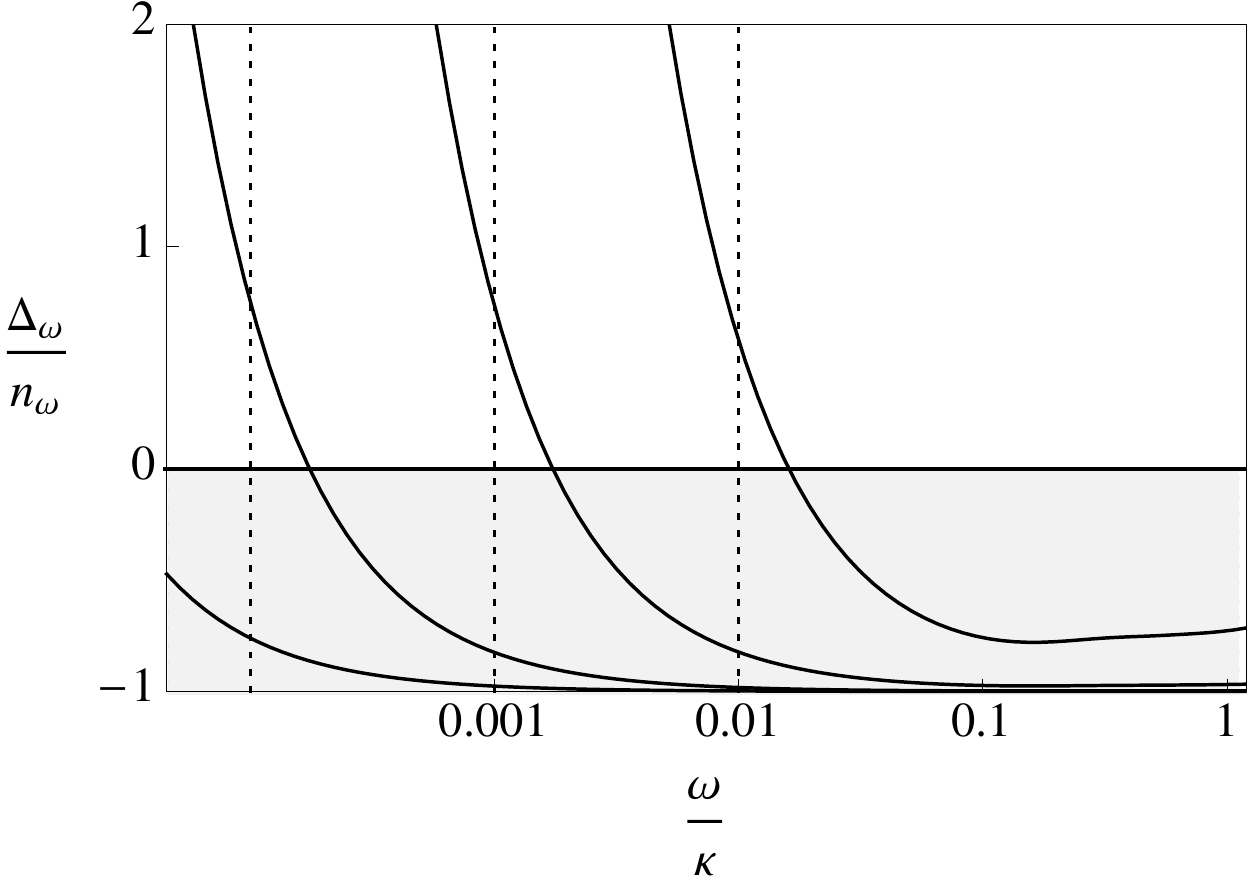}}
\caption{\textsc{Results for $P$-independent $\Gamma$}: On the left is plotted the relative change in the spectrum from the standard Hawking prediction, for quasiparticles emitted in the subsonic (solid curves) and supersonic (dashed curves) regions, for four different values of $\Gamma_{\mathrm{hor}}/\kappa$: $10^{-5}$, $10^{-4}$, $10^{-3}$ and $10^{-2}$.
On the right is shown the corresponding value of $\Delta_{\omega}/n_{\omega}$, for the same values of $\Gamma_{\mathrm{hor}}$; the vertical dotted lines show where $\omega = \Gamma$, to show that the nonseparability threshold where $\Delta_{\omega}$ vanishes varies as $\Gamma$, while the shaded region shows where $\Delta_{\omega}<0$ and thus where the state is nonseparable. 
For all plots, the geometry is fixed at $D = 0.2$ and the environment temperature at $T_{\Psi}/\kappa = 10^{-2}$.
\label{fig:GammaConst}}
\end{figure}

Results for a low environment temperature $T_{\Psi}/\kappa = 10^{-2}$ are shown in Figure \ref{fig:GammaConst}.  
Interestingly, just as for the $\Gamma \propto P^{2}$ model, we find that the corrections to the Hawking spectrum amount to a global change of temperature in the low-frequency regime $\omega/\kappa \ll 1$, where the relative change in the spectrum (in fact a reduction) is proportional to $\Gamma_{\mathrm{hor}}/\kappa$.
The coefficient is approximately 20.
Thus the robustness of the spectrum is again recovered when the following inequality is satisfied: $\Gamma_{\mathrm{hor}}/\kappa \lesssim 1/20$. 

The differences from the $\Gamma \propto P^{2}$ model show up clearly when we look at the impact of dissipation on the strength of the correlations.
When studying the parameter $\Delta_{\omega}$, we find that the pairs of $u$-quanta are nonseparable at {\it high} frequencies and separable at {\it low} frequencies, with the changeover occurring around $\omega \sim \Gamma_{\mathrm{hor}}$. 
This result is relevant for experiments in polariton systems which aim to observe the spontaneous channel~\cite{Gerace-Carusotto-2012,prl-Marcoussis}.

\bibliography{bibliopubli}

\end{document}